\documentclass[pre,aps,showpacs,preprint]{revtex4}

\usepackage{amsmath}
\usepackage{amsfonts}
\usepackage{graphicx}
\usepackage{mathrsfs}
\usepackage{rotating}
\usepackage{setspace}
\usepackage{supertabular}
\usepackage{amsthm}
\usepackage{subfigure}
\usepackage{multirow}
\usepackage{mciteplus}

\begin{document}
\newtheorem{Theorem1}{Theorem}

\title{Densest binary sphere packings}

\author{Adam B. Hopkins and Frank H. Stillinger}
\affiliation{Department of Chemistry, Princeton University, Princeton, New 
Jersey 08544}

\author{Salvatore Torquato}
\affiliation{Department of Chemistry, Department of Physics, Princeton Center
for Theoretical Science, 
Princeton Institute for the Science and Technology of Materials, Program in
Applied and Computational Mathematics, Princeton University, Princeton, New 
Jersey, USA, 08544}

\begin{abstract}
The densest binary sphere packings in the $\alpha$-$x$ plane of small to large
sphere radius ratio $\alpha$ and small sphere relative concentration $x$ have
historically been very difficult to determine. Previous research had led to the
prediction that these packings were composed of a few known ``alloy'' phases
including, for example, the AlB$_2$ (hexagonal $\omega$), HgBr$_2$, and AuTe$_2$
structures, and to XY$_n$ structures composed of close-packed large spheres with
small spheres (in a number ratio of $n$ to $1$) in the interstices, {\it e.g.},
the NaCl packing for $n=1$. However, utilizing an implementation of the
Torquato-Jiao sphere-packing algorithm [S. Torquato and Y. Jiao, Phys. Rev. E
{\bf 82}, 061302 (2010)], we have discovered that many more structures appear in
the densest packings. For example, while all previously known densest structures
were composed of spheres in small to large number ratios of one to one, two to
one, and very recently three to one, we have identified densest structures with
number ratios of seven to three and five to two. In a recent work [A. B.
Hopkins, Y. Jiao, F. H. Stillinger, and S. Torquato, Phys. Rev. Lett. {\bf 107},
125501 (2011)], we summarized these findings. In this work, we present the
structures of the densest-known packings and provide details about their
characteristics. Our findings demonstrate that a broad array of different
densest mechanically stable structures consisting of only two types of components can form
without any consideration of attractive or anisotropic interactions. In addition, 
the novel structures that we have identified may correspond to currently 
unidentified stable phases of certain binary atomic and molecular systems, 
particularly at high temperatures and pressures.
\end{abstract}


\maketitle

\section{Introduction}
A packing is defined as a set of nonoverlapping objects arranged in a space of
given dimension $d$, and its packing fraction $\phi$ is the fraction of space
that the objects cover. Packings of spheres in $d$-dimensional Euclidean space
${\mathbb R}^d$ are frequently used as starting points in modeling atomic,
molecular, and granular materials consisting of particles exhibiting strong
repulsive pair interactions at small particle separations. In particular, the
densest sphere packings in ${\mathbb R}^d$, or packings with maximal packing
fraction $\phi_{max}$, often correspond to ground states of systems of particles
with pairwise interactions dominated by steep isotropic pairwise repulsion
\cite{TS2010a}. For example, see the ground states in Refs.
\cite{Pollack1964a,Sanders1980a,BOP1992a,VHVE1997a,MRMW2001a,AG2010a}.

Packings of identical spheres have been employed in ${\mathbb R}^3$ to describe 
the structures and some fundamental properties of a diverse range of substances 
from crystals and colloids to liquids, amorphous solids and glasses 
\cite{TorquatoRHM2002,CLPCMP1995,HMTSL2006,ZallenPAS1983}. Despite their 
simplicty, complicated structures and interesting properties can arise in packings 
of identical spheres through simple principles like density maximization in a 
confined space \cite{HST2009a,HST2010a,*HST2010b,*HST2011a}. Packings of identical 
nonspherical objects in ${\mathbb R}^3$ have also been studied, though not to the 
extent of sphere packings, and have applications, for example, in the 
self-assembly of colloids and nanoparticles \cite{DSCT2004a,JST2009a,*BST2010a,TJ2009a,*TJ2009b,*TJ2010b,Chen2008a,*CEG2010a,KEG2010a,GRD2011a,JT2011a}. 
In structural biology, molecular dynamics simulations of interactions between large 
numbers of molecules employ chains of identical nonoverlapping spheres as models 
for various biological structures such as proteins and lipids 
\cite{DD2005a,*Dokholyan2006a,*DND2007a}, and packing of nonspherical objects of 
different sizes are used, for example, in tumor growth modeling \cite{DH2005a,GGT2008a,*GT2008a}.

Packings of spheres of many different sizes have been employed as structural 
models for materials such as solid propellants and concrete, among others 
\cite{KJB2001a,*KBJH2001a,*MSJ2008a,KRW2010a}. Dense binary packings of spheres, 
packings of spheres of only two sizes, have long been employed as models for 
the structures of a wide range of alloys \cite{SMA1969,Sanders1980a,MS1980a,SVC1982a,DA1990a,EMF1993a,*EMF1993b,*EMF1993c,CM1993a,*CM1995a,WM2005a}. 
In this work, we focus on binary sphere packings. Though they are relatively 
simple models, much about them is still unknown due in part to the enormous binary 
sphere-packing parameter space of sphere size ratio and relative concentration.
Recently, we presented the most comprehensive determination to date, to the best of our knowledge, of the ``phase diagram'' in small to large sphere size ratio $\alpha$ and small sphere relative
concentration $x$ for the densest binary sphere packings in ${\mathbb R}^3$
\cite{HJST2011a}. In the present paper, we extend these results to produce a new
phase diagram, and we present a theorem and proof concerning the number of
phases present in a densest packing with $\eta$ different types of objects.
Additionally, we present detailed structural descriptions of the alloy phases
present in the densest packings, identified using the Torquato-Jiao (TJ)
determinental sphere-packing algorithm \cite{TJ2010a}, including descriptions of
many that were heretofore unknown. Here we use the term ``alloy'' in a general
sense to mean a structure composed of two or more distinguishable components
that are not phase-separated.

\begin{figure}[b]
\centering
\subfigure[]{\includegraphics[width=3.0in]{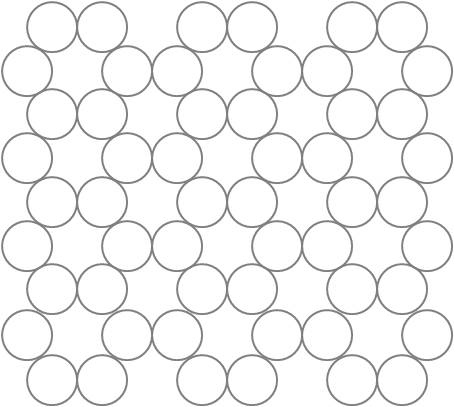}} \\
\subfigure[]{\includegraphics[width=3.0in]{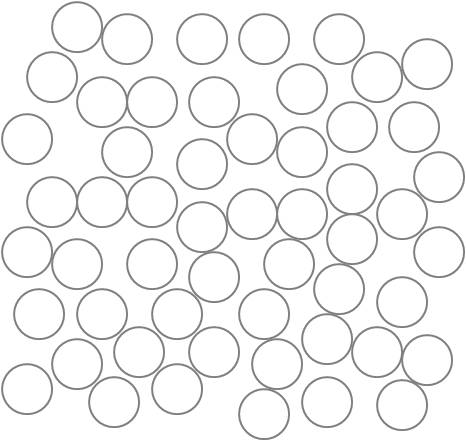}}
\caption{Illustrations of subsections of packings of disks that are a) periodic
and b) aperiodic.}
\label{orderedVSdisordered}
\end{figure}

Structures, or configurations of points, can be classified as either periodic or
aperiodic. Roughly defined, a {\it periodic} structure (packing) is one
consisting of a certain number of points (sphere centers), called the {\it
basis}, placed in a defined region of space, the {\it unit cell}, replicated
many times over such that the cells cover all space without any overlap between
cells (or spheres). A {\it fundamental cell} is one with {\it minimal basis},
i.e., such that a smaller fundamental cell and basis with the same periodic
structure does not exist. An {\it aperiodic} structure is one with an infinite
(or in practice, very large finite) minimal basis. By definition, it is clear
that a binary sphere packing must have at least two spheres in its minimal
basis. Figure \ref{orderedVSdisordered} is an illustration that compares
subsections of periodic and aperiodic packings of monodisperse disks. 

The packing fraction $\phi_{max}(\alpha,x)$ of the densest binary packings of
spheres in ${\mathbb R}^3$ is a function only of small to large sphere radius ratio
$\alpha = R_S/R_L$, with $R_S$ and $R_L$ the respective radii of the small and
large spheres, and small sphere relative concentration $x$. Specifically, one
has,
\begin{equation}
x \equiv \frac{N_S}{N_S+N_L},
\label{concentration}
\end{equation}
with $N_S$ and $N_L$ the respective numbers of small and large spheres in the
packing, and when $N_S + N_L \rightarrow \infty$ in the infinite volume limit,
$x$ remains constant. It is implicit in the definition of $\phi_{max}(\alpha,x)$ 
that radii are additive, {\it i.e.}, two spheres of different radii can be no 
closer than distance $R_S + R_L$ from one another. In this work, we focus on the 
additive case, though we note that the TJ algorithm can be trivially modified to study 
sphere-packings with non-additive diameters. Past studies of certain sphere 
packings with non-additive diameters can be found, for example, in Ref. \cite{SV1987a}.

The densest packings in ${\mathbb R}^3$ are composed of a countable number of
distinct phase-separated alloy and monodisperse phases. Previously, the only
alloys thought to be present in the densest packings for $\alpha > \sqrt{2} -1 =
0.414213\dots$ corresponded to structures such as the A$_3$B, AlB$_2$ (hexagonal
$\omega$), HgBr$_2$, and AuTe$_2$ structures
\cite{OH2011a,FD2009a,LH1993a,MS1980a}, and to a structure composed of equal
numbers of small and large spheres \cite{MH2010a}. For $\alpha \leq \sqrt{2}-1$,
the alloys thought to be present were XY$_n$ structures of close-packed large
spheres with small spheres (in a ratio of $n$ to $1$) in the interstices, {\it
e.g.}, the NaCl packing for $n=1$. However, in a recent work \cite{HJST2011a},
we showed that in addition to these alloys, many more are present in the
putative densest packings, including several with heretofore unknown
structures.

The densest binary packings of spheres can be directly relevant to atomic and 
molecular phases in binary solids and compounds. For example, structures such 
as that exhibited by AlB$_2$ have been predicted and observed to be present in 
high temperature and pressure phases of various binary intermetallic and 
rare-gas compounds \cite{DLRS2006a,CES2009a}. Furthermore, phase separation of 
alloys like that present in the densest binary packings has been observed in 
certain binary atomic and molecular solids at high temperatures and pressures, 
conditions where relaxation time scales are fast and diffusion rates high 
\cite{Degtyareva2005a}. These observations indicate that the hard-sphere 
additive-radii interactions that lead to the densest binary sphere packings may 
be sufficient to determine the stable phases of many binary atomic and molecular 
solids and compounds at high temperatures and pressures. In addition, we believe 
that there are binary atomic and molecular systems, particularly at high 
temperatures and pressures, that will exhibit stable phases with structures 
corresponding to the heretofore unknown alloys described in Sec. \ref{thePackings}.

In the past, finding the densest packings has been difficult in part due to the
complexity of proving that a packing of spheres in ${\mathbb R}^d$ is the
densest possible, evident in that Kepler's Conjecture concerning the densest
packings of monodisperse spheres in ${\mathbb R}^3$ was only recently proved by
Hales \cite{Hales2005a}. In the $\alpha$-$x$ plane, the monodisperse case
corresponds to the Kepler limit $\alpha = 1$; in this limit, the packing
fraction of the densest packings is $\phi = \pi/\sqrt{18} = 0.740480\dots$. This
fraction is achieved by any of the infinite number of Barlow packings
\cite{Barlow1883a}, which include the well-known fcc and hcp sphere packings.
Some efforts have been made to identify the densest alloy packings away from the
limit $\alpha =1$ by using simple crystallographic techniques
\cite{HH2008a,MS1980a}; however, these have been limited to only a small subset
of possible periodic alloy packings.

Methods such as Monte Carlo \cite{KHH2008a,FMOPSD2009a,MH2010a} and genetic algorithms
\cite{FD2009a} have also been employed, with limited success, to attempt to find
densest binary sphere packings over certain ranges of $\alpha$ and $x$. In part
due to the enormous number of different initial sphere spatial configurations
required by these methods to identify densest packings of a large number of
spheres, high resolution searches in $\alpha$ and $x$ would previously have
required an enormous amount of computational time. As a result, the densest
alloys found by past efforts were limited to minimal bases with relatively
smaller numbers of spheres, and, aside from the XY$_n$ packings, relative
compositions of $x=1/2$, $x=2/3$, and very recently, $x=3/4$ \cite{OH2011a}. In
contrast, employing an implementation of the TJ sphere-packing algorithm
\cite{TJ2010a}, we have found several new alloys for $\alpha > \sqrt{2}-1$
including three with $12$, $10$, and $7$ spheres in their minimal bases, in
respective small to large number ratios of one to one, seven to three, and five
to two.

\begin{figure}[ht]
\centering
\includegraphics[width = 6.45in,viewport = 0 250 600 540,clip]{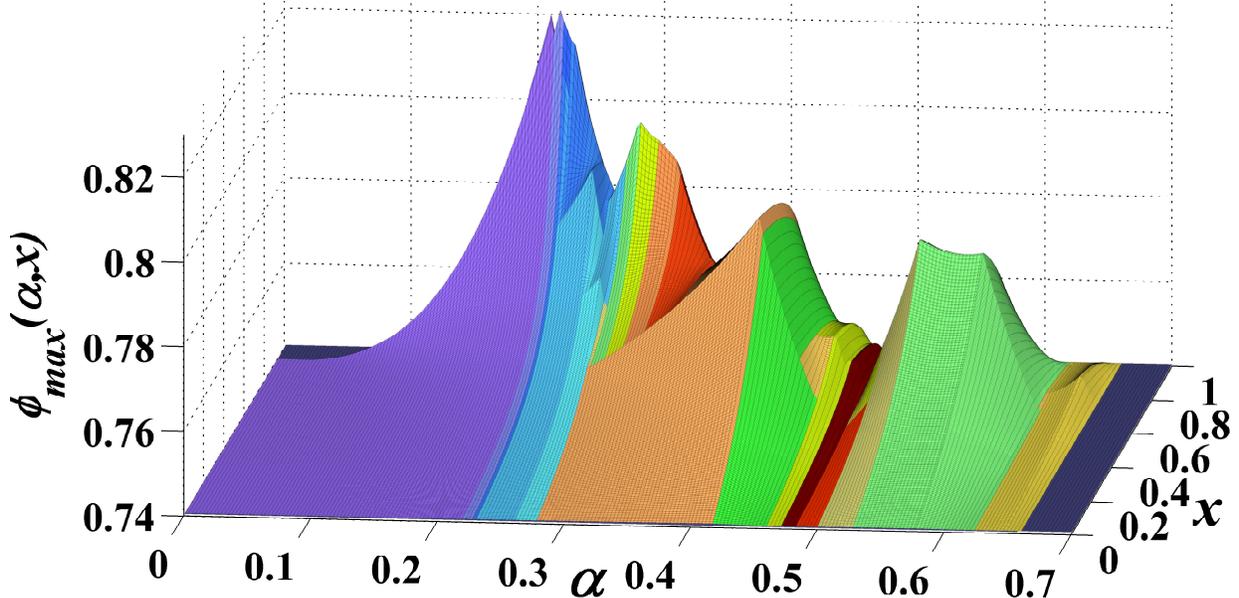}
\caption{(Color online) The most comprehensive determination to date of the phase
diagram and maximal packing fraction surface $\phi_{max}(\alpha,x)$ of the
densest binary sphere packings in ${\mathbb R}^3$. The highest point is
$\phi_{max}(0.224744\dots,10/11) = 0.824539\dots$, and all packings for
approximately $\alpha > 0.660\dots$ consist of two phase-separated monodisperse
Barlow phases. Note that we have excluded the rectangular region $\alpha <
0.20$, $x > 11/12$. Different shading indicates a different phase composition,
as specified in Fig. \ref{phaseDiagram}.}
\label{surface1}
\end{figure}

Using the TJ algorithm, an overview of which is given in Sec. \ref{algorithm},
we have systematically surveyed the parameter space $(\alpha,x) \in
[0,1]\times[0,1]$, omitting the rectangular area $\alpha < 0.2$ and $x > 11/12$
for reasons mentioned below, to find the putative densest binary packings at
high resolution in $\alpha$ and $x$ for bases of up to $12$ spheres. From this
survey, we are able to construct the most comprehensive determination to date of
the phase diagram of the densest binary packings in ${\mathbb R}^3$ in the
$\alpha$-$x$ plane, and the best known lower bound on the function
$\phi_{max}(\alpha,x)$ for the values of $\alpha$ that we survey. Though the
representation we construct is technically a lower bound on
$\phi_{max}(\alpha,x)$ as it excludes densest aperiodic packings and periodic
packings with bases greater than $12$, we contend (for reasons described in Sec.
\ref{algorithm}) that for the vast majority of $(\alpha,x)$, it is a precise
representation of $\phi_{max}(\alpha,x)$.

\begin{figure}[ht]
\centering
\includegraphics[width = 6.45in,viewport = 0 250 590 545,clip]{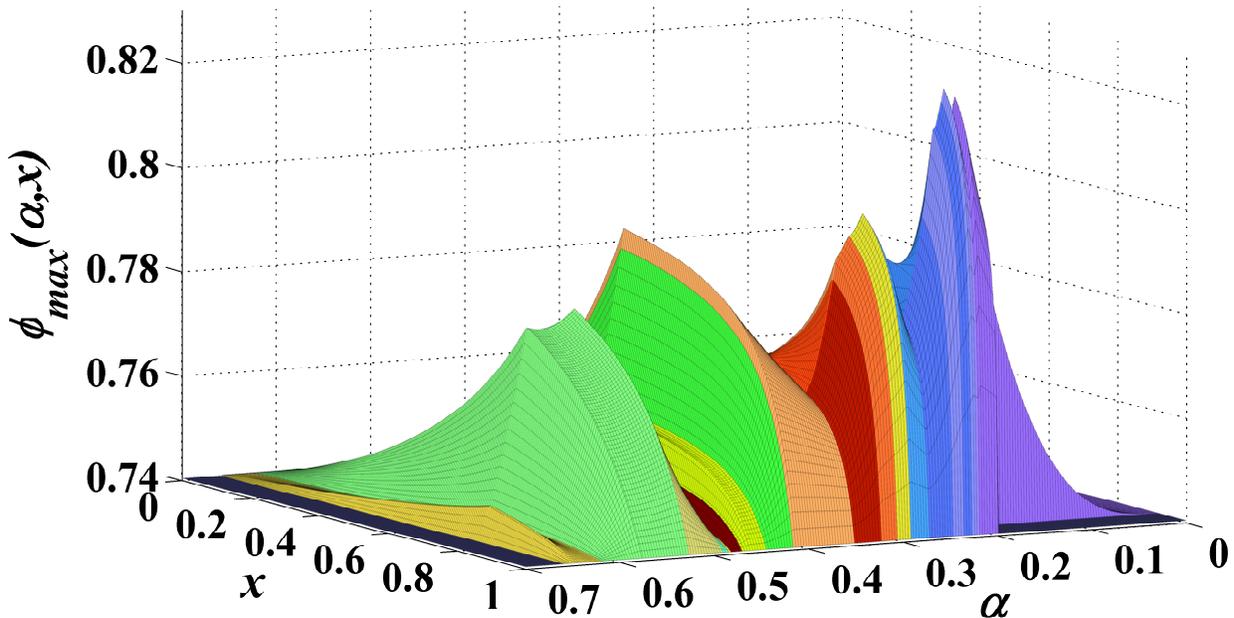}
\caption{(Color online) The most comprehensive determination to date of the phase
diagram and maximal packing fraction surface $\phi_{max}(\alpha,x)$ of the
densest binary sphere packings in ${\mathbb R}^3$. Note that we have excluded
the rectangular region $\alpha < 0.20$, $x > 11/12$. Different shading indicates
a different phase composition, as specified in Fig. \ref{phaseDiagram}.}
\label{surface2}
\end{figure}

\begin{figure*}[ht]
\centering
\includegraphics[width = 6.45in,viewport = 12 185 711 397,clip]{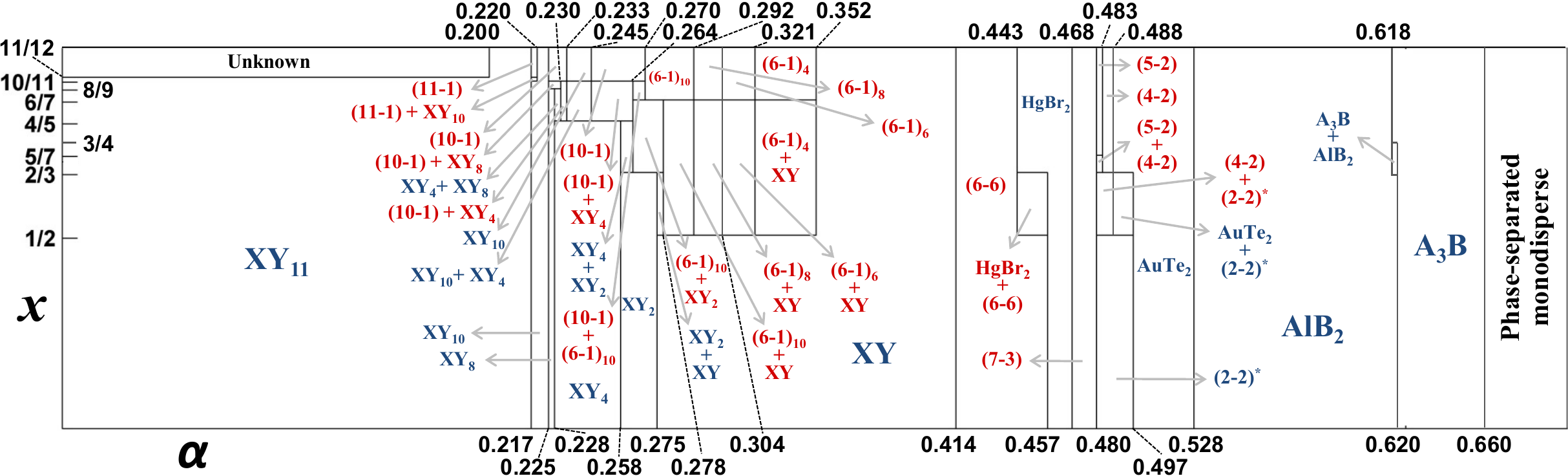}
\caption{(Color online) Phase diagram in $(\alpha,x)$, excluding the region
$\alpha < 0.2$ and $x > 11/12$, of the densest-known binary sphere packings in
${\mathbb R}^3$ considering periodic packings with minimal bases of $12$ or
fewer spheres. Figs. \ref{phaseDiagramLeft} and \ref{phaseDiagramRight} are
larger versions of this image.}
\label{phaseDiagram}
\end{figure*}

\begin{figure*}[ht]
\centering
\includegraphics[width = 6.45in,viewport = 5 138 712 460,clip]{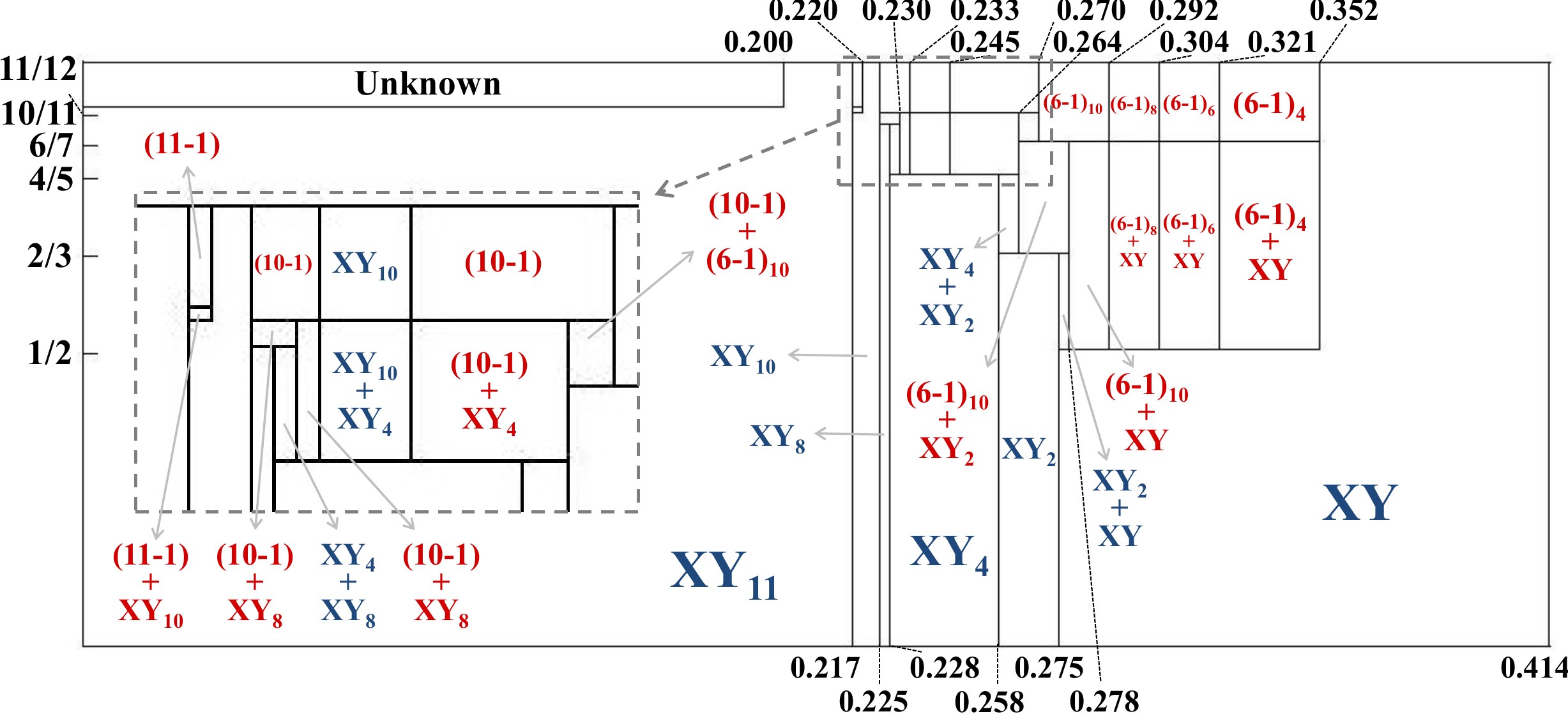}
\caption{(Color online) Phase diagram in $(\alpha,x)$ for $0 \leq \alpha \leq
\sqrt{2}-1$, excluding the region $\alpha < 0.2$ and $x > 11/12$, of the
densest-known binary sphere packings in ${\mathbb R}^3$ considering periodic
packings with minimal bases of $12$ or fewer spheres. This image is a blow-up of
the left hand side of Fig. \ref{phaseDiagram}.}
\label{phaseDiagramLeft}
\end{figure*}

\begin{figure*}[ht!]
\centering
\includegraphics[width = 6.45in,viewport = 15 150 710 450,clip]{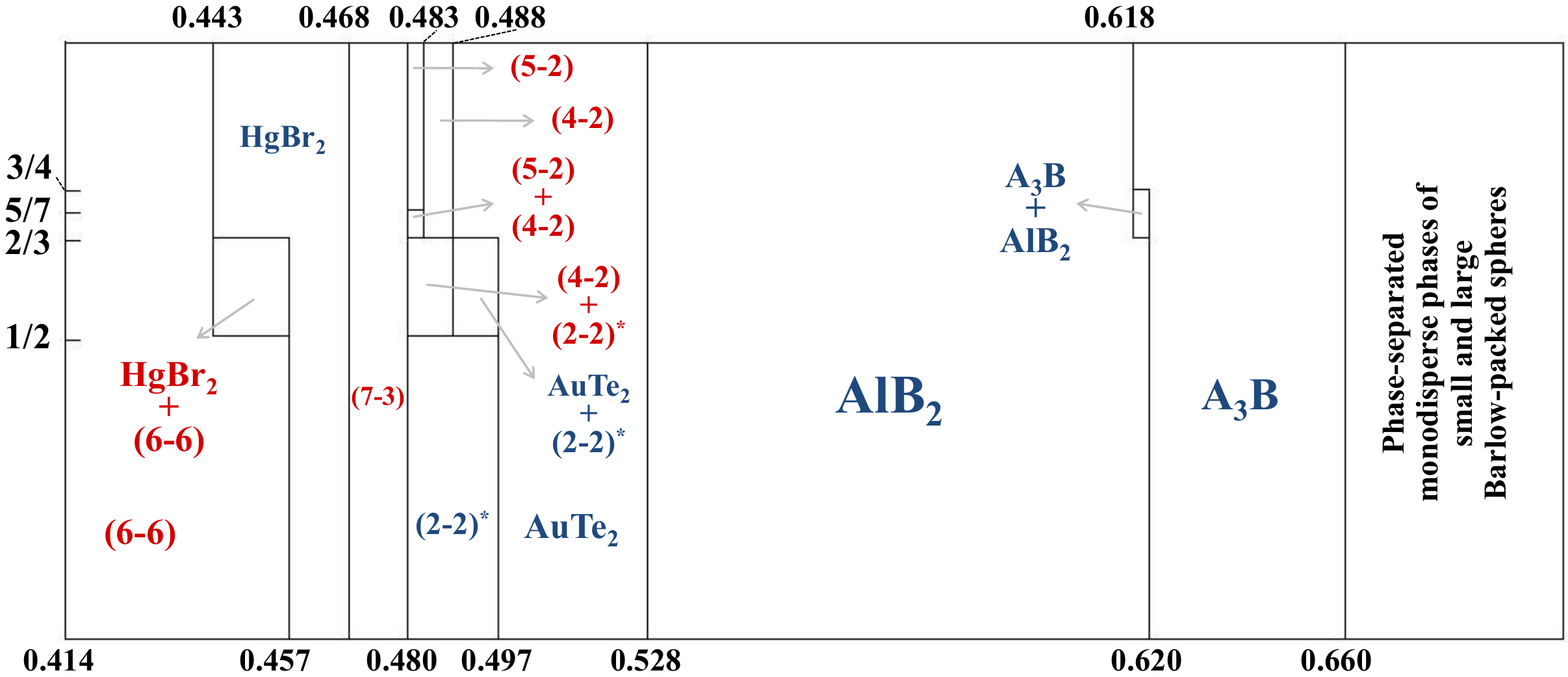}
\caption{(Color online) Phase diagram in $(\alpha,x)$ for $\sqrt{2} \leq \alpha
\leq 1$ of the densest-known binary sphere packings in ${\mathbb R}^3$
considering periodic packings with minimal bases of $12$ or fewer spheres. This
image is a blow-up of the right hand side of Fig. \ref{phaseDiagram}.}
\label{phaseDiagramRight}
\end{figure*}

We present different representations of the most comprehensive determination to
date of the phase diagram in Figs. \ref{surface1}, \ref{surface2},
\ref{phaseDiagram}, \ref{phaseDiagramLeft}, and \ref{phaseDiagramRight}. Similar
versions of Figs. \ref{surface1} and \ref{surface2} were presented in Ref.
\cite{HJST2011a}. In Fig. (\ref{phaseDiagram}), we describe heretofore unknown
alloys by the number of small and large spheres in their minimal bases, {\it
e.g.}, ($6$-$6$) for the alloy with six small and six large spheres. In both
figures, phase boundaries are drawn between packings with {\it distinct} alloys,
where each distinct alloy exhibits a unique lattice system characterization of
its fundamental cell and composition of spheres in its minimal basis. In the
figures, points (lines) where the composition of phase-separated phases changes
from alloy plus monodisperse packing of small spheres to the same alloy plus a
monodisperse packing of large spheres are not drawn. In Fig. \ref{phaseDiagram},
where only one alloy is listed, it is assumed that the densest packing consists
of a monodisperse phase and an alloy phase, except at points such that $x =
S_i/(S_i+L_i)$, with $S_i$ and $L_i$ the respective numbers of small and large
spheres in the minimal basis of the alloy phase listed, where only the alloy phase is present.

\begin{figure}[ht]
\centering
\subfigure[]{\includegraphics[width=3.5in,viewport = 0 35 790 535,clip]{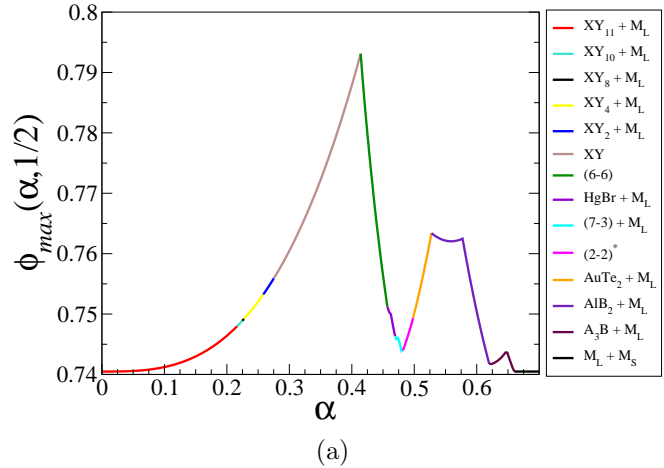}}
\\
\subfigure[]{\includegraphics[width=3.5in,viewport = 0 35 790 535,clip]{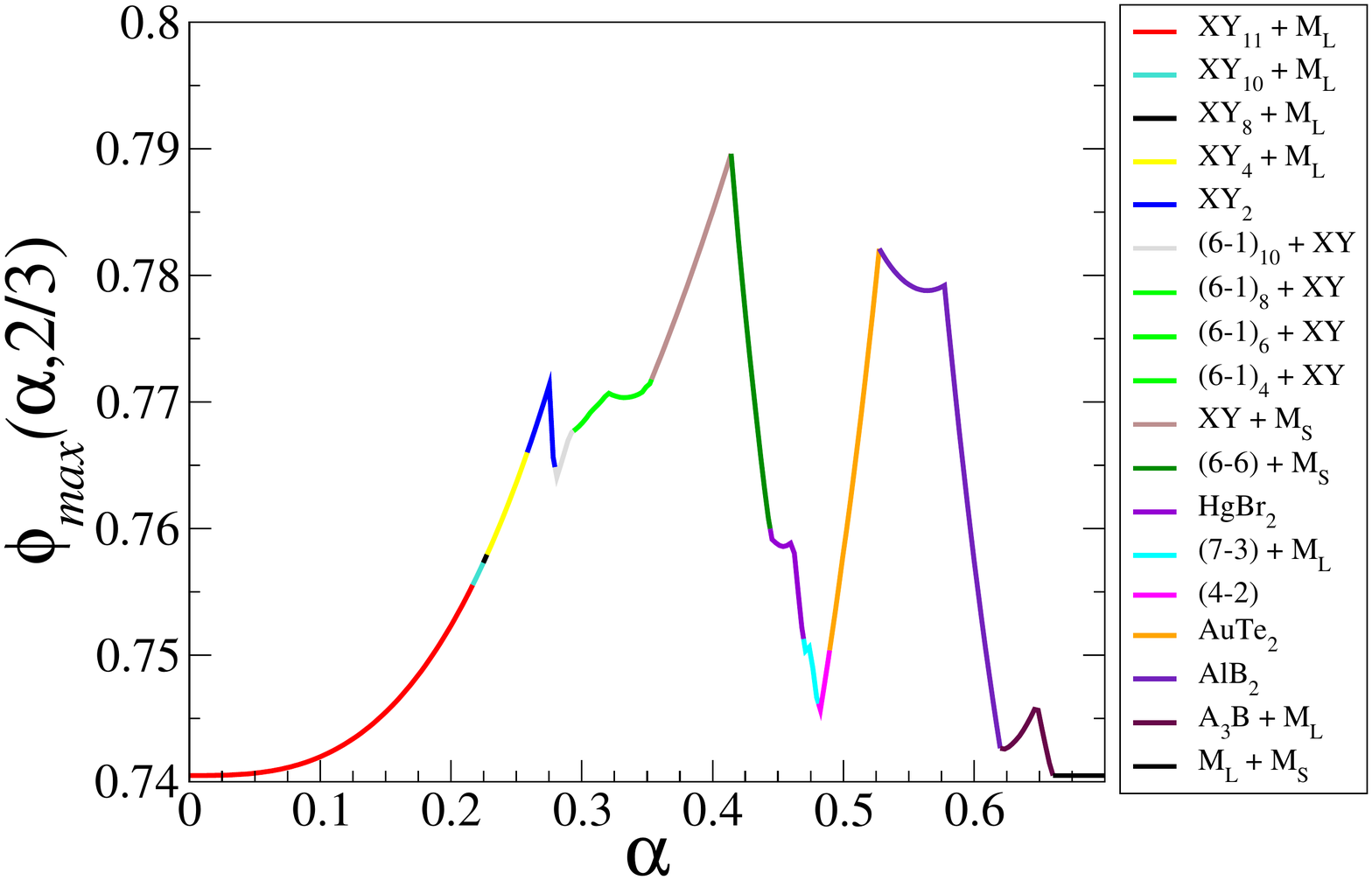}}
\caption{(Color online) Plots of the best-known lower bounds on
$\phi_{max}(\alpha,x)$ for $0\leq \alpha \leq 0.7$ and (a) $x=1/2$, (b) $x=2/3$,
considering minimal bases up to $12$ spheres. Coloration (shading) in these plots
indicates the phases present in the densest packings, as described in the
legend.}
\label{DensestBinarySameX}
\end{figure}

Each distinct alloy appears in the densest packings over a range of $\alpha$.
However, most alloys exhibit varying contact networks over this range, where a
contact network describes the numbers of small and large sphere contacts for
each sphere in the minimal basis. Each distinct alloy could thus be subdivided
into ``suballoys'' based on contact networks, as was done for the periodic disk
alloys in Ref. \cite{LH1993a}. Though we do not make these subdivisions in this
work due to the very large number of suballoys that would be identified, we note
that there are practical reasons for the subdivision. In particular, subdivision
by contact network would allow for characterization of all of the line segments
in the $\alpha$-$x$ plane along which the slope of the surface changes
discontinuously. These discontinuities can be clearly seen in Fig.
\ref{DensestBinarySameX}, plots of the best-known lower bounds on
$\phi_{max}(\alpha,1/2)$ and $\phi_{max}(\alpha,2/3)$ for $0 \leq \alpha \leq
0.7$. In Fig. \ref{DensestBinarySameX}, M$_S$ and M$_L$ refer to monodisperse,
Barlow-packed phases of small and large spheres, respectively. 

One example of a discontinuity in slope occurring along a line segment inside a
phase region is for the highest point in Figs. \ref{surface1} and
\ref{surface2}, at $\phi_{max}(0.224744\dots,10/11) = 0.824539\dots$. If this
phase region was subdivided according to the contact networks of the alloys that
compose the phase, this point would occur along a phase boundary. Another
example is found in the alloy with triclinic lattice system characterization
with six small spheres and one large sphere in its minimal basis, present in the
densest packings over the range $0.292 \leq \alpha \leq 0.344$. In Figs.
\ref{surface1}, \ref{surface2}, and \ref{phaseDiagram} to serve as an example,
the phase region including this alloy has been subdivided according to the
number of large-large sphere contacts in the alloy. There are eight, six, and
four large-large sphere contacts, respectively, in the ($6$-$1$)$_8$,
($6$-$1$)$_6$, and ($6$-$1$)$_4$ ``suballoys'', and close inspection of the
bottom plot in Fig. \ref{DensestBinarySameX} reveals discontinuous changes in
slope in $\alpha$ at each of the boundaries between them.

In general, the surface presented in Figs. \ref{surface1} and \ref{surface2} is
continuous and piecewise differentiable, though as $\alpha\rightarrow 0$ and
$x\rightarrow 1$, the density of curves along which the surface is not
differentiable approaches infinity due to the number of XY$_n$ and similar
packings. For this reason, we exclude the region $\alpha < 0.2$, $x > 11/12$
from our study, truncating $\alpha$ at $0.2$ because it is close to the maximum
value $\alpha = 0.216633\dots$ at which $11$ small spheres fit in the
interstices of a close-packed Barlow packing of large spheres. For $\alpha >
0.660\dots$, we have been unable to find any periodic packing with basis of $12$
or fewer spheres that exceeds the packing fraction $\pi/\sqrt{18}$ of two
phase-separated monodisperse Barlow-packed phases.

Away from the point $(\alpha,x) = (0,1)$, we represent the surface at given
$\alpha$ piecewise-analytically in $x$. This is possible because densest
packings can be constructed from a finite number of phase-separated alloy and
Barlow-packed monodisperse phases, where phase-separated packings are the
densest since the interfacial volume of the boundaries between phases is
negligible in the infinite volume limit. Further, we prove in Sec.
\ref{identifying} that at any point $(\alpha,x)$ with $\alpha > 0$, $x < 1$,
there is at least one densest binary packing that consists of no more than two
phases. However, this does not preclude the possibility of a densest packing
consisting of more than two distinct phases, nor the possibility of ``mixing''
of phases in certain specific cases where there is no boundary cost.

In Sec. \ref{definitions}, differences between periodic, aperiodic, disordered,
and quasicrystalline packings are described. We discuss which types of packings
are found to be the densest in ${\mathbb R}^2$, and which we have been able to
find (and predict to find) in ${\mathbb R}^3$. In Sec. \ref{algorithm}, an
overview of the TJ algorithm is provided that includes reasons why the algorithm
is particularly successful in identifying dense alloys. In Sec.
\ref{identifying}, we describe precisely the method used to construct the
maximal-density surface $\phi_{max}$ and present a proof of why for $\eta$
different sizes of spheres in any dimension $d$, there is always a densest
packing consisting of only $\eta$ phase-separated phases. In Sec.
\ref{thePackings}, the structural details of the putative densest binary sphere
packings in ${\mathbb R}^3$ for minimal bases of $12$ or fewer spheres are
presented and their properties discussed. In Sec. \ref{conclusions}, we conclude
with the goals of related future research and the implications and applications
of our findings.

\section{Periodic Packings, Aperiodic Packings, and Jamming}
\label{definitions}
The space of binary sphere packings in ${\mathbb R}^d$ can be formed from a
configuration of points where there are $N_S$ points designated type $S$ and
$N_L$ points designated type $L$. Each pair of points of type $S$ must be
separated by at least a distance $2R_S$, each pair of points of type $L$ by at
least a distance $2R_L$, and each pair including one point of each type by at
least distance $R_S + R_L$. With the points representing sphere centers, this
construct describes the space of all binary packings of small (S) and large (L)
spheres in ${\mathbb R}^d$ with radius ratio $\alpha = R_S/R_L$ and small sphere
relative concentration $x = N_S/(N_S + N_L)$.

Configurations of points can be categorized by their spatial distributions. In
this way, all packings can be described as lattice, periodic, or aperiodic. A
{\it lattice} ${\bf \Lambda}$ in ${\mathbb R}^d$ is a subgroup consisting of the
integer linear combinations of a given set of $d$ {\it lattice vectors} that
span ${\mathbb R}^d$. This is referred to as a {\it Bravais} lattice in the
physical sciences and engineering. A {\it lattice packing} of identical
nonoverlapping objects is one in which the objects have centers located at the
points of ${\bf \Lambda}$. In a lattice packing, the space ${\mathbb R}^d$ can
be geometrically divided into identical regions called {\it fundamental cells},
each of which contains the center of only one object. For a lattice packing of
spheres, which are necessarily identical, the packing fraction $\phi$ in
${\mathbb R}^d$ is given by,
\begin{equation}
\phi = v_1(R)/V_F,
\label{latticePhi}
\end{equation}
with $R$ the radius of the spheres, $V_F$ the volume of the fundamental cell,
and 
\begin{equation}
v_1(R) = \pi^{d/2}R^d/\Gamma(1+d/2)
\label{v1}
\end{equation}
the volume of a single sphere, where $\Gamma(x)$ is the Euler gamma function.

An extension of the notion of a lattice packing is a {\it periodic packing}. A
periodic packing is obtained by placing a fixed configuration of the centers of
$N \geq 1$ objects, called the {\it basis}, in one cell of a lattice ${\bf
\Lambda}$, and then replicating this fixed configuration, without overlap, in
all other cells. The cell in this case is termed a {\it unit cell}, and is also
a fundamental cell with corresponding {\it minimal basis} if and only if no
smaller unit cell (with fewer than $N$ object centers) can describe the same
packing. In either case, the packing fraction for a packing of spheres, not
necessarily identical in size, is given by,
\begin{equation}
\phi = \frac{\sum_{i=1}^Nv_1(R_i)}{V_U},
\label{periodicPhi}
\end{equation}
with $R_i$ the radius of each sphere in the basis and $V_U$ the volume of the
cell. It is critical to note that no binary packing with $0 < x < 1$ can be a
lattice packing, since the basis of a binary packing by definition must consist
of at least two objects.

An aperiodic packing is one that exhibits no long-range translational order;
{\it i.e.}, its minimal basis is equal to the number of particles in the
packing. Practically speaking though, a packing with a very large number of
objects might be called aperiodic if it is comprised of only a few translations
of the minimal basis. Both periodic and aperiodic structures are found widely in
nature; examples of the former include crystalline solids like many metals and
salts, and examples of the latter include liquids, glasses, gels, gases, plasmas
and quasicrystals \cite{BSYL2009a}.

A quasicrystal is an aperiodic structure that nonetheless exhibits bond
orientational order in symmetries ({\it e.g.}, five-fold) forbidden to periodic
crystals. We refer the reader to Ref. \cite{LS1986a} for a more precise
definition. We describe a {\it directionally periodic} structure as an aperiodic
structure that exhibits a period along at least one spatial axis but never
simultaneously along $d$ lattice vectors, {\it e.g.}, a random stacking Barlow
packing.

The Barlow packings are constructed by stacking layers, each consisting of
contacting spheres in ${\mathbb R}^3$ with centers on a plane in a triangular
lattice configuration, on top of one another. There are two ways (layers in
positions $B$ and $C$) to stack such a layer on top of another (layer in
position $A$) such that all spheres in each layer are in contact with three
spheres in the adjacent layer. Any packing that is composed of an infinite
number of $A$, $B$, and $C$ layers, with no adjacent layers the same, is a
Barlow packing and achieves the maximal packing fraction of identical spheres in
${\mathbb R}^3$. A random stacking of layers according to these rules is a
random stacking Barlow packing, which exhibits periodicity parallel to the
planes of the layers but not perpendicular to them.

For the purposes of this paper, we describe a {\it disordered} structure as an
aperiodic structure that does not exhibit perfect long-range symmetry,
translational, rotational, or reflective, of any kind.

In ${\mathbb R}^2$, periodic, quasicrystalline, and directionally periodic
structures can all be found among the putative densest binary disk packings
\cite{LH1993a,TothRF1964,LHC1989a,Widom1993a}. We believe that all of these
types of structures may be found among the densest binary sphere packings in
${\mathbb R}^3$ as well, though we have identified only periodic and
directionally periodic structures in this work. Due to computational constraints
attributable in large part to the scope and resolution of our survey in
$(\alpha,x)$, we have limited our survey in this work to investigating the
densest periodic packings with bases of $12$ spheres or fewer. This limitation
substantially increases the difficulty of identifying aperiodic packings, which
most often cannot be approximated well by a fundamental cell with a basis of
only $12$ spheres, though densest directionally periodic packings often can
still be identified by extrapolating from periodic packings.

One property of a densest packing is that a subset of its objects is {\it
collectively jammed}, or in the case of a densest periodic packing, that the
objects within each unit cell are {\it strictly jammed} under periodic boundary
conditions. None of the nonoverlapping objects in a packing that is {\it locally
jammed} can be continuously displaced (displaced by infinitesimally small
movements) without displacing one of its neighbors \cite{TTD2000a,*TS2001a}. In a
collectively jammed packing, a subset of the objects (the {\it backbone}) is
locally jammed and no continuous collective motion of any subset of objects can
lead to {\it unjamming}, {\it i.e.}, a packing that is no longer locally jammed.
The objects that can be displaced without displacing their neighbors in a
collectively jammed packing are termed {\it rattlers}. A strictly jammed packing
is collectively jammed, and no collective motion of any subset of objects,
coupled with a continuous deformation of the boundary that does not increase
volume, can result in unjamming. If a subset of the objects in a packing is not
collectively or strictly jammed, then by definition a uniform continuous motion
exists that results in free space around the objects, i.e., the packing is not a
densest packing because its volume can be reduced.

One reason that the TJ algorithm is particularly effective at finding densest
periodic packings is that it guarantees that final packings are strictly jammed
\cite{TJ2010a}. If there is a collective continuous motion of the spheres,
coupled with a uniform continuous deformation of the unit cell, that will
decrease the volume of the cell, then the exact linear programming method
employed by the TJ algorithm will find that motion.

With respect to jamming, a certain group of packings that we call ``host-guest''
packings deserves further attention. In such a packing, a subset (usually the
larger spheres) of the total packing are packed as a jammed periodic packing,
and a mutually exclusive subset (usually the smaller spheres) sit in the
interstices formed by the jammed spheres but without contacting the jammed
spheres. Clearly, the spheres within the interstices can be continually
displaced without altering the density of the packing. In this way, a host-guest
packing can be periodic or aperiodic. If the rattlers are placed in the same
place relative to one another throughout every cell of the packing, then the
packing is periodic. If they are placed at random, then the packing is
aperiodic. A host-guest packing is similar in nature to an interstitial solid solution 
(ISS) phase, such as the stable binary colloidal phases described in Ref. \cite{FHNVKCSVBD2011a}. 
We are not aware of any densest packings where the large spheres are
free to move, as rattlers, within interstices created by jammed small spheres. 

The XY$_n$ packings where small spheres are not required to contact the jammed
large spheres are good examples of host-guest sphere packings, where an XY$_n$
packing consists of a Barlow-packed structure of contacting large spheres with
small spheres placed in the interstices. Among the host-guest packings, one can
distinguish topologically between packings where the small spheres are small
enough to move freely about all of the interstices and packings where they are
confined to a given interstice or finite set of interstices. In the former case,
the space available to the small spheres is topologically connected, where in
the latter case it is not. Though we find densest packings ({\it e.g.}, XY$_n$
packings) where small spheres are confined to a single interstice, we do not
find for $\alpha \geq 0.2$ any where they are free to move about all of the
interstices. However, it is the case for small enough $\alpha$ that there are
densest host-guest packings where the space available to the small spheres is
topologically connected.

\section{The Torquato-Jiao Sphere-Packing Algorithm}
\label{algorithm}
There are several factors that have limited past algorithmic techniques in
finding densest packings. As previously discussed, one of these is the immensity
of the parameter space in $(\alpha,x)$ of densest binary packings. Another is
the strong initial condition dependence of many algorithms, which requires that
a very large number of simulations beginning with different initial spatial
conditions be undertaken to find a densest packing. This required number of
simulations can increase exponentially with the number of spheres simulated. 

Past algorithmic techniques have also been limited by the vast multitude of
local minima in ``energy'', defined as the negative of the packing fraction,
present in the phase space of a periodic sphere packing with unit cell of
indeterminate size and shape with even a small basis. Methods that move only one
sphere at a time cannot escape from local minima where a collective motion of
spheres is required to decrease energy, and methods that move several spheres at
once rely on chance, and hence a very large number of steps, to find the right
collective movement. Many molecular dynamics algorithms rely upon principles of
equilibrium thermodynamics, which necessitate that large numbers of
time-consuming steps (individual sphere movements) be taken at each stage of
sphere growth or compression of the unit cell in order to reduce the chance of
becoming ``stuck'' in a local minimum.

In this paper, we are able to overcome these time and computational limitations
by using the Torquato-Jiao linear programming algorithm \cite{TJ2010a}. The TJ
sphere packing algorithm does not easily become ``stuck'' in local minima
because it is inherently designed to find the simultaneous collective linear
motion of both the spheres and the unit cell geometry that maximizes the density
of the packing.   

The TJ algorithm approaches the problem of generating dense, periodic packings
of nonoverlapping spheres with an arbitrary size distribution as an optimization
problem to be solved using linear programming techniques. In particular, the
objective function is chosen to be the negative of the packing fraction (as
indicated above), which is minimized with respect to particle (sphere center)
positions and the shape and size of the unit cell, subject to sphere nonoverlap
conditions. The use of a deformable unit cell, defined in terms of $d$,
$d$-dimensional lattice vectors ${\bf M}_\Lambda = \{{\bf \lambda}_1;\dots;{\bf
\lambda}_d\}$, was first introduced by Torquato and Jiao \cite{TJ2009a,*TJ2009b}
to obtain dense packings of hard polyhedral particles including Platonic and
Archimedean solids.

For nonoverlapping spheres, the optimization problem can be efficiently solved
by linearizing the objective function and nonoverlap constraints. This approach
has the advantage of being rigorously exact near the jamming limit of the
spheres and unit cell \cite{TJ2010a}.

The TJ algorithm begins with an initial packing of $N$ spheres, and obtains a
new, denser packing by solving the following linear programming problem:
\begin{align}
&\mbox{minimize:}\,\, Tr(\boldsymbol\varepsilon) =
\varepsilon_{11}+\cdots+\varepsilon_{dd} \notag \\
&\mbox{subject to:} \notag \\
& \qquad {\bf M}_\Lambda \cdot {\bf r}^\lambda_{ij}
\cdot{\boldsymbol\varepsilon}\cdot {\bf M}_\Lambda \cdot {\bf r}^\lambda_{ij} +
\Delta{\bf r}^\lambda_{ij}\cdot {\bf G} \cdot {\bf r}^\lambda_{ij} \notag \ge
\frac{1}{2}(\overline{D}_{ij}^2 - {\bf r}^\lambda_{ij}\cdot {\bf G}\cdot {\bf
r}^\lambda_{ij})+{\cal R}, \notag \\
&~\mbox{for all neighbor pairs}~ (i,j)~ \mbox{of interest, and} \notag \\
& \qquad \Delta{\bf x}^{\lambda, lower}_i \le \Delta{\bf x}^\lambda_i \le
\Delta{\bf x}^{\lambda, upper}_i, ~\forall ~i = (1,\dots, N), \notag \\
& \qquad \varepsilon^{lower}_{kl}\le \varepsilon_{kl} \le
\varepsilon^{upper}_{kl},~\forall ~k,l=(1,\dots,d).
\label{LP}
\end{align}

\noindent where $\overline{D}_{ij} = (D_i+D_j)/2$ is the average diameter of
spheres $i$ and $j$; ${\bf r}^\lambda_{ij} = {\bf x}^\lambda_i - {\bf
x}^\lambda_j$ is the displacement vector between spheres $i$ and $j$ in the
initial packing with respect to the lattice vectors ${\bf M}_\Lambda$;
$\Delta{\bf r}^\lambda_{ij} = \Delta{\bf x}^\lambda_i - \Delta{\bf x}^\lambda_j$
is the relative change in displacement of spheres $i$ and $j$ in coordinates
with respect to the ${\bf M}_\Lambda$; ${\boldsymbol\varepsilon} = \{
\varepsilon_{kl}\}$ is the strain tensor associated with the unit cell, with
lower and upper bounds $\{\varepsilon_{kl}^{lower}\}$ and
$\{\varepsilon_{kl}^{upper}\}$, respectively; ${\bf G} = {\bf M}_\Lambda^T \cdot
{\bf M}_\Lambda$ is the \textit{Gram matrix} of the lattice and ${\cal R}$ is a
scalar representing higher order terms, which are given in Appendix
\ref{AlgMods}. The {\it neighbor} pairs are determined by an {\it influence
sphere} with radius $\gamma_{ij}$, i.e., two spheres $i$ and $j$ are considered
neighbors if their pair distance is smaller than $\gamma_{ij}$. We note that the
choice of the influence sphere radius can significantly affect the density and
degree of disorder in the final packing. For densest packings, a large
$\gamma_{ij}$ should be used, as detailed in Ref. \cite{TJ2010a}. We also note 
that to study packings of spheres with non-additive diameters, one simply sets 
$\overline{D}_{ij}$ to the desired value when spheres $i$ and $j$ are not of the 
same type.

Our tests of the TJ algorithm, for which the implementation for binary sphere
packings is described in Appendix \ref{AlgMods}, indicate that the algorithm is
both particularly efficient and particularly robust in finding densest packings
across the entire range of $(\alpha,x)$ for a sufficiently small basis of
spheres. For example, we have recovered all of the alloys known to be present in
the densest packings, including the A$_3$B \cite{OH2011a}, AlB$_2$, HgBr$_2$,
AuTe$_2$, and ``Structure 2'' \cite{MH2010a} alloys, and XY$_n$ packings for
certain values of $n$. In addition, we have found many areas in the $\alpha$-$x$
plane where the densest packings are denser than previously predicted and
include heretofore unknown alloys. Overall, in the area of the $\alpha$-$x$
plane searched, we always recover a packing fraction $\phi_{max}(\alpha,x)$ that
is either greater than or equal to that previously known.

\section{Identifying the densest binary sphere packings}
\label{identifying}
To identify the densest binary packings in ${\mathbb R}^3$, we begin with the
obvious statement that at all given $(\alpha,x)$, there is a densest packing
that consists of a finite number of phase-separated alloy and
monodisperse phases. In the infinite volume limit in ${\mathbb R}^d$ for $\eta$
different sizes of spheres, the packing fraction of a phase-separated collection
of $\beta$ monodisperse and distinct alloy phases can be written,
\begin{equation}
\phi =
\frac{\left(\pi^{d/2}\Gamma(1+d/2)\right)\left(\sum_{j=1}^{\eta}X^{(j)}(R^{(j)}
)^d\right)}{\sum_{i=1}^{\beta}\frac{C_i}{S_i^{F_i}}x_i^{F_i}},
\label{packingFractionFull}
\end{equation}
with $X^{(j)}$ and $R^{(j)}$ the relative fraction and radii, respectively, of
spheres of type (size) $j$, $x_i^j$ the relative fraction of spheres of type $j$
distributed in phase $i$ ($\sum_{i=1}^{\beta}x_i^j=X^{(j)}$), and $C_i$ the
volume of a unit cell of phase $i$ containing, for each sphere type $j$, $S_i^j$
spheres. The index $j=F_i$ in $S_i^{F_i}$ and $x_i^{F_i}$ is the first index $j$
in phase $i$ such that $S_i^j \neq 0$. For $\eta$ sphere sizes, there are $\eta$
monodisperse phases with relative fraction $x_i^{F_i}$, $F_i=1\dots\eta$
respectively and $S_i^j = 0$ for all $j \neq F_i$, representing packings with
spheres packed as densely as possible in ${\mathbb R}^d$.

To find the densest packing $\phi_{max}(R^{(1)}\dots R^{(\eta)},X^{(1)}\dots
X^{(\eta)})$ at specified sphere sizes $R^{(j)}$ and relative compositions
$X^{(j)}$ from among $\beta$ known alloy and close-packed monodisperse phases,
Eq. (\ref{packingFractionFull}) must be maximized. This is accomplished by
minimizing the denominator of Eq. (\ref{packingFractionFull}) over the relative
fractions $x_i^{F_i}$. As the denominator is linear in the variables
$x_i^{F_i}$, it can be formulated as the objective function in a linear
programming problem. This formulation allows us to state the following theorem,
the proof of which is found in Appendix \ref{proof1}. We note that Theorem 1
applies not only to spheres, but in fact to any packing of $\eta$ different
types of objects in ${\mathbb R}^d$ with specified volumes and relative
fractions $X^{(j)}$.

\begin{Theorem1}
Consider any sphere packing in ${\mathbb R}^d$ composed of $\beta$
phase-separated alloy and monodisperse phases. For $\eta$ different sizes of
spheres, there is always a densest packing consisting of no more than $\eta$
phase-separated phases.
\end{Theorem1}

For binary sphere packings with $\eta=2$ in ${\mathbb R}^3$, it is thus clear
that there is at least one densest packing at every $(\alpha,x)$ consisting of
no more than two phases. For simplicity in this case, we set $X^{(1)} = 1-x$ to
be the relative fraction of large spheres and $X^{(2)} = x$ the relative
fraction of small spheres, and rewrite Eq. (\ref{packingFractionFull}) as,
\begin{equation}
\phi = \frac{\frac{4\pi}{3}\left((1-x)R_L^3 + xR_S^3\right)}{xC_{B}^S +
(1-x)C_{B}^L + \sum_{i=1}^{\beta}x_i^L\left(\frac{C_i}{L_i} -
\frac{S_i}{L_i}C_{B}^S - C_{B}^L\right)},
\label{packingFraction1}
\end{equation}
with $C_{B}^S$ and $C_{B}^L$ the volume per sphere, respectively, in a
close-packed Barlow packing of small and large spheres, $x_i^L$ the relative
fraction of large spheres distributed in alloy phase $i$, and $C_i$ the volume
of a fundamental cell of alloy phase $i$ containing $L_i$ large and $S_i$ small
spheres.

Using Eq. (\ref{packingFraction1}) and considering a fixed value of $\alpha$,
the densest binary packings in ${\mathbb R}^3$ can be found for all values of
$x$ by solving for all globally optimal vertices considering all dense alloy
phases at $\alpha$. Since we limit ourselves in this work to minimal bases of no
more than $12$ spheres, we must assume that all of these alloy phases can be
constructed from repetitions of local structures consisting of $12$ spheres or
fewer. Though we recognize that this assumption is most likely false for some
values of $\alpha$, especially near the point $(\alpha,x) = (0,1)$, we contend
that for the majority of the area of the parameter space studied, it is correct.
We therefore employ the TJ algorithm to search the space of fundamental cells in
${\mathbb R}^3$ containing all combinations of positive integer $L_i$ and $S_i$
such that $L_i + S_i = 2,3,\dots,12$. We have solved these problems (putatively)
to accuracy of about $10^{-4}$ in $\phi$ for $\alpha$ spaced $0.025$ apart, and
on a finer grid with $\alpha$ spaced about $0.0028$ apart for certain values of
$S_i$ and $L_i$ where particularly dense packings were identified.

\section{Packing Results}
\label{thePackings}
In the following subsections, the densest binary sphere packings that we have
identified are presented, including both those that were previously known and
those that, to the best of our knowledge, were not. For those alloys that were
not previously known, in Table I in Appendix \ref{packingFractions} we list packing
fractions across the range of $\alpha$ where they appear in the densest
packings. In subsection \ref{packingProperties}, we discuss some of the
properties of these packings, including, for example, the lattice symmetry of
alloy phases present in the packings and classification of these phases as
periodic or aperiodic. The properties that we discuss do not serve as an
exhaustive list, nor are the examples we provide the only examples of densest
binary packings with these properties. Instead, we discuss packing properties
with the intent of providing an overview of the diversity of the different types
of densest packings.

The figures presented in this section are illustrations of unit cells of the
alloys that appear in the densest packings, where the unit cells are chosen to
highlight packing symmetry. Contacts between two large spheres are indicated by
dotted lines, as are contacts between two small spheres; however, large-small
contacts are not drawn. Unit cell boundaries are also indicated by dotted lines,
and unit cell boundary lengths are displayed in units of the diameter of the
large spheres. Angles are given in degrees.

\subsection{Packing properties}
\label{packingProperties}
The densest binary sphere packings that we have identified exhibit a broad array
of properties. For example, there are densest packings that consist of only a
single phase, and many that consist of multiple phases. In the majority of these
phases, all spheres are collectively jammed. In others, each fundamental cell
contains rattlers, {\it i.e.}, there are an infinite number of sphere
configurations in each cell that exhibit the same packing fraction. In others
still, there are only a finite number of sphere configurations in each cell that
exhibit identical packing fraction.

Concerning rattlers in densest packings, it is intuitive that for the majority
of the volume of the parameter space of densest packings of $\eta$ different
sizes of spheres, packings with rattlers would dominate as $\eta$ increases.
This is because for densest packings with substantial size disparity between the
largest and smallest spheres, except for at selected values of ${X^{j}}$, it
seems clear that the small spheres would move about as rattlers in the
interstices formed by jammed larger spheres. It is interesting therefore that we
have found that there are densest binary packings over the majority of the
$\alpha-x$ plane that are composed of phases that do not include rattlers.

For the rectangular region $\alpha < 0.2$, $x > 11/12$, we have not attempted to
identify densest packings. This is because the number of distinct densest
packings, for example, those of type XY$_n$, approaches infinity as
$\alpha\rightarrow 0$ with $x \rightarrow 1$. In this latter limit, the densest
packings are the XY$_{\infty}$ packings consisting of infinitesimally small
Barlow-packed small spheres located within the spaces between Barlow-packed
large spheres. The packing fraction of this packing is \cite{TorquatoRHM2002},
\begin{equation}
\lim_{(\alpha,x)\rightarrow (0,1)}{\phi_{max}(\alpha,x)} = 1 - (1-
\pi/\sqrt{18})^2 = 0.932649\dots.
\label{densestBinary}
\end{equation}

At all points $(\alpha,x)$ outside of the rectangular region $\alpha < 0.2$, $x
> 11/12$, we have identified putatively densest packings that consist only of
periodic alloy phases. However, due to the infinity of densest Barlow packings
(which are, in general, directionally periodic), all densest packings that
contain a periodic monodisperse phase are degenerate in density with packings
exhibiting an aperiodic monodisperse phase. Additionally, there are alloys that
can be paired with monodisperse Barlow packings without a boundary cost, {\it
e.g.}, the AlB$_2$ alloy. This pairing allows a ``mixed-phase'' densest packing,
or a densest packing composed of a single aperiodic (directionally periodic)
phase. There are also pairs of alloys that appear to exhibit this property.

For the majority of the area in the $\alpha$-$x$ plane, the densest packings
consist of two phase-separated phases, one alloy and one monodisperse. For
$\alpha > \alpha^* \approx 0.660$, the known densest binary packings consist
simply of two phase-separated monodisperse Barlow-packed phases. For $(\alpha,x)
\in (0,\alpha^*)\times(0,1)$, either a single alloy phase or a combination of an
alloy and a monodisperse phase is known to be denser than two phase-separated
monodisperse phases.

\begin{figure}[ht!]
\centering
\includegraphics[width = 3.5in,viewport = 310 60 1000 710,clip]{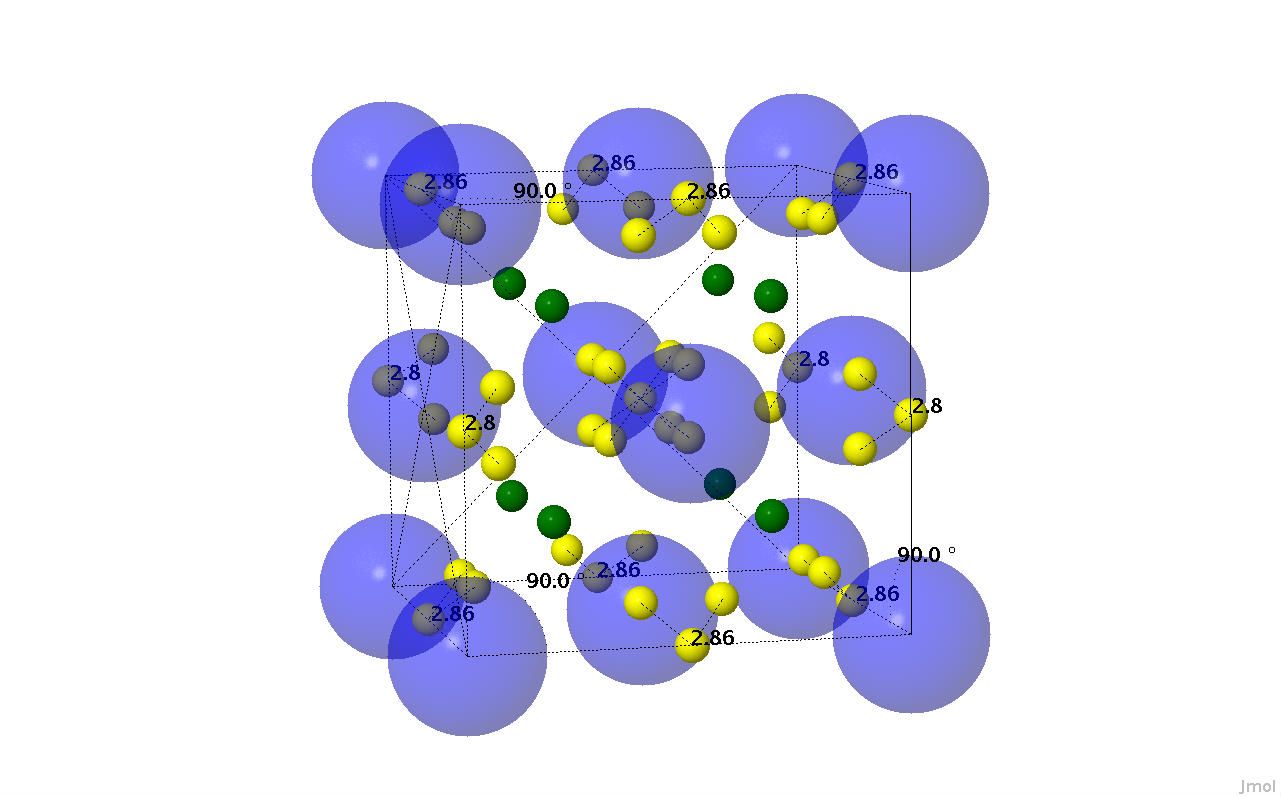}
\caption{(Color online) Putative densest packing ($\phi = 0.818$) at radius
ratio $\alpha=0.2195$ of 11 small spheres and 1 large sphere in a periodic
fundamental cell. In this image, small spheres in tetrahedral interstices are
shown in green (dark grey) and small spheres in octahedral interstices are shown in yellow (light grey).
The fundamental cell of this ($11$-$1$) alloy belongs to the tetragonal lattice
system.} 
\label{11-1}
\end{figure}

At certain points, densest packings consisting of three distinct phases can
exist. Examples of this case include; at $\alpha = \alpha^*$, $0 < x < 1$, where
two monodisperse phases and one A$_3$B phase exhibit packing fraction
$\pi/\sqrt{18}$; at about $\alpha = 0.292$, $6/7 < x < 1$ where one monodisperse
(small spheres), one ($6$-$1$)$_{10}$, and one ($6$-$1$)$_8$ phase exhibit a
peak packing fraction of about $0.801$; and at the same $\alpha$ with $1/2 < x <
6/7$ where one ($6$-$1$)$_{10}$, one ($6$-$1$)$_8$, and one XY phase exhibit a
peak packing fraction of about $0.801$. One example of a packing with four
distinct phases occurs at about $\alpha = 0.480$, $1/2 < x < 2/3$, where one
monodisperse (large spheres), one ($7$-$3$), one AuTe$_2$, and one ($2$-$2$)$^*$
phase exhibit a peak packing fraction of $0.748$.

\subsection{Densest packings for $\alpha \leq \sqrt{2}-1$}

Many of the densest binary alloys for $\alpha < \sqrt{2}-1$ are of the XY$_n$
type. In an XY$_n$ alloy, the small spheres, of which there are $n$ for every
large sphere, are distributed in the octahedral and sometimes tetrahedral
interstices, of which there are one and two, respectively, for each large
sphere, of strictly jammed Barlow-packed large spheres. The XY$_n$ alloys are
present in the densest binary packings that we have found for $n=1$, $2$, $4$,
$8$, $10$ and $11$. In these packings, the small spheres occupy the interstices
as rattlers, except for at ``magic'' \cite{LH1993a} $\alpha$ where they are
jammed between large spheres, {\it e.g.}, at $\alpha = \sqrt{2}-1$ for the NaCl
alloy. All XY$_n$ alloys belong to the cubic lattice system.

\begin{figure}[ht!]
\centering
\includegraphics[width = 3.2in,viewport = 265 50 1005 712,clip]{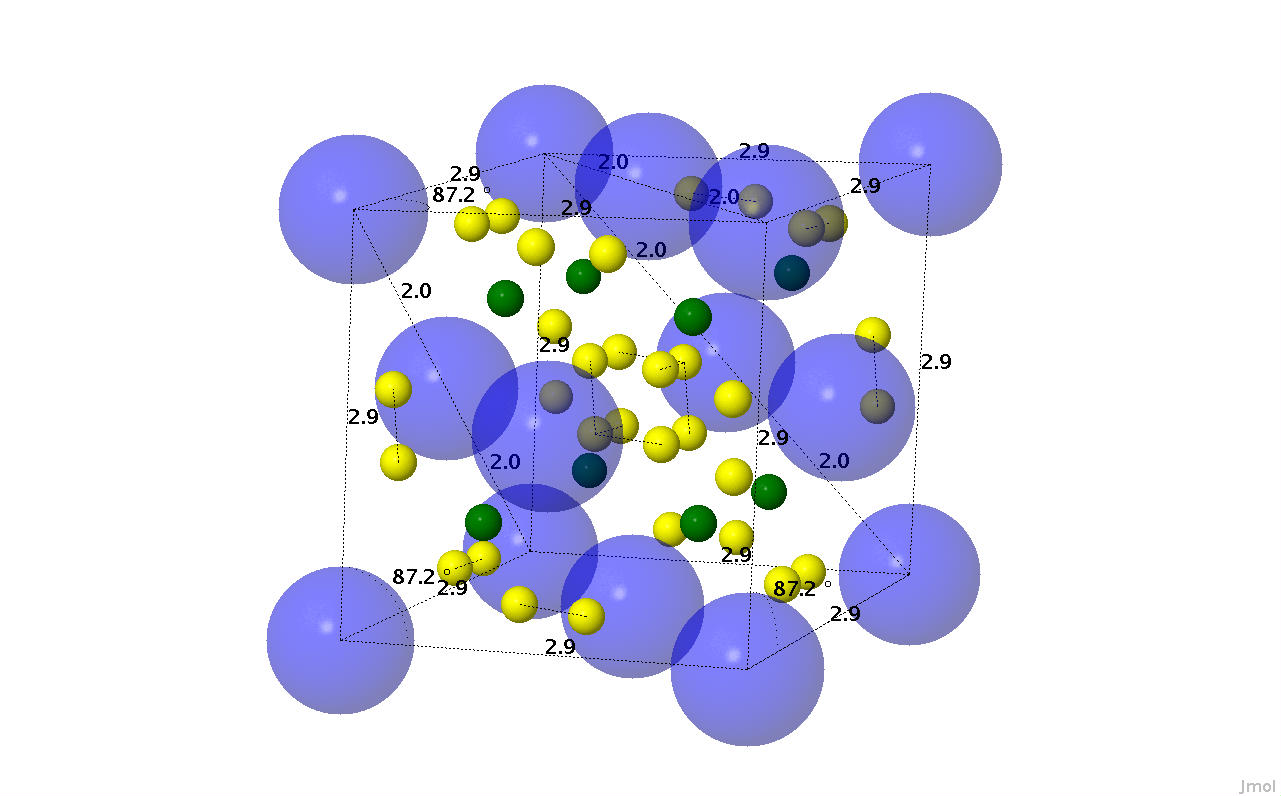}
\caption{(Color online) Putative densest packing ($\phi = 0.797$) at radius
ratio $\alpha=0.2503$ of 10 small spheres and 1 large sphere in a periodic
fundamental cell. In this image, small spheres in tetrahedral interstices are
shown in green (dark grey) and small spheres in octahedral interstices are shown in yellow (light grey).
The ($10$-$1$) alloy belongs to the rhombohedral lattice system.}
\label{10-1}
\end{figure}

For $n=1$, $2$, $4$, and $8$, these small spheres occupy only the octahedral
interstice, where for $n=4$ and $n=8$, each set of $4$ and $8$ spheres in an
interstice at the ``magic'' radii form a perfect tetrahedron and cube,
respectively. For $n=10$, a small sphere occupies each of the tetrahedral
interstices, and at the magic radius the remainder form a perfect cube; for
$n=11$, there is one extra small sphere in the cube's center. Additionally, for
$n=2$, $4$, $8$ and $10$, there are XY$_n$ alloys for $\alpha$ greater than the
magic radius ratios. These packings consist of large spheres arranged as in a
Barlow packing (with cubic symmetry) but not in contact, with interstitial
jammed small spheres arranged as was the case for the magic $\alpha$.

\begin{figure}[ht!]
\centering
\includegraphics[width = 3.2in,viewport = 220 15 1100 740,clip]{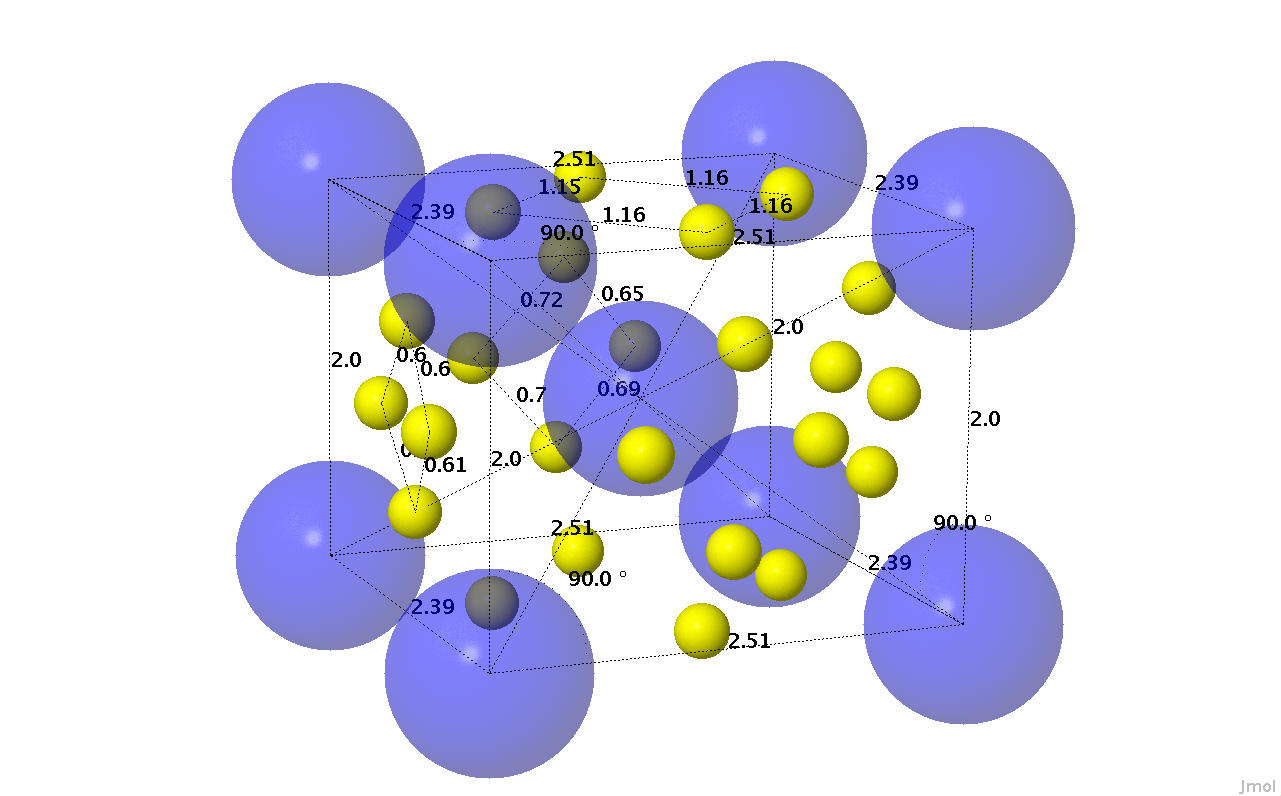}
\caption{(Color online) Putative densest packing ($\phi = 0.789$) at radius
ratio $\alpha=0.2781$ of six small spheres and one large sphere in a periodic
fundamental cell. In this image, dotted lines between small spheres are drawn to
indicate distances and do not indicate contacts. The ($6$-$1$)$_{10}$ alloy
belongs to the orthorhombic lattice system.}
\label{6-1-10}
\end{figure}

In the case of XY$_n$ packings where the large spheres are in contact, it is
clear that there is no boundary cost between monodisperse layers of contacting
spheres packed in a triangular lattice (which we recall is one layer of a Barlow
packing) and an XY$_n$ packing. Furthermore, for $x < n/(n+1)$, different
numbers of small spheres can be distributed among the interstices of the
Barlow-packed large spheres. Consequently, wherever a densest packing consists
of any XY$_n$ phase and a Barlow-packed monodisperse phase of large spheres,
mixed phase packings and aperiodic alloy phase packings are also densest
packings.

\begin{figure}[ht!]
\centering
\includegraphics[width = 3.2in,viewport = 250 15 1060 740,clip]{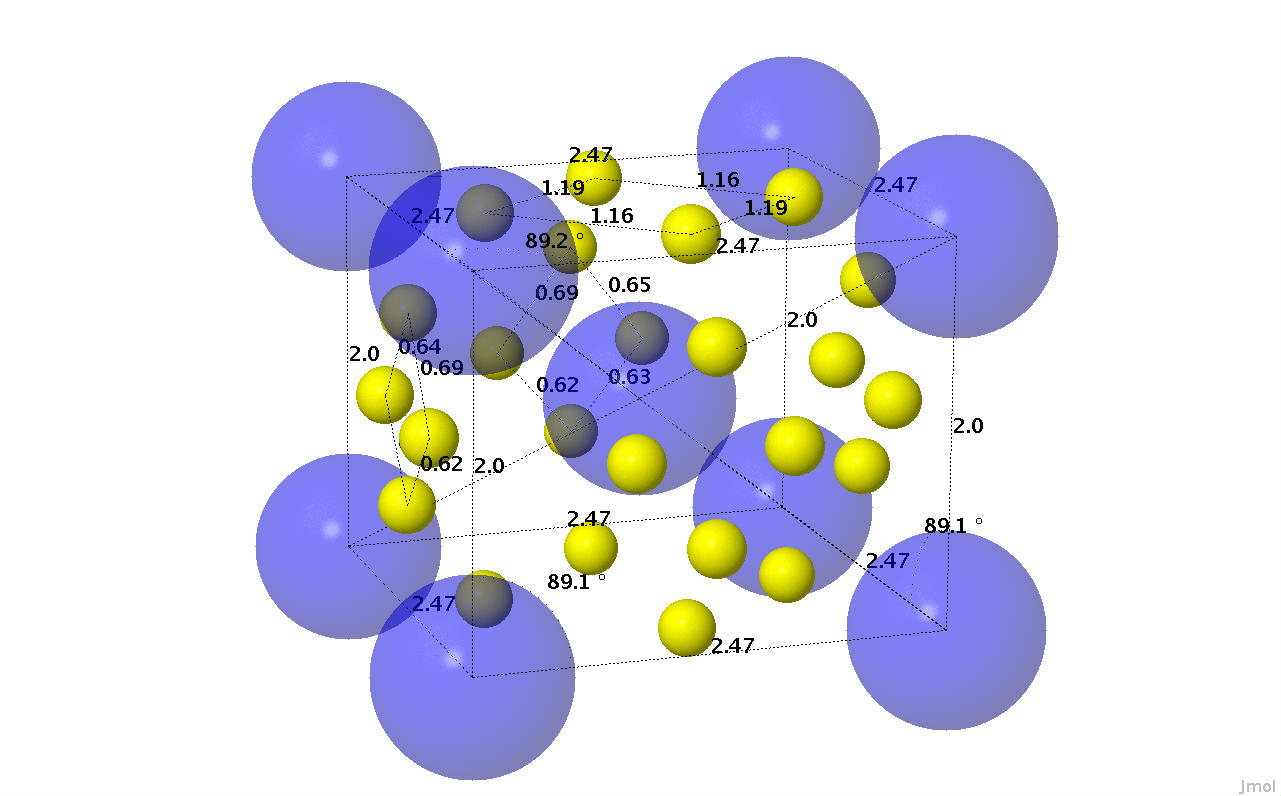}
\caption{(Color online) Putative densest packing ($\phi = 0.799$) at radius
ratio $\alpha=0.2980$ of six small spheres and one large sphere in a periodic
fundamental cell. In this image, dotted lines between small spheres are drawn to
indicate distances and do not indicate contacts. The ($6$-$1$)$_{8}$ suballoy
belongs to the triclinic lattice system.}
\label{6-1-8}
\end{figure}

The ($11$-$1$) and ($10$-$1$) alloys are similar to the XY$_{11}$ and XY$_{10}$
alloys, respectively, except that they belong to the tetragonal and rhombohedral
lattice systems. The unit cells of these packings can be viewed respectively as
``stretched'' and ``skewed'' versions of the XY$_{11}$ and XY$_{10}$ unit cells,
and the packings of large spheres can be seen this way as well. However, the
clusters of contacting small spheres that occupy the octahedral interstices in
these packings do not stretch and skew in the same fashion as the large spheres.
Figures \ref{11-1} and \ref{10-1} are images that highlight the symmetry present
in these alloys.

\begin{figure}[ht!]
\centering
\includegraphics[width = 3.2in,viewport = 230 45 1025 712,clip]{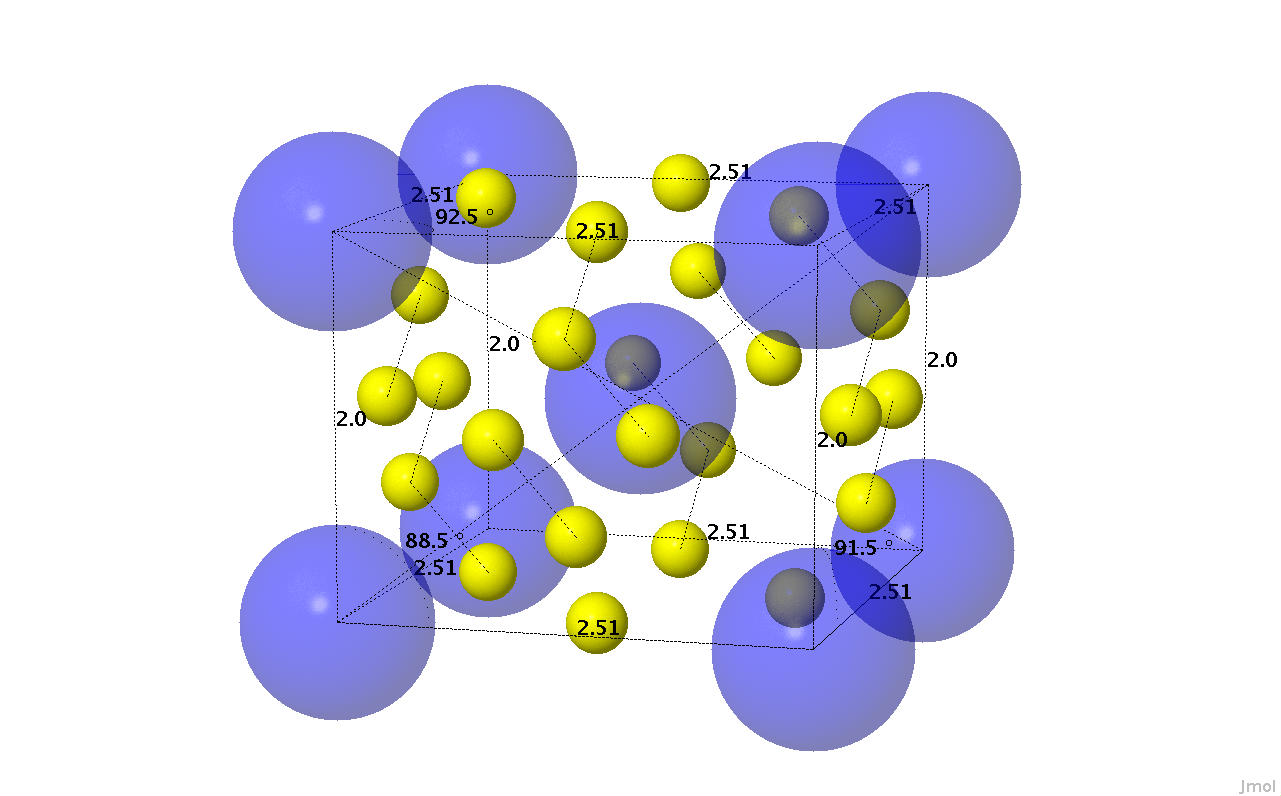}
\caption{(Color online) Putative densest packing ($\phi = 0.794$) at radius
ratio $\alpha=0.3151$ of six small spheres and one large sphere in a periodic
fundamental cell. The ($6$-$1$)$_{6}$ suballoy belongs to the triclinic lattice
system.}
\label{6-1-6}
\end{figure}

The ($6$-$1$)$_{10}$ alloy can be described as a body-centered orthorhombic
packing of large spheres with four small spheres present on each of the faces.
The small spheres on each face are not in contact, and they are not equidistant
from one another owing to the different lengths of the three lattice vectors
comprising the orthorhombic unit cell. Figure \ref{6-1-10} illustrates the lack
of contact between small spheres, in that the distances depicted between small
spheres are all greater than $2\alpha = 0.5562$ large sphere radii. In Fig.
\ref{6-1-10}, the central large sphere contacts all surrounding spheres and the
central large spheres in the adjacent unit cells on the $z$-axis, where the
$z$-axis is along the basis vectors that are $2.0000$ sphere diameters in
length.

\begin{figure}[ht!]
\centering
\includegraphics[width = 3.2in,viewport = 210 40 1050 720,clip]{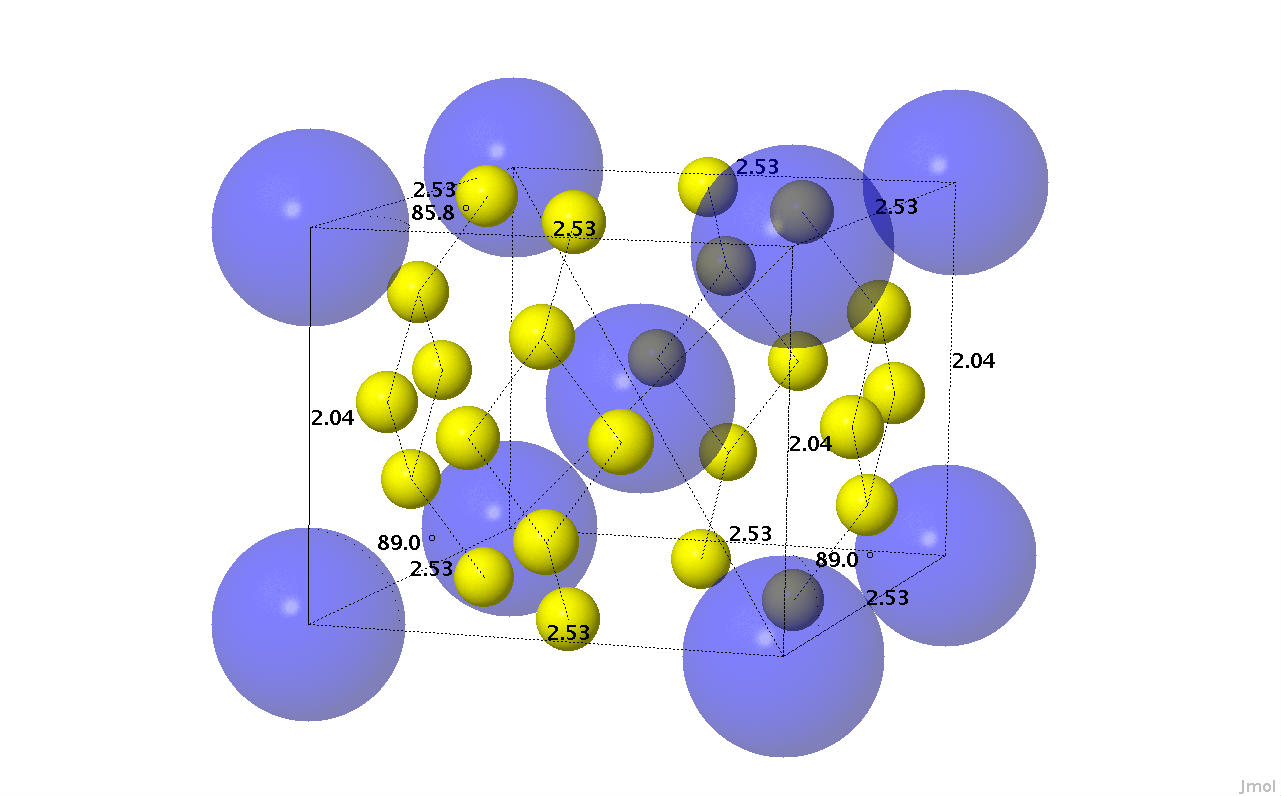}
\caption{(Color online) Putative densest packing ($\phi = 0.783$) at radius
ratio $\alpha=0.3293$ of six small spheres and one large sphere in a periodic
fundamental cell. The ($6$-$1$)$_{4}$ suballoy belongs to the triclinic lattice
system.}
\label{6-1-4}
\end{figure}

The ($6$-$1$)$_8$, ($6$-$1$)$_6$, and ($6$-$1$)$_4$ suballoys are a subdivision
of the alloy exhibiting triclinic lattice system characterization with six small
and one large spheres in its minimal basis. This alloy is present in the densest
packings over the approximate range $0.292 \leq \alpha \leq 0.352$, and it is
very similar to the ($6$-$1$)$_{10}$ alloy. Figures \ref{6-1-8}, \ref{6-1-6},
and \ref{6-1-4} illustrate large-large and small-small sphere contacts in the
($6$-$1$)$_8$, ($6$-$1$)$_6$, and ($6$-$1$)$_4$ suballoys, respectively. The
small spheres in the ($6$-$1$)$_8$ suballoy do not contact one another, though
many of those in the ($6$-$1$)$_6$ and ($6$-$1$)$_4$ alloys do. Additionally,
the centers of the small spheres in the ($6$-$1$)$_4$ alloy do not lie on the
faces of the triclinic unit cell. 

\begin{figure}[ht!]
\centering
\includegraphics[width = 3.2in,viewport = 350 70 930 705,clip]{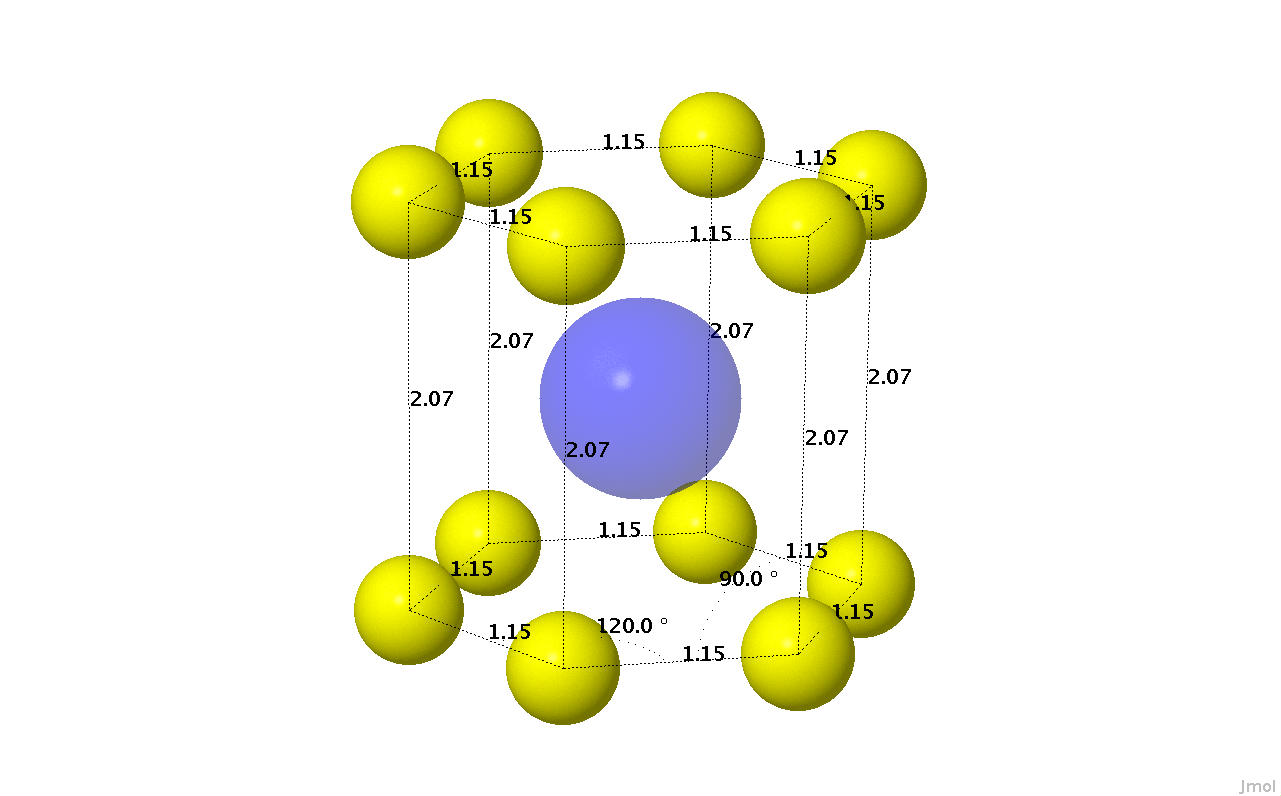}
\caption{(Color online) Putative densest packing ($\phi = 0.7793$) at radius
ratio $\alpha=0.5500$ of two small spheres and one large sphere in a periodic
fundamental cell. The AlB$_2$ alloy belongs to the hexagonal lattice system.}
\label{AlB2}
\end{figure}

It is also the case that over the entire range that the ($6$-$1$) alloys are
present in the densest packings, they can form mixed phase packings with large
spheres packed as in a Barlow packing, or with large spheres packed as in a
Barlow packing with one small sphere (a rattler) in the octahedral interstice.
These mixed phase packings exhibit only slightly smaller packing fractions than
their phase-separated counterparts. This is possible due to the nearly cubic
symmetry of the ($6$-$1$) packings, which allows only a small boundary cost
between a ($6$-$1$) unit cell and a Barlow-packed unit cell.

\subsection{Densest packings for $\alpha > \sqrt{2}-1$}

The previously known densest alloys for $\alpha > \sqrt{2}-1$ all exhibit ratios
of small to large spheres of either one to one or two to one. Perhaps the best
known is the AlB$_2$ alloy, Fig. \ref{AlB2}, with two small and one large
spheres in its minimal basis, present in the densest packings over the range
$\sqrt{7/3}-1 \leq \alpha \leq 0.620$. This alloy phase can be described as
alternating layers of contacting small spheres packed in a honeycomb lattice and
large spheres packed in a triangular lattice. The alloy exhibits two maxima in
packing density, one at $\alpha = \sqrt{7/3} -1$ and the other at $\alpha =
1/\sqrt{3}$, with $\phi_{max}(\sqrt{7/3}-1,2/3) = 0.782112\dots$ and
$\phi_{max}(1/\sqrt{3},2/3) = 0.779205\dots$. At both maxima, the large spheres
are in contact both parallel and perpendicular to the planes of the layers.

\begin{figure}[ht!]
\centering
\includegraphics[width = 2.5in,viewport = 470 45 820 750,clip]{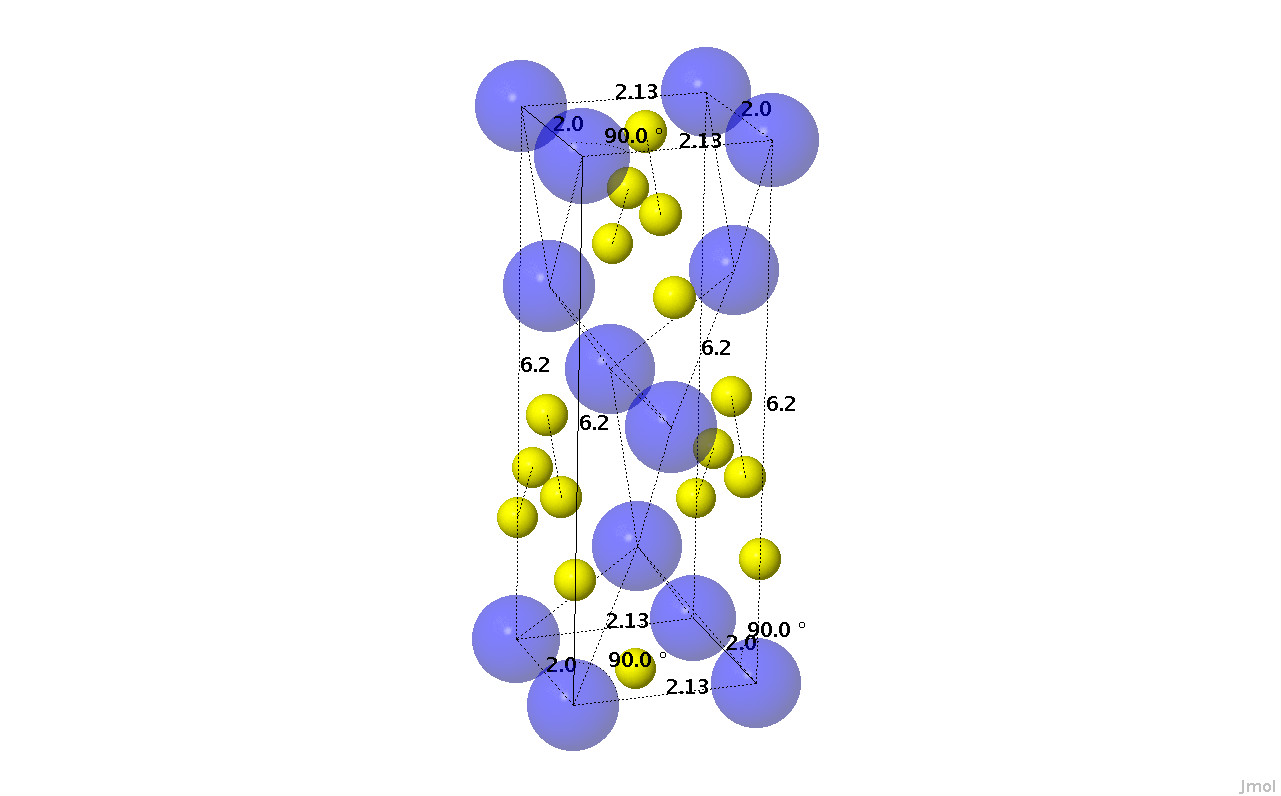}
\caption{(Color online) Putative densest packing ($\phi = 0.759$) at radius
ratio $\alpha=0.4597$ of four small and two large spheres in a periodic
fundamental cell. The HgBr$_2$ alloy belongs to the orthorhombic lattice
system.}
\label{HgBr2}
\end{figure}

The HgBr$_2$ alloy was recently discovered to be present in the densest packings
\cite{FD2009a}, a finding that our results support. The alloy phase belongs to
the orthorhombic lattice system, and its minimal basis is composed of four small
and two large spheres. It exhibits many local maxima and minima in packing
fraction over the approximate range $0.443 \leq \alpha \leq 0.468$ that it
appears in the densest packings, suggesting that its contact network changes
many times over this range. Due to these many local extrema, the packing
fraction of the pure HgBr$_2$ alloy phase does not vary much, with $0.752 \leq
\phi_{max}(\alpha,2/3) \leq 0.760$ over the aforementioned range in $\alpha$.
Figure \ref{HgBr2} highlights the large-large and small-small sphere contacts in
the alloy at $\alpha = 0.4597$.

\begin{figure}[ht!]
\centering
\includegraphics[width = 3.2in,viewport = 210 28 1095 740,clip]{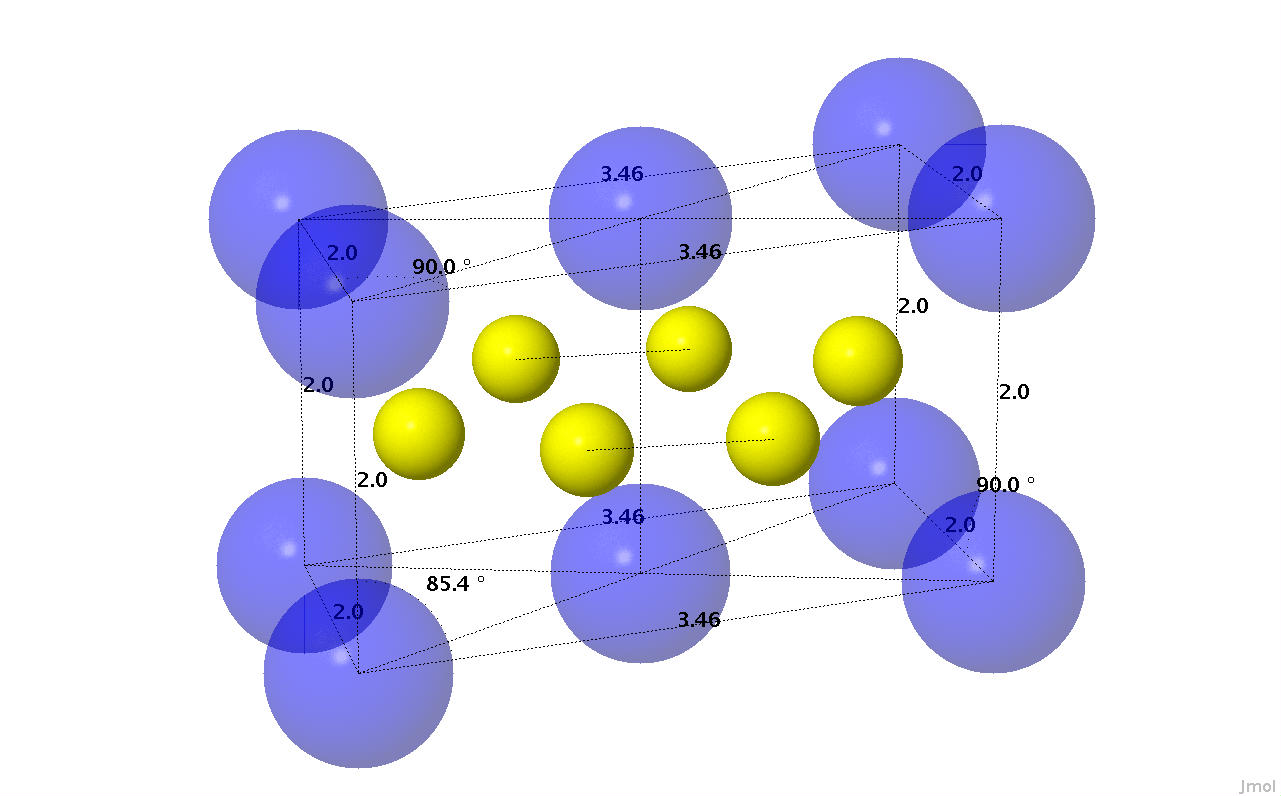}
\caption{(Color online) Putative densest packing ($\phi = 0.758$) at radius
ratio $\alpha=0.4995$ of two small spheres and one large sphere in a periodic
fundamental cell. The fundamental cell of this AuTe$_2$ alloy belongs to the
monoclinic lattice system.}
\label{AuTe2}
\end{figure}

The AuTe$_2$ alloy was also recently discovered to be present in the densest
packings \cite{FD2009a}. The alloy phase belongs to the monoclinic lattice
system, and its minimal basis is composed of two small spheres and one large
sphere. In a previous work \cite{HJST2011a}, our phase diagram indicated that
this alloy was present in the densest packings over the approximate range $0.480
\leq \alpha \leq 0.528$, the same range that was reported in Ref.
\cite{FD2009a}. However, we correct this statement to read that the alloy
appears to be present in the densest packings over the range $0.488 \leq \alpha
\leq 0.528$, with the ($4$-$2$) alloy present over the approximate range $0.480
\leq \alpha \leq 0.488$. Figure \ref{AuTe2} highlights the large-large and
small-small sphere contacts in the AuTe$_2$ alloy at $\alpha = 0.4995$.

\begin{figure}[ht!]
\centering
\includegraphics[width = 3.2in,viewport = 215 5 1100 840,clip]{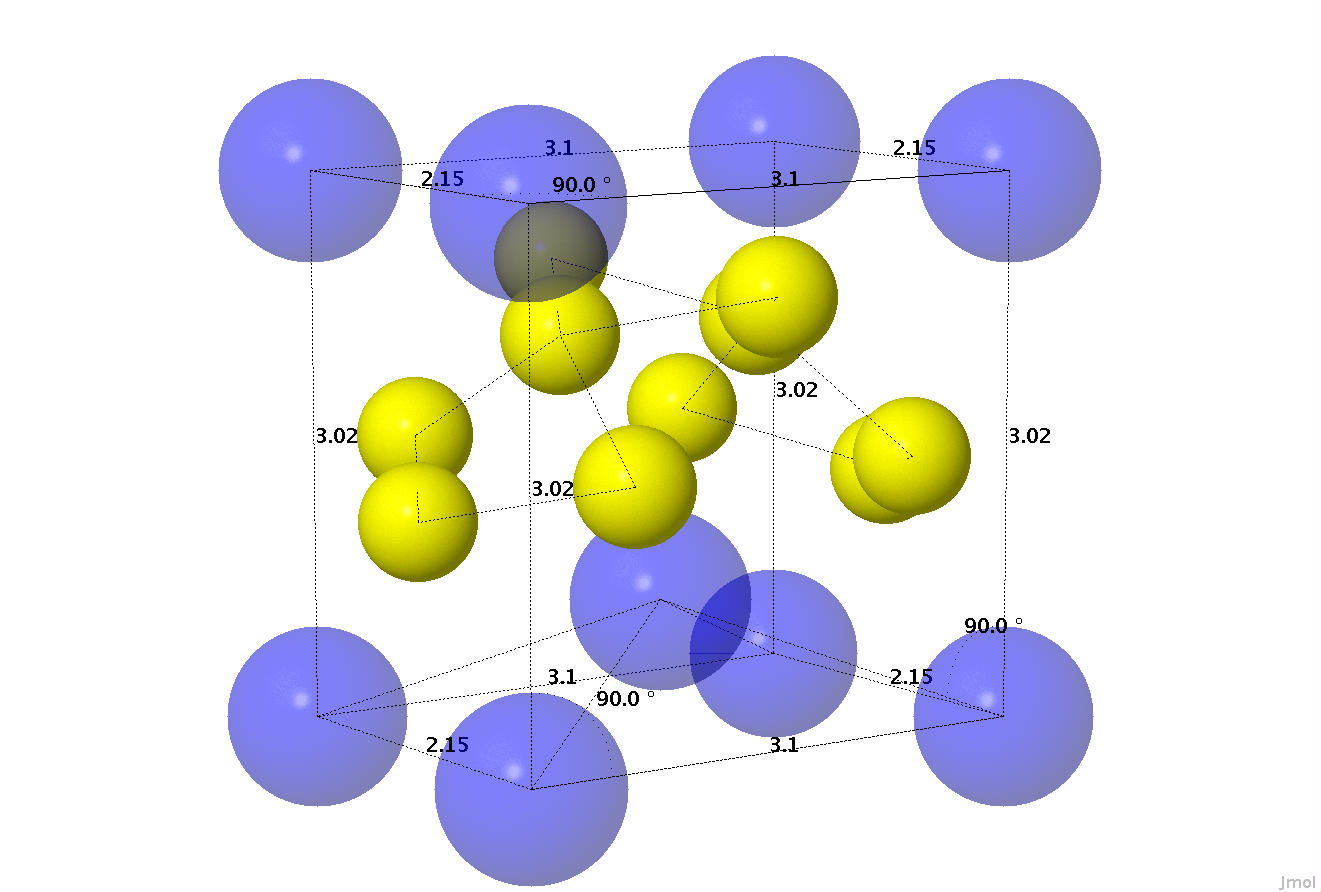}
\caption{(Color online) Putative densest packing ($\phi = 0.746$) at radius
ratio $\alpha=0.643$ of six small and two large spheres in a periodic
fundamental cell. The A$_3$B alloy belongs to the orthorhombic lattice system.}
\label{A3B}
\end{figure}

The A$_3$B alloy was recently discovered \cite{OH2011a} to be the densest-known
binary alloy over the approximate range $0.619 < \alpha < \alpha^*$. Previously,
the AlB$_2$ alloy was thought to be the only alloy denser than phase-separated
monodisperse small and large spheres for high enough $\alpha$, where it becomes
less dense than a phase-separated monodisperse packing for all $\alpha >
0.623387\dots$. The fundamental cell of the A$_3$B alloy, depicted in Fig.
\ref{A3B}, belongs to the orthorhombic lattice system, and it contains six small
and two large spheres.

\begin{figure}[ht!]
\centering
\includegraphics[width = 3.2in,viewport = 275 30 1020 740,clip]{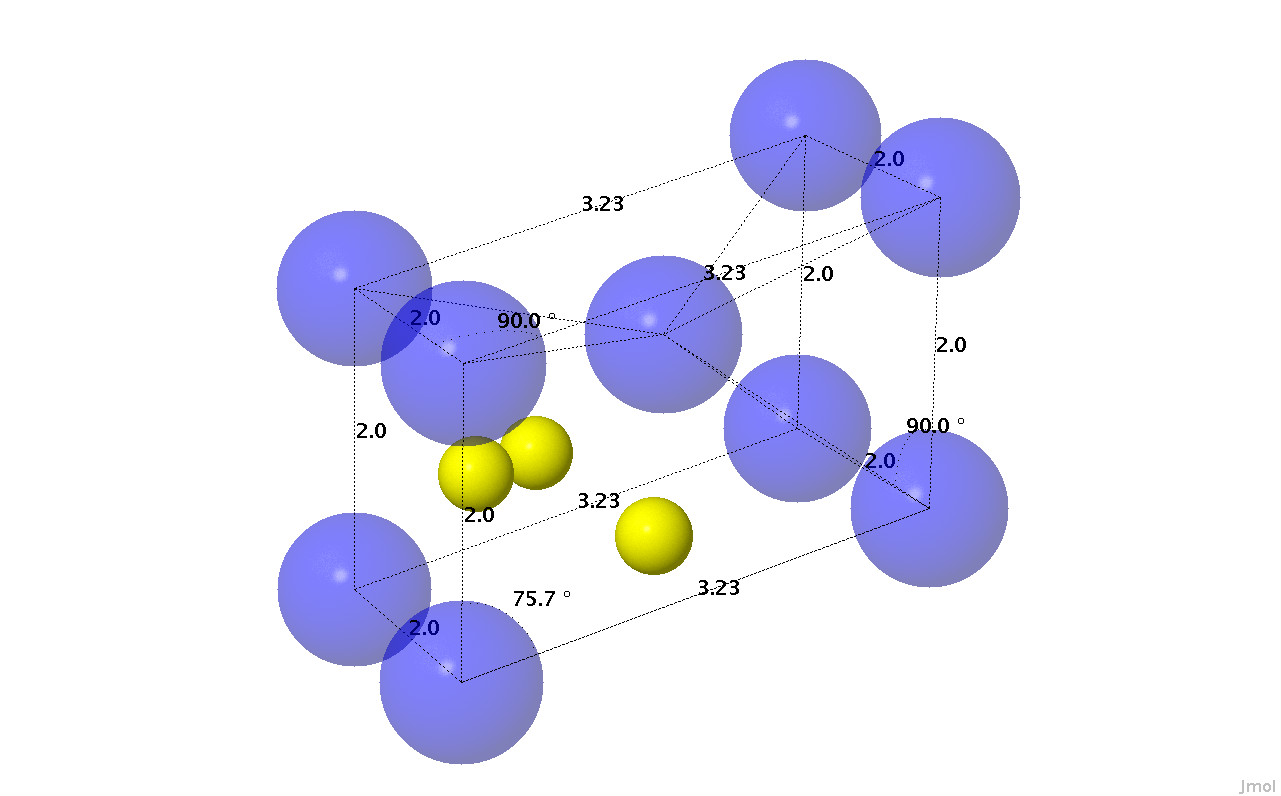}
\caption{(Color online) Putative densest packing ($\phi = 0.746$) at radius
ratio $\alpha=0.4881$ of two small and two large spheres in a periodic
fundamental cell. The ($2$-$2$)$^*$ alloy belongs to the monoclinic lattice
system.}
\label{2-2star}
\end{figure}

In our investigations, the ($2$-$2$)$^*$ alloy exhibits the same packing
fraction (error of less than $10^{-4}$) as the ``Structure 2'' alloy described
in Ref. \cite{MH2010a}, which has four small and four large spheres in its
minimal basis. Within the error, we were unable to determine whether doubling
the minimal basis from four to eight spheres, just as increasing the minimal
basis from four small and four large to six small and six large for the
($6$-$6$) alloy, results in an increased packing fraction. We suggest that it
might result in an increase of less than $10^{-4}$ over the approximate range
$0.480 \leq \alpha \leq 0.497$ that the alloy appears in the densest packings,
though we would need to run simulations at higher accuracy in order to confirm
or reject this hypothesis. Figure \ref{2-2star} depicts the fundamental cell of
the ($2$-$2$)$^*$ packing, which belongs to the monoclinic lattice system, at
$\alpha = 0.4881$.

\begin{figure}[ht!]
\centering
\includegraphics[width = 3.2in,viewport = 290 5 1010 770,clip]{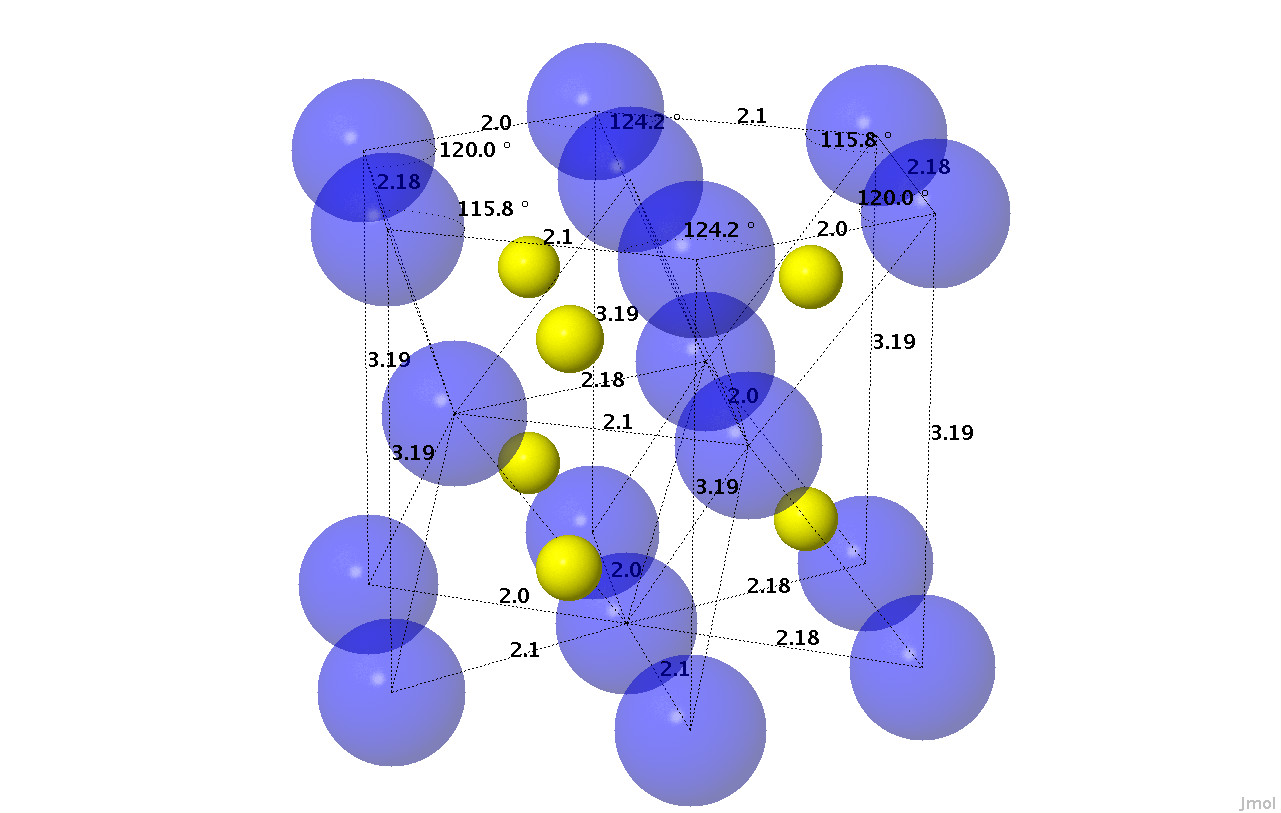}
\caption{(Color online) Putative densest packing ($\phi = 0.758$) at radius
ratio $\alpha=0.4483$ of six small and six large spheres in a periodic
fundamental cell. The ($6$-$6$) alloy belongs to the triclinic lattice system.}
\label{6-6}
\end{figure}

The ``Structure 1'' alloy described in Ref. \cite{MH2010a} has a minimal basis
of four small and four large spheres. However, our simulations show that
increasing the minimal basis to six small and six large spheres results in
identification of a denser alloy. Specifically, we find (error of less than
$10^{-4}$) that the ($6$-$6$) alloy exhibits the same packing fraction as the
``Structure 1'' alloy over the approximate range $0.414 \leq \alpha < 0.428$,
and a denser packing fraction over the approximate range $0.428 \leq \alpha \leq
0.457$. The ($6$-$6$) alloy phase, illustrated in Fig. \ref{6-6}, belongs to the
triclinic lattice system and is similar to a skewed and stretched version of the
NaCl alloy. It is unclear if increasing the minimal basis beyond $12$ spheres
will result in a slightly denser alloy over the approximate range $0.414 \leq
\alpha \leq 0.457$ that the ($6$-$6$) alloy appears to be present in the densest
packings.

\begin{figure}[ht!]
\centering
\includegraphics[width = 2.5in,viewport = 470 10 820 800,clip]{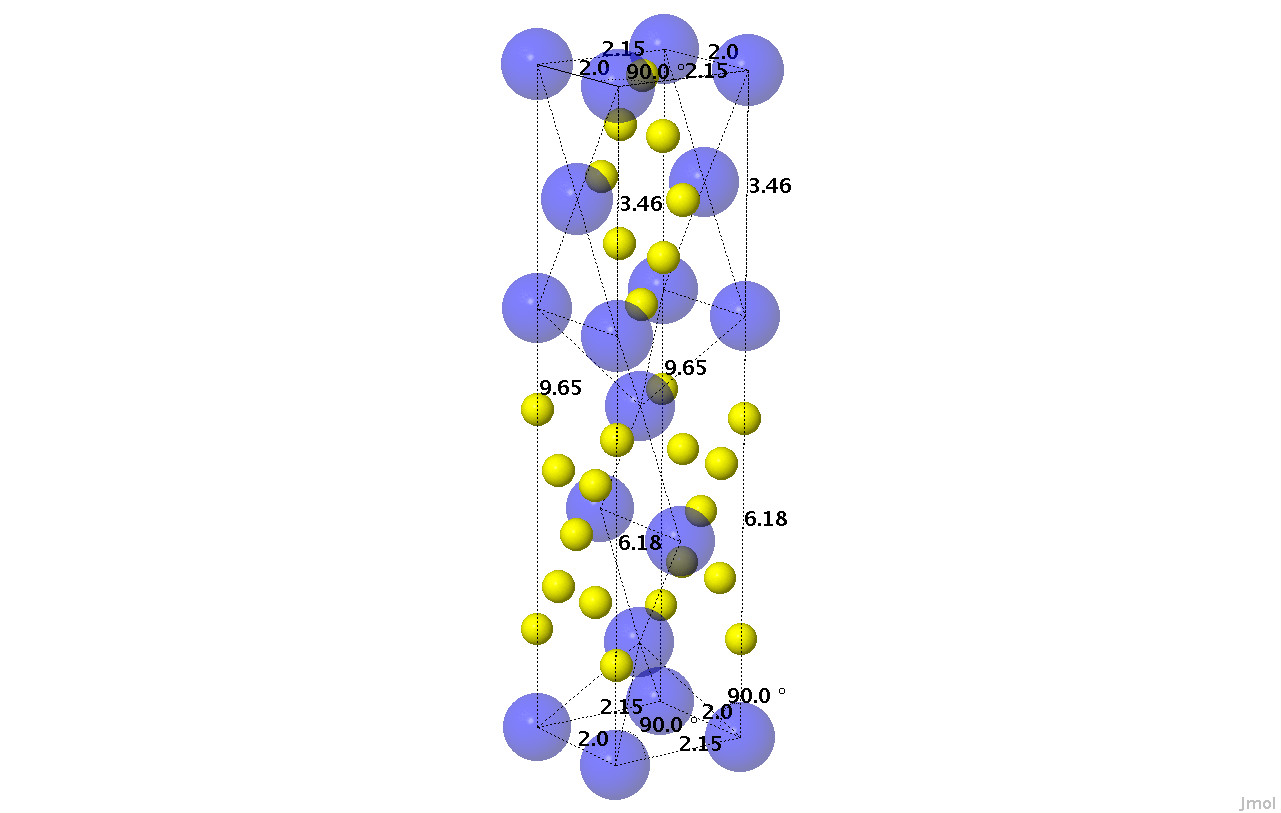}
\caption{(Color online) Putative densest packing ($\phi = 0.752$) at radius
ratio $\alpha=0.4682$ of seven small and three large spheres in a periodic
fundamental cell. The ($7$-$3$) alloy belongs to the orthorhombic lattice
system.}
\label{7-3}
\end{figure}

The ($7$-$3$) alloy appears in the densest packings over the approximate range
$0.468 \leq \alpha \leq 0.480$. Its minimal basis contains seven small and three
large spheres, and the alloy belongs to the orthorhombic lattice system. It
exhibits a local maximum at $\phi_{max}(0.474568\dots,7/10) = 0.752189\dots$.
For $\alpha < 0.474568$ where the alloy appears in the densest packings, it is
stabilized along the $x$-axis and $y$-axis (the axes parallel to the basis
vectors in Fig. \ref{7-3} shown as having lengths $2.0$ and $9.65$,
respectively) by large-large contacts, and along the $z$-axis it is stabilized
by small-large contacts.  For $\alpha > 0.474568\dots$ where the alloy appears
in the densest packings, the $y$-axis is stabilized both by large-large and
small-large contacts. 

\begin{figure}[ht!]
\centering
\includegraphics[width = 2.5in,viewport = 445 5 845 855,clip]{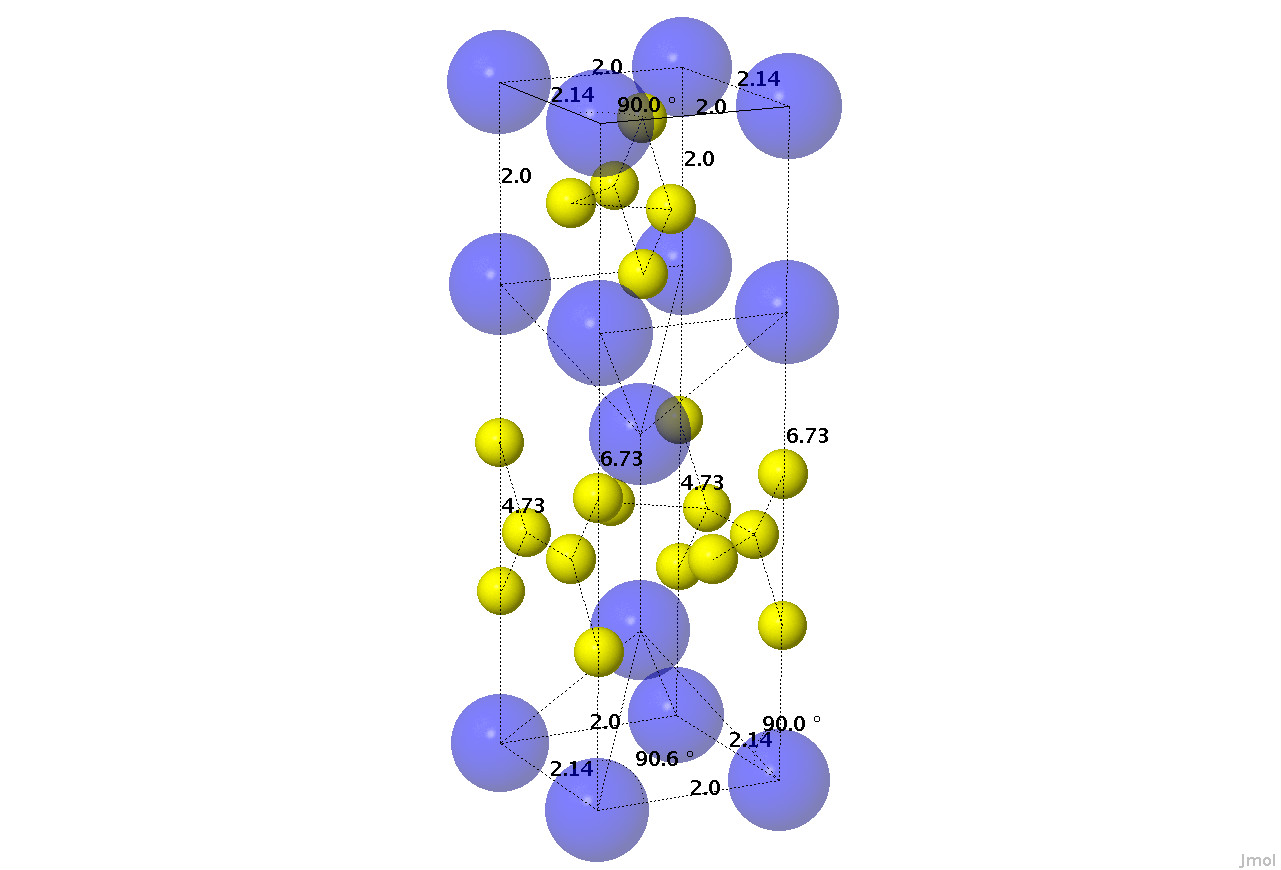}
\caption{(Color online) Putative densest packing ($\phi = 0.746$) at radius
ratio $\alpha=0.4824$ of five small and two large spheres in a periodic
fundamental cell. The ($5$-$2$) alloy belongs to the monoclinic lattice system.}
\label{5-2}
\end{figure}

The ($5$-$2$) alloy belongs to the monoclinic lattice system and appears in the
densest packings only over a very short range of $\alpha$, approximately $0.480
\leq \alpha \leq 0.483$. This brief appearance is not due to a local maximum in
packing fraction in the ($5$-$2$) alloy, but rather to the fact that other dense
binary alloys, such as the ($4$-$2$), HgBr$_2$, AuTe$_2$, and ($7$-$3$), all
happen to exhibit relatively lower packing fractions over this range of
$\alpha$. One notable feature of the ($5$-$2$) alloy is a large void space,
visible in Fig. \ref{5-2} just above the four large spheres at the top right of
the image.

\begin{figure}[ht!]
\centering
\includegraphics[width = 2.5in,viewport = 450 5 820 860,clip]{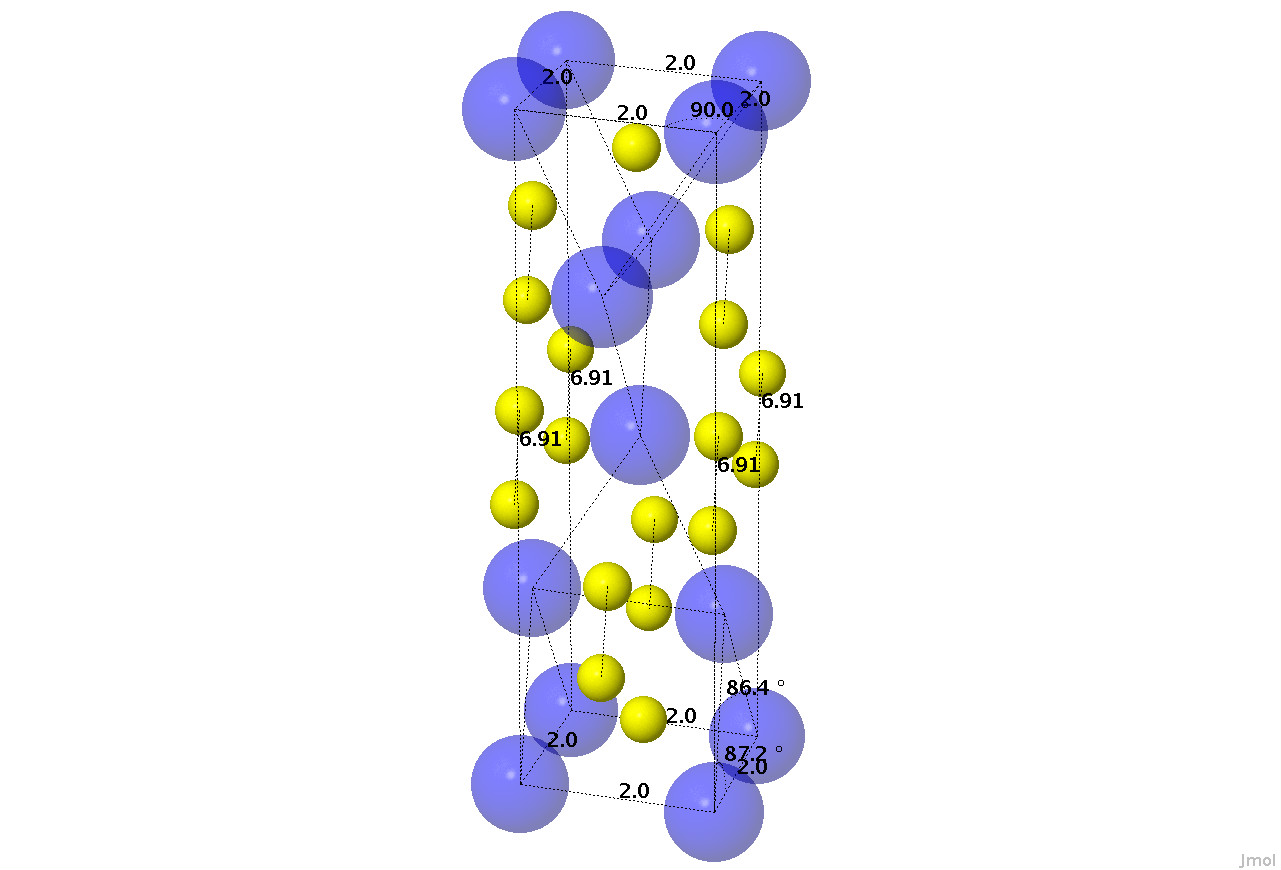}
\caption{(Color online) Putative densest packing ($\phi = 0.748$) at radius
ratio $\alpha=0.4853$ of four small and two large spheres in a periodic
fundamental cell. The ($4$-$2$) alloy belongs to the triclinic lattice system.}
\label{4-2}
\end{figure}

The ($4$-$2$) alloy, illustrated in Fig. \ref{4-2}, is present in the densest
packing over the approximate range $0.480 \leq \alpha \leq 0.488$. It has four
small and two large spheres in its minimal basis and belongs to the triclinic
lattice system. The alloy is very similar in packing fraction to both the
AuTe$_2$ and HgBr$_2$ alloys over the range of $\alpha$ in which it appears in
the densest packings. However, the ($4$-$2$) alloy fundamental cell has less
symmetry, and the distortion in the cell away from a monoclinic arrangement
allows for a slightly increased packing fraction over the range $0.480 \leq
\alpha \leq 0.488$ as compared to the AuTe$_2$ alloy.

\subsection{Another dense packing}
\begin{figure}[ht!]
\centering
\includegraphics[width = 3.2in,viewport = 185 10 1130 835,clip]{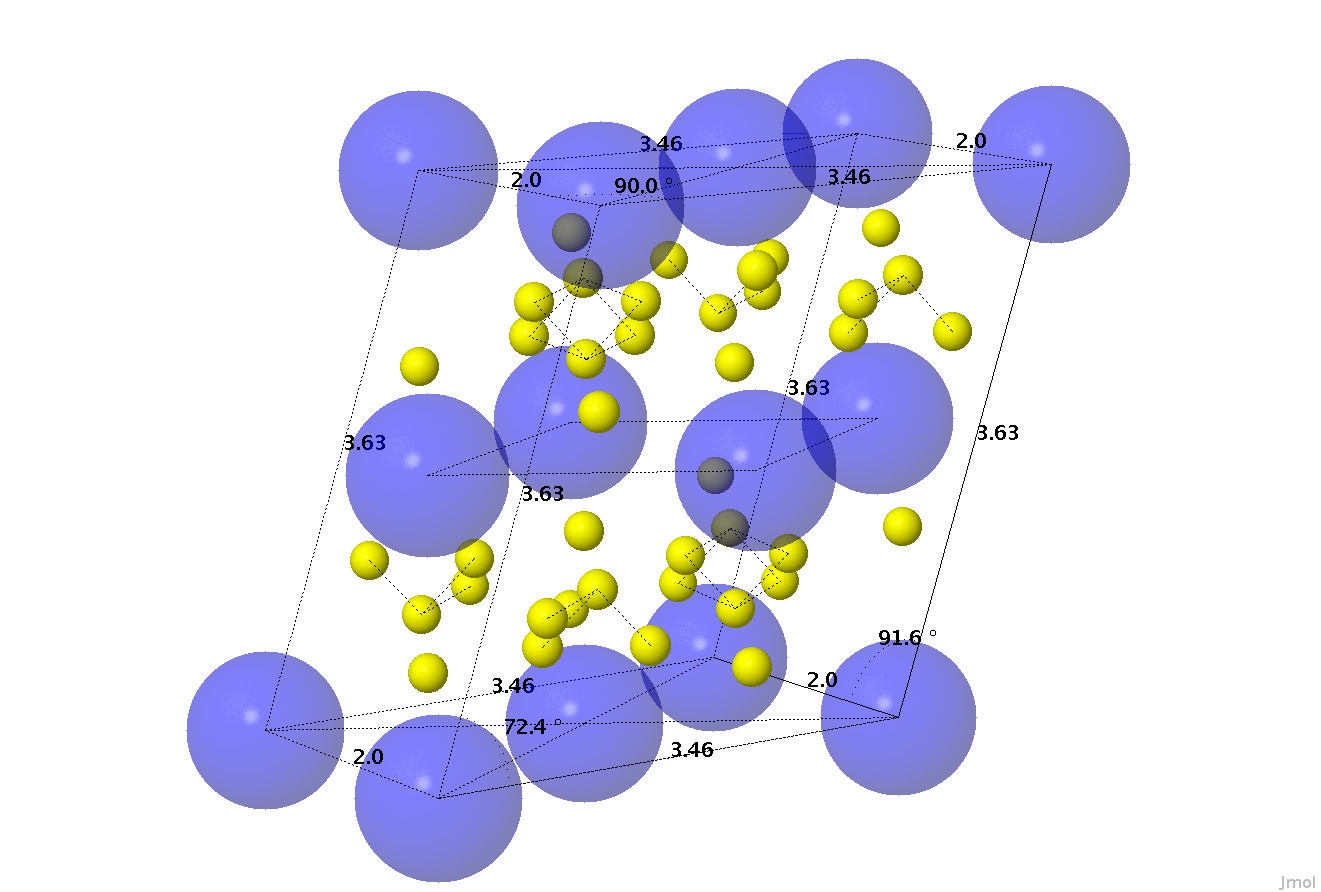}
\caption{(Color online) Very dense packing ($\phi = 0.787$) at radius ratio
$\alpha=0.2503$ of eight small and one large spheres in a periodic fundamental
cell. This ($8$-$1$) alloy did not appear in the densest packings in our
simulations; though it is denser over the approximate range $0.244 < \alpha <
0.253$ at $x = 8/9$ than the combination of two phase-separated phases where one is any other known alloy and the other is a monodisperse phase, the combination of two
phase-separated alloy phases (the ($10$-$1$) and XY$_4$ phases) is denser. The
($8$-$1$) alloy belongs to the triclinic lattice system.}
\label{8-1}
\end{figure}

The ($8$-$1$) alloy is particularly dense over the range $0.244 \leq \alpha \leq
0.253$, but it does not appear in the densest packings. As can be seen in Fig.
\ref{8-1}, the ($8$-$1$) alloy consists of large spheres arranged in a unit cell
belonging to the triclinic lattice system, with small spheres in the primary and
secondary interstices. There are six small spheres arranged in a skewed
octahedron in each primary interstice, and a small sphere in each secondary
interstice, where there are twice as many secondary interstices as primary
interstices. Interestingly, at $x=8/9$, the ($8$-$1$) alloy is denser than any
phase-separated combination of a different alloy phase and a monodisperse phase.
Nevertheless, a phase-separated combination of the ($10$-$1$) and XY$_4$ alloy
phases permits a denser packing at $x=8/9$ than a single ($8$-$1$) phase. 

\section{Conclusions}
\label{conclusions}
We have surveyed the $\alpha$-$x$ plane at high resolution in both $\alpha$ and
$x$ in order to find the densest binary sphere packings in ${\mathbb R}^d$ for
minimal bases of up to $12$ spheres. Employing an implementation of the TJ
algorithm, we have identified all of the previously known alloys present in the
densest packings and found many new alloys as well. We have described and
presented the structure and symmetries of these densest alloys, and we have
classified them by the composition of their minimal bases and lattice system
characterization of their unit cells. Taken together, these results demonstrate
that there is a broad diversity of alloys present in the densest packings,
including alloys with large minimal bases and uncommon small to large sphere
number ratios, {\it e.g.,} the ($6$-$6$), ($7$-$3$) and ($5$-$2$) alloys.

We have also discussed the densest binary packings in ${\mathbb R}^3$, which are
composed of phase-separated densest Barlow and alloy phases. We have proved that
for any packing consisting of $\eta$ different types of objects in ${\mathbb
R}^d$, there is a densest packing that consists of no more than $\eta$ different
phases. Additionally, we have discussed the properties of the function
$\phi_{max}(\alpha,x)$ and described its features in the limit as
$(\alpha,x)\rightarrow(0,1)$.  

One implication of our findings is that that entropic (free-volume maximizing) 
particle interactions contribute to the structural diversity of mechanically 
stable and ground-state structures of atomic, molecular, and granular solids. 
Additionally, the structures we have identified can be useful as known points 
of departure when investigating experimentally the properties of binary solids 
composed of particles that exhibit steep isotropic pair-repulsion, including 
of dense atomic and molecular phases at high temperatures and pressures. In 
particular, some of the novel structures that we have identified could 
correspond to currently unidentified stable atomic and molecular states of 
matter. We note though that these novel structures would be most relevant to 
binary phases, perhaps at high temperatures and pressures, where the two 
particle-types exhibit roughly additive pair repulsion. 

Though we have limited ourselves in this work to minimal bases of $12$ or fewer
spheres, the discovery of the ($7$-$3$), ($6$-$6$), and ($5$-$2$) alloys
suggests that periodic structures with minimal bases larger than $12$, further
directionally-periodic, quasicrystalline and disordered structures might be
present among the densest packings. This may also include a densest alloy,
composed of a minimal basis of greater than $12$ spheres, for $\alpha >
\alpha^*$.

It is possible that the densest packings we have identified have a relation to
the structures of glassy binary sphere solids and/or to those of binary sphere
liquids near the freezing point. For example, recent work \cite{Zeng2011a}
demonstrates that a binary metallic glass, Ce$_{75}$Al$_{25}$ ($x = 1/4$, with
the atomic volume of Ce about twice that of Al at ambient pressure), exhibiting
long range structural order can be created by a melt-spinning process. Further
investigation suggests that the long range fcc order is derived from the binary
material's densely packed configuration during spinning. It is possible that
other binary glasses could exhibit similar long range order related to the
densest packings of binary spheres at specified ($\alpha$, $x$). 

In future work, we will employ the TJ algorithm to study maximally random jammed
(MRJ) packings of binary spheres at specified $(\alpha,x)$. This is a
complicated problem, in part due to the difficulty in defining a MRJ state for
binary packings. Nevertheless, a thorough investigation may reveal similarities
between the MRJ packings and the densest binary packings at the same radius
ratios $\alpha$ and small sphere relative concentrations $x$.

\bigskip
\noindent {\bf ACKNOWLEDGEMENTS:}
\medskip

This work was supported by the MRSEC Program of the National Science Foundation
under Award Number DMR-0820341. The simulations of dense binary sphere packings
using the TJ algorithm were performed primarily on computational resources
supported by the Princeton Institute for Computational Science and Engineering
(PICSciE) and the Office of Information Technology's High Performance Computing
Center and Visualization Laboratory at Princeton University.

\appendix
\section{Implementation of the Algorithm}
\label{AlgMods}
The implementation of the TJ algorithm for binary spheres requires a multistep
process. This multistep process is run many times with many different initial
configurations of spheres, in particular over initial configurations where the
positions of large and small spheres are interchanged within a periodic unit
cell.

In the first step of the process, an initial configuration of nonoverlapping
small and large spheres is generated. This is accomplished through a random
sequential addition \cite{TorquatoRHM2002} process of adding large spheres to a
randomly generated unit cell, then choosing which spheres will be designated
small and which will be large. The cell is randomly generated in ${\mathbb R}^3$
using five random variables which represent the angles between the three lattice
vectors and the relative lengths of the second and third lattice vectors to the
first, which is normalized. The angles and lattice vector lengths are required
to fall within certain bounds, and the total volume of the cell is bounded from
below. These bounds are useful algorithmically in preventing overlap between the
spheres in the unit cell and their images in non-adjacent unit cells.

The next step is to solve the linear programming problem (\ref{LP}) assuming
that ${\cal R} = 0$, where we recall that ${\cal R}$ represents the higher order
terms in the linear expansion of the packing problem and its constraints. After
solving the problem (\ref{LP}), it is necessary to check if any overlap is
present between spheres in the unit cell, or between the spheres in the unit
cell and their images in surrounding cells. This step is necessary because
${\cal R}$ is sometimes negative, which can result in a solution to the problem
(\ref{LP}) with overlap between spheres. The higher order terms ${\cal R}$ can
be written,
\begin{equation}
{\cal R} = 3\Delta {\bf r}_{ij}\cdot{\boldsymbol\varepsilon}^2\cdot\Delta {\bf
r}_{ij} + 2|{\boldsymbol\varepsilon}\cdot{\bf r}_{ij} + \Delta {\bf r}_{ij}|^2 -
|{\boldsymbol\varepsilon}\cdot({\bf r}_{ij} - \Delta {\bf r}_{ij})|^2 - |\Delta
{\bf r}_{ij} - {\boldsymbol\varepsilon}\cdot\Delta {\bf r}_{ij}|^2.
\label{reformHigherTerms}
\end{equation}

The first two terms in Eq. (\ref{reformHigherTerms}) are positive semi-definite
and the last two negative semi-definite. When the last two terms are greater in
absolute value than the first two, this indicates that overlap between spheres
can exist in a correctly-solved linearized programming problem.

If overlap is present between any two spheres, the lattice vectors
$\{\lambda_i\}$ are resized by a constant such that all previously overlapping
spheres are now out of contact by at least some small $\delta > 0$. All spheres
in the packing are subsequently displaced in random directions by small random
distances such that all distances are less than or equal to $\delta$. This step
is necessary so that the algorithm does not become stuck in a loop.
Additionally, the randomness introduced by this step aids in finding densest
packings. If no overlap is present, then no resize of the lattice vectors is
necessary. In either case, the algorithm repeats until there is no solution to
the problem (\ref{LP}) that reduces packing fraction, {\it i.e.}, the packing is
strictly jammed. 

\section{Proof of Theorem 1}
\label{proof1}
To prove theorem 1, we will utilize the fundamental theorem of linear
programming, which states that there is always a globally optimal solution to a
linear programming problem (that has at least one solution) located at (at
least) one of the vertices of the convex polytope that is the problem's feasible
region. In this case, the objective function of the linear programming problem
is the denominator of Eq. (\ref{packingFractionFull}), and the feasible region
is defined by the constraints on the variables $x_i^{F_i}$ that result from
conservation of particle numbers. We first outline these constraints, then apply
the fundamental theorem of linear programming to prove that for a packing of
$\eta$ different sphere sizes in ${\mathbb R}^d$, there is at least one densest
packing composed of no more than $\eta$ phase-separated phases.

As the $X^{(j)}$ are relative fractions, $\sum_{j=1}^{\eta}X^{(j)} = 1$ and
$X^{(j)} \geq 0$ for all $j$ due to conservation of particle numbers. For the
same reason, summing up all the relative fractions $x_i^j$ of spheres of type
$j$ in each alloy type $i$, we have,
\begin{equation}
\sum_{i=1}^{\beta}\frac{S_i^j}{S_i^{F_i}}x_i^{F_i} = X^{(j)},
\label{etaTypesConstraints}
\end{equation}
where we note that each term in the sum on the left hand side of Eqs.
(\ref{etaTypesConstraints}) is equal to $x_i^j$. It follows from Eqs.
(\ref{etaTypesConstraints}) and the nonnegativity of the $x_i^{F_i}$ that
\begin{equation}
\frac{S_i^j}{S_i^{F_i}}x_i^{F_i} \leq X^{(j)},
\label{posHyperRectPrismBounds}
\end{equation}
or for each phase $i$ and sphere type $j$, the relative fraction $x_i^j$ of
spheres of type $j$ in phase $i$ is less than or equal to the total relative
fraction of spheres $X^{(j)}$ of type $j$.

For each variable $x_i^{F_i}$, there is one or more indices $j$ with nonzero
$S_i^j$ that yields the most restrictive constraint on $x_i^{F_i}$ from Eqs.
(\ref{posHyperRectPrismBounds}), {\it i.e.}, one that yields the minimum value
of $X^{(j)}/S_i^j$ from among all $j$. To be consistent with previous index
nomenclature, we term this first such index $J_i$. Employing this form, the Eqs.
(\ref{posHyperRectPrismBounds}) reduce to $\beta$ inequalities that together
with the nonnegativity of the $x_i^{F_i}$ produce the $\beta$ bounds,
\begin{equation}
0 \leq x_i^{F_i} \leq \frac{S_i^{F_i}}{S_i^{J_i}}X^{(J_i)},
\label{hyperRectPrism}
\end{equation}
which describe a hyper-rectangular prism in $\beta$-dimensional Euclidean space
in the variables $x_i^{F_i}$.

The $\eta$ equations (\ref{etaTypesConstraints}) describe $\eta$ hyperplanes in
$\beta$-dimensional Euclidean space. The convex polytope that forms the feasible
region consists of their intersection with each other and the hyper-rectangular
prism described by inequalities (\ref{hyperRectPrism}). We can thus formulate a
linear programming problem corresponding to the maximum of Eq.
(\ref{packingFractionFull}) as,
\begin{equation}
\min{\sum_{i=1}^{\beta}\frac{C_i}{S_i^{F_i}}x_i^{F_i}},
\label{linProgProb}
\end{equation}
subject to the equality constraints (\ref{etaTypesConstraints}) and the
inequality constraints (\ref{hyperRectPrism}). This allows us to prove Theorem
1, as follows.

\begin{proof}
Due to the monodisperse alloys, the hyperplanes defined by Eqs.
(\ref{etaTypesConstraints}) have at least one point of mutual intersection.
Terming the relative relative fractions of the monodisperse alloys $x_1^1\dots
x_{\eta}^{\eta}$, this point is $\{x_1^1\dots x_{\eta}^{\eta}\} =
\{X^{(1)}...X^{(\eta)}\}$ and $\{x_i^{F_i} = 0;\,\, i > \eta\}$, namely the
packing where all phases are monodisperse. As this point is contained within the
feasible region, we can apply the fundamental theorem of linear programming to
state that there is a globally optimal solution to the problem
(\ref{linProgProb}) located at (at least) one of the vertices of the convex
polytope formed by the intersection of the $\eta$ hyperplanes described by Eqs.
(\ref{etaTypesConstraints}) and the hyper-rectangular prism described by
inequalities (\ref{hyperRectPrism}).

To prove that there is a globally optimal solution that consists of no more than
$\eta$ phase-separated phases, we will show that there can be no more than
$\eta$ positive values of $x_i^{F_i}$ at any of the vertices of the feasible
region. To this end, we will first describe the vertices of the feasible region
where the $\eta$ hyperplanes intersect the hyper-rectangular prism only at the
lower bounds of inequalities (\ref{hyperRectPrism}). Immediately following, we
will describe the vertices when the $\eta$ hyperplanes intersect at one or more
of the upper bounds of inequalities (\ref{hyperRectPrism}). 

The vertices where the $\eta$ hyperplanes intersect only at the lower bounds of
the hyper-rectangular prism consist of all combinations of no more than $\eta$
positive $x_i^{F_i}$ and no fewer than $\beta - \eta$ zero values of $x_i^{F_i}$
such that Eqs. (\ref{etaTypesConstraints}) are satisfied. There can be no more
than $\eta$ nonzero $x_i^{F_i}$ at a vertex because the vertices must occur
where at least $\beta$ equality constraints are satisfied, {\it i.e.}, where
$x_i^{F_i} = 0$ on at least $\beta - \eta$ of the halfplanes described by the
inequalities (\ref{hyperRectPrism}), and at the intersection of the $\eta$
hyperplanes described by Eqs. (\ref{etaTypesConstraints}). There can be fewer
than $\eta$ positive $x_i^{F_i}$, in particular if the intersection of the
$\eta$ $d$-dimensional hyperplanes describes a hyperplane embedded in $\beta$
dimensions that is more than $(\beta - \eta)$-dimensional, or if one of the
$\eta$ hyperplanes happens to intersect with the halfplane described by one of
the lower bounds of the inequalities (\ref{hyperRectPrism}) at a given vertex
where all nonzero $x_i^{F_i}$ are positive.

If the intersection of the $\eta$ hyperplanes simultaneously intersects $t$
faces of the hyper-rectangular prism at the upper bounds of inequalities
(\ref{hyperRectPrism}), then for a set of $t$ indices $\{q\}_t$, $x_q^{F_q} =
(S_q^{F_q}/S_q^{J_q})X^{(J_q)}$. At all points contained within this
intersection, all spheres of types $J_q$ are present only in their respective
phases $q$. This means that for all other $x_i^{F_i}$ where phase $i$ includes
any spheres of types $J_q$, $x_i^{F_i} = 0$. Excluding these phases and the
phases $\{q\}_t$, the problem is reduced to the remaining $\beta_t$ variables
with indices $\{i\}_t$ in $\beta_t$ dimensions, and where there are $\eta-t$
constraints of the form,
\begin{equation}
\sum_{\{i\}_t}\frac{S_i^j}{S_i^{F_i}}x_i^{F_i} = X^{(j)} -
\sum_{\{q\}_t}\frac{S_q^j}{S_q^{F_q}}x_q^{F_q}.
\label{reducedEtaTypesConstraints}
\end{equation}
In the Eqs. (\ref{reducedEtaTypesConstraints}), the index $j$ is for all
$j=1\dots\eta$ excluding the $j=\{J_q\}$, the sum on the left hand side runs
over the $\beta_t$ indices included in the set $\{i\}_t$, and the sum on the
right hand side runs over the $t$ indices included in the set $\{q\}_t$.

In this reduced case, the Eqs. (\ref{reducedEtaTypesConstraints}) imply new
upper bounds on the $\beta_t$ remaining $x_i^{F_i}$ which describe a new
hyper-rectangular prism in $\beta_t$ variables. However, the lower bounds on the
$\beta_t$ variables $x_i^{F_i}$ over the indices $\{i\}_t$ are the same;
consequently, the vertices of the feasible region where the $\eta-t$ hyperplanes
described by Eqs. (\ref{reducedEtaTypesConstraints}) intersect the reduced
hyper-rectangular prism only at its lower bounds include no more than $\eta -t$
positive $x_i^{F_i}$ and no fewer than $\beta_t-\eta +t$ zero values of the 
remaining $x_i^{F_i}$. Coupled with the $\beta-\beta_t-t$ zero values of
$x_i^{F_i}$ that include spheres of types $J_q$ and the $t$ values $x_q^{F_q} =
(S_q^{F_q}/S_q^{J_q})X^{(J_q)}$, this gives no more than $\eta$ positive
$x_i^{F_i}$ and no fewer than $\beta-\eta$ zero values of $x_i^{F_i}$.

The value $t$ cannot exceed $\eta$ because at $t = \eta$, $\beta_{t} = 0$, as
all phases must include at least one type of sphere. It follows that all
vertices must include between $t=0$ and $t=\eta$ intersections at the upper
bounds of the hyper-rectangular prism, and so we have proved that there cannot
be more than $\eta$ positive $x_i^{F_i}$ at any vertex. It follows directly,
since there is at least one global optimum on a vertex, that there is always at
least one densest packing that includes no more than $\eta$ phase-separated
phases.
\end{proof}

We note that there can also be global minima that include more than $\eta$
phases. This occurs when there is more than one vertex that is a global optimum
such that there are more than $\eta$ distinct nonzero $x_i^{F_i}$ in total in
the optimal vertices. Specifically, when there is more than one vertex that is a
global optimum, all convex combinations of these vertices are also global
optima. This leads to the conclusion that of $\beta$ possible phases, there
could potentially be $\beta$ phases present in a densest packing, with $\beta$
unbounded.

\section{Packing Fractions}
\label{packingFractions}
The packing fractions in this Appendix are for the periodic alloys that comprise
the densest packings at given values of $\alpha$. By comparing the alloys found
by the algorithm to densest alloys for which we have explicit representations of
the packing fraction, we have discovered that the packing fractions are in
general about $1.0\times10^{-4}$ too low, though sometimes as much as
$2.5\times10^{-4}$ too low. For this reason, and due to the resolution of our
survey in $\alpha$, we report packing fractions only to the third decimal place.
Additionally, we note that our algorithm, being constructive, always produces
packing fractions for real packings, and therefore the fractions presented are
all lower bounds on the true maximal fractions. Finally, we note that although
the comparisons we have made in the text to explicitly-known packing fractions
are for the precise values of $\alpha$ that we employed in our survey, the
values of $\alpha$ reported in this table have been rounded to the third decimal
place.

\begin{center}
\tablefirsthead{
\hline
\hline
$\alpha$  & \, ($11$-$1$) \, & \, ($10$-$1$) \, & \, ($6$-$1$)$_{10}$ \, & \,
($6$-$1$)$_{4,6,8}$ & \,\, $\alpha$ \,\, & \, ($6$-$6$) \, & \, ($7$-$3$) \, &
\, ($5$-$2$) \, & \, ($4$-$2$) \, & \, ($2$-$2$)$^*$  \\
\hline
}
\tablehead{
\hline
\multicolumn{11}{l}{\textit{continued from previous page}} \\
\hline
$\alpha$  & \, ($11$-$1$) \, & \, ($10$-$1$) \, & \, ($6$-$1$)$_{10}$ \, & \,
($6$-$1$)$_{8,6,4}$ & \,\, $\alpha$ \,\, & \, ($6$-$6$) \, & \, ($7$-$3$) \, &
\, ($5$-$2$) \, & \, ($4$-$2$) \, & \, ($2$-$2$)$^*$  \\
\hline}
\tabletail{
\hline
\multicolumn{11}{r}{\textit{continued on next page}} \\
\hline}
\tablelasttail{
\hline}
\tablecaption{Selected densest binary alloy packing fractions.}
\begin{supertabular}{c c c c c | c c c c c c}
\label{tabPackFrac}
0.217\, & 0.823  &   &   &   &  \,0.414\,  & 0.793  &   &   &   \\ 
0.220\, & 0.818  &   &   &   &  \,0.417\,  & 0.789  &   &   &   \\
$\dots$\, &   &   &   &   &  \,0.420\,  & 0.786  &   &   &   \\
0.225\, &   & 0.825  &   &   &  \,0.423\,  & 0.783  &   &   &   \\
0.228\, &   & 0.822  &   &   &  \,0.426\,  & 0.779  &   &   &   \\
0.230\, &   & 0.820  &   &   &  \,0.428\,  & 0.776  &   &   &   \\
0.233\, &   & 0.818  &   &   &  \,0.431\,  & 0.773  &   &   &   \\
$\dots$\, &   &   &   &   &  \,0.434\,  & 0.771  &   &   &   \\
0.245\, &   & 0.804  &   &   &  \,0.437\,  & 0.768  &   &   &   \\
0.247\, &   & 0.800  &   &   &  \,0.440\,  & 0.765  &   &   &   \\
0.250\, &   & 0.797  &   &   &  \,0.443\,  & 0.763  &   &   &   \\
0.253\, &   & 0.794  &   &   &  \,0.445\,  & 0.760  &   &   &   \\
0.256\, &   & 0.791  &   &   &  \,0.448\,  & 0.758  &   &   &   \\
0.258\, &   & 0.789  &   &   &  \,0.451\,  & 0.756  &   &   &   \\
0.261\, &   & 0.787  &   &   &  \,0.454\,  & 0.754  &   &   &   \\
0.264\, &   & 0.785  & 0.781  &   &  \,0.457\,  & 0.751  &   &   &   \\
0.267\, &   & 0.783  & 0.782  &   &  \,0.460\,  &   &   &   &   \\
0.270\, &   & 0.781  & 0.784  &   &  \,0.463\,  &   &   &   &   \\
0.273\, &   &   & 0.785  &   &  \,0.465\,  &   &   &   &   \\
0.275\, &   &   & 0.787  &   &  \,0.468\,  &   & 0.752  &   &   \\
0.278\, &   &   & 0.789  &   &  \,0.471\,  &   & 0.752  &   &   \\
0.281\, &   &   & 0.792  &   &  \,0.474\,  &   & 0.752  &   &   \\
0.284\, &   &   & 0.794  &   &  \,0.477\,  &   & 0.750  &   &   \\
0.287\, &   &   & 0.797  &   &  \,0.480\,  &   & 0.748  & 0.746  & 0.744  &
0.744  \\
0.289\, &   &   & 0.800  &   &  \,0.483\,  &   &   & 0.746  & 0.746  & 0.744  \\
0.292\, &   &   & 0.801  &   &  \,0.485\,  &   &   &   & 0.748  & 0.745  \\
0.295\, &   &   &   & 0.800  &  \,0.488\,  &   &   &   & 0.750  & 0.746  \\
0.298\, &   &   &   & 0.799  &  \,0.491\,  &   &   &   &   & 0.747 \\
0.301\, &   &   &   & 0.798  &  \,0.494\,  &   &   &   &   & 0.748 \\
0.304\, &   &   &   & 0.797  &  \,0.497\,  &   &   &   &   & 0.749 \\
0.307\, &   &   &   & 0.797  &   &   &   &   &   & \\
0.309\, &   &   &   & 0.796  &   &   &   &   &   & \\
0.312\, &   &   &   & 0.795  &   &   &   &   &   & \\
0.315\, &   &   &   & 0.794  &   &   &   &   &   & \\
0.318\, &   &   &   & 0.793  &   &   &   &   &   & \\
0.321\, &   &   &   & 0.792  &   &   &   &   &   & \\
0.324\, &   &   &   & 0.788  &   &   &   &   &   & \\
0.326\, &   &   &   & 0.786  &   &   &   &   &   & \\
0.329\, &   &   &   & 0.783  &   &   &   &   &   & \\
0.332\, &   &   &   & 0.780  &   &   &   &   &   & \\
0.335\, &   &   &   & 0.778  &   &   &   &   &   & \\
0.338\, &   &   &   & 0.775  &   &   &   &   &   & \\
0.341\, &   &   &   & 0.773  &   &   &   &   &   & \\
0.343\, &   &   &   & 0.771  &   &   &   &   &   & \\
0.346\, &   &   &   & 0.769  &   &   &   &   &   & \\
0.349\, &   &   &   & 0.768  &   &   &   &   &   & \\
0.352\, &   &   &   & 0.766  &   &   &   &   &   & \\
\hline
\end{supertabular}
\end{center}

\ifx\mcitethebibliography\mciteundefinedmacro
\PackageError{apsrevM.bst}{mciteplus.sty has not been loaded}
{This bibstyle requires the use of the mciteplus package.}\fi


\begin{mcitethebibliography}{68}
\expandafter\ifx\csname natexlab\endcsname\relax\def\natexlab#1{#1}\fi
\expandafter\ifx\csname bibnamefont\endcsname\relax
  \def\bibnamefont#1{#1}\fi
\expandafter\ifx\csname bibfnamefont\endcsname\relax
  \def\bibfnamefont#1{#1}\fi
\expandafter\ifx\csname citenamefont\endcsname\relax
  \def\citenamefont#1{#1}\fi
\expandafter\ifx\csname url\endcsname\relax
  \def\url#1{\texttt{#1}}\fi
\expandafter\ifx\csname urlprefix\endcsname\relax\def\urlprefix{URL }\fi
\providecommand{\bibinfo}[2]{#2}
\providecommand{\eprint}[2][]{\url{#2}}

\bibitem[{\citenamefont{Torquato and Stillinger}(2010)}]{TS2010a}
\bibinfo{author}{\bibfnamefont{S.}~\bibnamefont{Torquato}} \bibnamefont{and}
  \bibinfo{author}{\bibfnamefont{F.~H.} \bibnamefont{Stillinger}},
  \bibinfo{journal}{Rev. Mod. Phys.} \textbf{\bibinfo{volume}{82}},
  \bibinfo{pages}{2633} (\bibinfo{year}{2010})\relax
\mciteBstWouldAddEndPuncttrue
\mciteSetBstMidEndSepPunct{\mcitedefaultmidpunct}
{\mcitedefaultendpunct}{\mcitedefaultseppunct}\relax
\EndOfBibitem
\bibitem[{\citenamefont{Pollack}(1964)}]{Pollack1964a}
\bibinfo{author}{\bibfnamefont{G.~L.} \bibnamefont{Pollack}},
  \bibinfo{journal}{Rev. Mod. Phys.} \textbf{\bibinfo{volume}{36}},
  \bibinfo{pages}{748} (\bibinfo{year}{1964})\relax
\mciteBstWouldAddEndPuncttrue
\mciteSetBstMidEndSepPunct{\mcitedefaultmidpunct}
{\mcitedefaultendpunct}{\mcitedefaultseppunct}\relax
\EndOfBibitem
\bibitem[{\citenamefont{Sanders}(1980)}]{Sanders1980a}
\bibinfo{author}{\bibfnamefont{J.~V.} \bibnamefont{Sanders}},
  \bibinfo{journal}{Phil. Mag. A} \textbf{\bibinfo{volume}{42}},
  \bibinfo{pages}{705} (\bibinfo{year}{1980})\relax
\mciteBstWouldAddEndPuncttrue
\mciteSetBstMidEndSepPunct{\mcitedefaultmidpunct}
{\mcitedefaultendpunct}{\mcitedefaultseppunct}\relax
\EndOfBibitem
\bibitem[{\citenamefont{Bartlett et~al.}(1992)\citenamefont{Bartlett, Ottewill,
  and Pusey}}]{BOP1992a}
\bibinfo{author}{\bibfnamefont{P.}~\bibnamefont{Bartlett}},
  \bibinfo{author}{\bibfnamefont{R.~H.} \bibnamefont{Ottewill}},
  \bibnamefont{and} \bibinfo{author}{\bibfnamefont{P.~N.} \bibnamefont{Pusey}},
  \bibinfo{journal}{Phys. Rev. Lett.} \textbf{\bibinfo{volume}{68}},
  \bibinfo{pages}{3801} (\bibinfo{year}{1992})\relax
\mciteBstWouldAddEndPuncttrue
\mciteSetBstMidEndSepPunct{\mcitedefaultmidpunct}
{\mcitedefaultendpunct}{\mcitedefaultseppunct}\relax
\EndOfBibitem
\bibitem[{\citenamefont{Vlot et~al.}(1997)\citenamefont{Vlot, Huitema,
  de~Vooys, and van~der Eerden}}]{VHVE1997a}
\bibinfo{author}{\bibfnamefont{M.~J.} \bibnamefont{Vlot}},
  \bibinfo{author}{\bibfnamefont{H.~E.~A.} \bibnamefont{Huitema}},
  \bibinfo{author}{\bibfnamefont{A.}~\bibnamefont{de~Vooys}}, \bibnamefont{and}
  \bibinfo{author}{\bibfnamefont{J.~P.} \bibnamefont{van~der Eerden}},
  \bibinfo{journal}{J. Chem. Phys.} \textbf{\bibinfo{volume}{107}},
  \bibinfo{pages}{4345} (\bibinfo{year}{1997})\relax
\mciteBstWouldAddEndPuncttrue
\mciteSetBstMidEndSepPunct{\mcitedefaultmidpunct}
{\mcitedefaultendpunct}{\mcitedefaultseppunct}\relax
\EndOfBibitem
\bibitem[{\citenamefont{Middleton et~al.}(2001)\citenamefont{Middleton,
  Hernandez-Rojas, Mortenson, and Wales}}]{MRMW2001a}
\bibinfo{author}{\bibfnamefont{T.~F.} \bibnamefont{Middleton}},
  \bibinfo{author}{\bibfnamefont{J.}~\bibnamefont{Hernandez-Rojas}},
  \bibinfo{author}{\bibfnamefont{P.~N.} \bibnamefont{Mortenson}},
  \bibnamefont{and} \bibinfo{author}{\bibfnamefont{D.~J.} \bibnamefont{Wales}},
  \bibinfo{journal}{Phys. Rev. B} \textbf{\bibinfo{volume}{64}},
  \bibinfo{pages}{184201} (\bibinfo{year}{2001})\relax
\mciteBstWouldAddEndPuncttrue
\mciteSetBstMidEndSepPunct{\mcitedefaultmidpunct}
{\mcitedefaultendpunct}{\mcitedefaultseppunct}\relax
\EndOfBibitem
\bibitem[{\citenamefont{Amsler and Goedecker}(2010)}]{AG2010a}
\bibinfo{author}{\bibfnamefont{M.}~\bibnamefont{Amsler}} \bibnamefont{and}
  \bibinfo{author}{\bibfnamefont{S.}~\bibnamefont{Goedecker}},
  \bibinfo{journal}{J. Chem. Phys.} \textbf{\bibinfo{volume}{133}},
  \bibinfo{pages}{224104} (\bibinfo{year}{2010})\relax
\mciteBstWouldAddEndPuncttrue
\mciteSetBstMidEndSepPunct{\mcitedefaultmidpunct}
{\mcitedefaultendpunct}{\mcitedefaultseppunct}\relax
\EndOfBibitem
\bibitem[{\citenamefont{Torquato}(2002)}]{TorquatoRHM2002}
\bibinfo{author}{\bibfnamefont{S.}~\bibnamefont{Torquato}},
  \emph{\bibinfo{title}{Random Heterogeneous Materials}}
  (\bibinfo{publisher}{Springer-Verlag}, \bibinfo{address}{New York},
  \bibinfo{year}{2002})\relax
\mciteBstWouldAddEndPuncttrue
\mciteSetBstMidEndSepPunct{\mcitedefaultmidpunct}
{\mcitedefaultendpunct}{\mcitedefaultseppunct}\relax
\EndOfBibitem
\bibitem[{\citenamefont{Chaikin and Lubensky}(1995)}]{CLPCMP1995}
\bibinfo{author}{\bibfnamefont{P.~M.} \bibnamefont{Chaikin}} \bibnamefont{and}
  \bibinfo{author}{\bibfnamefont{T.~C.} \bibnamefont{Lubensky}},
  \emph{\bibinfo{title}{Principles of Condensed Matter Physics}}
  (\bibinfo{publisher}{Cambridge University Press},
  \bibinfo{address}{Cambridge}, \bibinfo{year}{1995})\relax
\mciteBstWouldAddEndPuncttrue
\mciteSetBstMidEndSepPunct{\mcitedefaultmidpunct}
{\mcitedefaultendpunct}{\mcitedefaultseppunct}\relax
\EndOfBibitem
\bibitem[{\citenamefont{Hansen and McDonald}(2006)}]{HMTSL2006}
\bibinfo{author}{\bibfnamefont{J.~P.} \bibnamefont{Hansen}} \bibnamefont{and}
  \bibinfo{author}{\bibfnamefont{I.~R.} \bibnamefont{McDonald}},
  \emph{\bibinfo{title}{Theory of Simple Liquids, 3rd ed.}}
  (\bibinfo{publisher}{Academic Press}, \bibinfo{address}{Amsterdam},
  \bibinfo{year}{2006})\relax
\mciteBstWouldAddEndPuncttrue
\mciteSetBstMidEndSepPunct{\mcitedefaultmidpunct}
{\mcitedefaultendpunct}{\mcitedefaultseppunct}\relax
\EndOfBibitem
\bibitem[{\citenamefont{Zallen}(1983)}]{ZallenPAS1983}
\bibinfo{author}{\bibfnamefont{R.}~\bibnamefont{Zallen}},
  \emph{\bibinfo{title}{The Physics of Amorphous Solids}}
  (\bibinfo{publisher}{John Wiley and Sons}, \bibinfo{address}{New York},
  \bibinfo{year}{1983})\relax
\mciteBstWouldAddEndPuncttrue
\mciteSetBstMidEndSepPunct{\mcitedefaultmidpunct}
{\mcitedefaultendpunct}{\mcitedefaultseppunct}\relax
\EndOfBibitem
\bibitem[{\citenamefont{Hopkins et~al.}(2009)\citenamefont{Hopkins, Stillinger,
  and Torquato}}]{HST2009a}
\bibinfo{author}{\bibfnamefont{A.~B.} \bibnamefont{Hopkins}},
  \bibinfo{author}{\bibfnamefont{F.~H.} \bibnamefont{Stillinger}},
  \bibnamefont{and} \bibinfo{author}{\bibfnamefont{S.}~\bibnamefont{Torquato}},
  \bibinfo{journal}{Phys. Rev. E} \textbf{\bibinfo{volume}{79}},
  \bibinfo{pages}{031123} (\bibinfo{year}{2009})\relax
\mciteBstWouldAddEndPuncttrue
\mciteSetBstMidEndSepPunct{\mcitedefaultmidpunct}
{\mcitedefaultendpunct}{\mcitedefaultseppunct}\relax
\EndOfBibitem
\bibitem[{\citenamefont{Hopkins
  et~al.}(2010{\natexlab{a}})\citenamefont{Hopkins, Stillinger, and
  Torquato}}]{HST2010a}
\bibinfo{author}{\bibfnamefont{A.~B.} \bibnamefont{Hopkins}},
  \bibinfo{author}{\bibfnamefont{F.~H.} \bibnamefont{Stillinger}},
  \bibnamefont{and} \bibinfo{author}{\bibfnamefont{S.}~\bibnamefont{Torquato}},
  \bibinfo{journal}{J. Math. Phys.} \textbf{\bibinfo{volume}{51}},
  \bibinfo{pages}{043302} (\bibinfo{year}{2010}{\natexlab{a}})\relax
\mciteBstWouldAddEndPuncttrue
\mciteSetBstMidEndSepPunct{\mcitedefaultmidpunct}
{\mcitedefaultendpunct}{\mcitedefaultseppunct}\relax
\EndOfBibitem
\bibitem[{\citenamefont{Hopkins
  et~al.}(2010{\natexlab{b}})\citenamefont{Hopkins, Stillinger, and
  Torquato}}]{HST2010b}
  \bibinfo{journal}{Phys. Rev. E} \textbf{\bibinfo{volume}{81}},
  \bibinfo{pages}{041305} (\bibinfo{year}{2010}{\natexlab{b}})\relax
\mciteBstWouldAddEndPuncttrue
\mciteSetBstMidEndSepPunct{\mcitedefaultmidpunct}
{\mcitedefaultendpunct}{\mcitedefaultseppunct}\relax
\EndOfBibitem
\bibitem[{\citenamefont{Hopkins
  et~al.}(2011{\natexlab{a}})\citenamefont{Hopkins, Stillinger, and
  Torquato}}]{HST2011a}
  \bibinfo{journal}{Phys. Rev. E} \textbf{\bibinfo{volume}{83}},
  \bibinfo{pages}{011304} (\bibinfo{year}{2011}{\natexlab{a}})\relax
\mciteBstWouldAddEndPuncttrue
\mciteSetBstMidEndSepPunct{\mcitedefaultmidpunct}
{\mcitedefaultendpunct}{\mcitedefaultseppunct}\relax
\EndOfBibitem
\bibitem[{\citenamefont{Donev et~al.}(2004)\citenamefont{Donev, Stillinger,
  Chaikin, and Torquato}}]{DSCT2004a}
\bibinfo{author}{\bibfnamefont{A.}~\bibnamefont{Donev}},
  \bibinfo{author}{\bibfnamefont{F.~H.} \bibnamefont{Stillinger}},
  \bibinfo{author}{\bibfnamefont{P.}~\bibnamefont{Chaikin}}, \bibnamefont{and}
  \bibinfo{author}{\bibfnamefont{S.}~\bibnamefont{Torquato}},
  \bibinfo{journal}{Phys. Rev. Lett.} \textbf{\bibinfo{volume}{92}},
  \bibinfo{pages}{255506} (\bibinfo{year}{2004})\relax
\mciteBstWouldAddEndPuncttrue
\mciteSetBstMidEndSepPunct{\mcitedefaultmidpunct}
{\mcitedefaultendpunct}{\mcitedefaultseppunct}\relax
\EndOfBibitem
\bibitem[{\citenamefont{Jiao et~al.}(2009)\citenamefont{Jiao, Stillinger, and
  Torquato}}]{JST2009a}
\bibinfo{author}{\bibfnamefont{Y.}~\bibnamefont{Jiao}},
  \bibinfo{author}{\bibfnamefont{F.~H.} \bibnamefont{Stillinger}},
  \bibnamefont{and} \bibinfo{author}{\bibfnamefont{S.}~\bibnamefont{Torquato}},
  \bibinfo{journal}{Phys. Rev. E} \textbf{\bibinfo{volume}{79}},
  \bibinfo{pages}{041309} (\bibinfo{year}{2009})\relax
\mciteBstWouldAddEndPuncttrue
\mciteSetBstMidEndSepPunct{\mcitedefaultmidpunct}
{\mcitedefaultendpunct}{\mcitedefaultseppunct}\relax
\EndOfBibitem
\bibitem[{\citenamefont{Batten et~al.}(2010)\citenamefont{Batten, Stillinger,
  and Torquato}}]{BST2010a}
\bibinfo{author}{\bibfnamefont{R.~D.} \bibnamefont{Batten}},
  \bibinfo{author}{\bibfnamefont{F.~H.} \bibnamefont{Stillinger}},
  \bibnamefont{and} \bibinfo{author}{\bibfnamefont{S.}~\bibnamefont{Torquato}},
  \bibinfo{journal}{Phys. Rev. E} \textbf{\bibinfo{volume}{81}},
  \bibinfo{pages}{061105} (\bibinfo{year}{2010})\relax
\mciteBstWouldAddEndPuncttrue
\mciteSetBstMidEndSepPunct{\mcitedefaultmidpunct}
{\mcitedefaultendpunct}{\mcitedefaultseppunct}\relax
\EndOfBibitem
\bibitem[{\citenamefont{Torquato and Jiao}(2009{\natexlab{a}})}]{TJ2009a}
\bibinfo{author}{\bibfnamefont{S.}~\bibnamefont{Torquato}} \bibnamefont{and}
  \bibinfo{author}{\bibfnamefont{Y.}~\bibnamefont{Jiao}},
  \bibinfo{journal}{Nature} \textbf{\bibinfo{volume}{460}},
  \bibinfo{pages}{876} (\bibinfo{year}{2009}{\natexlab{a}})\relax
\mciteBstWouldAddEndPuncttrue
\mciteSetBstMidEndSepPunct{\mcitedefaultmidpunct}
{\mcitedefaultendpunct}{\mcitedefaultseppunct}\relax
\EndOfBibitem
\bibitem[{\citenamefont{Torquato and Jiao}(2009{\natexlab{b}})}]{TJ2009b}
  \bibinfo{journal}{Phys. Rev E} \textbf{\bibinfo{volume}{80}},
  \bibinfo{pages}{041104} (\bibinfo{year}{2009}{\natexlab{b}})\relax
\mciteBstWouldAddEndPuncttrue
\mciteSetBstMidEndSepPunct{\mcitedefaultmidpunct}
{\mcitedefaultendpunct}{\mcitedefaultseppunct}\relax
\EndOfBibitem
\bibitem[{\citenamefont{Torquato and Jiao}(2010{\natexlab{a}})}]{TJ2010b}
  \bibinfo{journal}{Phys. Rev. E} \textbf{\bibinfo{volume}{81}},
  \bibinfo{pages}{041310} (\bibinfo{year}{2010}{\natexlab{a}})\relax
\mciteBstWouldAddEndPuncttrue
\mciteSetBstMidEndSepPunct{\mcitedefaultmidpunct}
{\mcitedefaultendpunct}{\mcitedefaultseppunct}\relax
\EndOfBibitem
\bibitem[{\citenamefont{Chen}(2008)}]{Chen2008a}
\bibinfo{author}{\bibfnamefont{E.}~\bibnamefont{Chen}}, \bibinfo{journal}{Disc.
  Comp. Geom.} \textbf{\bibinfo{volume}{40}}, \bibinfo{pages}{214}
  (\bibinfo{year}{2008})\relax
\mciteBstWouldAddEndPuncttrue
\mciteSetBstMidEndSepPunct{\mcitedefaultmidpunct}
{\mcitedefaultendpunct}{\mcitedefaultseppunct}\relax
\EndOfBibitem
\bibitem[{\citenamefont{Chen et~al.}(2010)\citenamefont{Chen, Engel, and
  Glotzer}}]{CEG2010a}
\bibinfo{author}{\bibfnamefont{E.}~\bibnamefont{Chen}},
  \bibinfo{author}{\bibfnamefont{M.}~\bibnamefont{Engel}}, \bibnamefont{and}
  \bibinfo{author}{\bibfnamefont{S.}~\bibnamefont{Glotzer}},
  \bibinfo{journal}{Disc. Comp. Geom.} \textbf{\bibinfo{volume}{44}},
  \bibinfo{pages}{253} (\bibinfo{year}{2010})\relax
\mciteBstWouldAddEndPuncttrue
\mciteSetBstMidEndSepPunct{\mcitedefaultmidpunct}
{\mcitedefaultendpunct}{\mcitedefaultseppunct}\relax
\EndOfBibitem
\bibitem[{\citenamefont{Kallus et~al.}(2010)\citenamefont{Kallus, Elser, and
  Gravel}}]{KEG2010a}
\bibinfo{author}{\bibfnamefont{Y.}~\bibnamefont{Kallus}},
  \bibinfo{author}{\bibfnamefont{V.}~\bibnamefont{Elser}}, \bibnamefont{and}
  \bibinfo{author}{\bibfnamefont{S.}~\bibnamefont{Gravel}},
  \bibinfo{journal}{Disc. Comp. Geom.} \textbf{\bibinfo{volume}{44}},
  \bibinfo{pages}{245} (\bibinfo{year}{2010})\relax
\mciteBstWouldAddEndPuncttrue
\mciteSetBstMidEndSepPunct{\mcitedefaultmidpunct}
{\mcitedefaultendpunct}{\mcitedefaultseppunct}\relax
\EndOfBibitem
\bibitem[{\citenamefont{de~Graaf et~al.}(2011)\citenamefont{de~Graaf, van Roij,
  and Dijkstra}}]{GRD2011a}
\bibinfo{author}{\bibfnamefont{J.}~\bibnamefont{de~Graaf}},
  \bibinfo{author}{\bibfnamefont{R.}~\bibnamefont{van Roij}}, \bibnamefont{and}
  \bibinfo{author}{\bibfnamefont{M.}~\bibnamefont{Dijkstra}},
  \bibinfo{journal}{Phys. Rev. Lett.} \textbf{\bibinfo{volume}{107}},
  \bibinfo{pages}{155501} (\bibinfo{year}{2011})\relax
\mciteBstWouldAddEndPuncttrue
\mciteSetBstMidEndSepPunct{\mcitedefaultmidpunct}
{\mcitedefaultendpunct}{\mcitedefaultseppunct}\relax
\EndOfBibitem
\bibitem[{\citenamefont{Jiao and Torquato}(2011)}]{JT2011a}
\bibinfo{author}{\bibfnamefont{Y.}~\bibnamefont{Jiao}} \bibnamefont{and}
  \bibinfo{author}{\bibfnamefont{S.}~\bibnamefont{Torquato}},
  \bibinfo{journal}{Phys. Rev. E} \textbf{\bibinfo{volume}{84}},
  \bibinfo{pages}{041309} (\bibinfo{year}{2011})\relax
\mciteBstWouldAddEndPuncttrue
\mciteSetBstMidEndSepPunct{\mcitedefaultmidpunct}
{\mcitedefaultendpunct}{\mcitedefaultseppunct}\relax
\EndOfBibitem
\bibitem[{\citenamefont{Ding and Dokholyan}(2005)}]{DD2005a}
\bibinfo{author}{\bibfnamefont{F.}~\bibnamefont{Ding}} \bibnamefont{and}
  \bibinfo{author}{\bibfnamefont{N.~V.} \bibnamefont{Dokholyan}},
  \bibinfo{journal}{Trends in Biotech.} \textbf{\bibinfo{volume}{23}},
  \bibinfo{pages}{450} (\bibinfo{year}{2005})\relax
\mciteBstWouldAddEndPuncttrue
\mciteSetBstMidEndSepPunct{\mcitedefaultmidpunct}
{\mcitedefaultendpunct}{\mcitedefaultseppunct}\relax
\EndOfBibitem
\bibitem[{\citenamefont{Dokholyan}(2006)}]{Dokholyan2006a}
\bibinfo{author}{\bibfnamefont{N.~V.} \bibnamefont{Dokholyan}},
  \bibinfo{journal}{Curr. Opin. Struct. Bio.} \textbf{\bibinfo{volume}{16}},
  \bibinfo{pages}{79} (\bibinfo{year}{2006})\relax
\mciteBstWouldAddEndPuncttrue
\mciteSetBstMidEndSepPunct{\mcitedefaultmidpunct}
{\mcitedefaultendpunct}{\mcitedefaultseppunct}\relax
\EndOfBibitem
\bibitem[{\citenamefont{Davis et~al.}(2007)\citenamefont{Davis, Nie, and
  Dokholyan}}]{DND2007a}
\bibinfo{author}{\bibfnamefont{C.~H.} \bibnamefont{Davis}},
  \bibinfo{author}{\bibfnamefont{H.}~\bibnamefont{Nie}}, \bibnamefont{and}
  \bibinfo{author}{\bibfnamefont{N.~V.} \bibnamefont{Dokholyan}},
  \bibinfo{journal}{Phys. Rev. E} \textbf{\bibinfo{volume}{75}},
  \bibinfo{pages}{051921} (\bibinfo{year}{2007})\relax
\mciteBstWouldAddEndPuncttrue
\mciteSetBstMidEndSepPunct{\mcitedefaultmidpunct}
{\mcitedefaultendpunct}{\mcitedefaultseppunct}\relax
\EndOfBibitem
\bibitem[{\citenamefont{Drasdo and H\"{o}hme}(2005)}]{DH2005a}
\bibinfo{author}{\bibfnamefont{D.}~\bibnamefont{Drasdo}} \bibnamefont{and}
  \bibinfo{author}{\bibfnamefont{S.}~\bibnamefont{H\"{o}hme}},
  \bibinfo{journal}{Phys. Bio.} \textbf{\bibinfo{volume}{2}},
  \bibinfo{pages}{133} (\bibinfo{year}{2005})\relax
\mciteBstWouldAddEndPuncttrue
\mciteSetBstMidEndSepPunct{\mcitedefaultmidpunct}
{\mcitedefaultendpunct}{\mcitedefaultseppunct}\relax
\EndOfBibitem
\bibitem[{\citenamefont{Gevertz et~al.}(2008)\citenamefont{Gevertz, Gillies,
  and Torquato}}]{GGT2008a}
\bibinfo{author}{\bibfnamefont{J.~L.} \bibnamefont{Gevertz}},
  \bibinfo{author}{\bibfnamefont{G.}~\bibnamefont{Gillies}}, \bibnamefont{and}
  \bibinfo{author}{\bibfnamefont{S.}~\bibnamefont{Torquato}},
  \bibinfo{journal}{Phys. Bio.} \textbf{\bibinfo{volume}{5}},
  \bibinfo{pages}{036010} (\bibinfo{year}{2008})\relax
\mciteBstWouldAddEndPuncttrue
\mciteSetBstMidEndSepPunct{\mcitedefaultmidpunct}
{\mcitedefaultendpunct}{\mcitedefaultseppunct}\relax
\EndOfBibitem
\bibitem[{\citenamefont{Gevertz and Torquato}(2008)}]{GT2008a}
\bibinfo{author}{\bibfnamefont{J.~L.} \bibnamefont{Gevertz}} \bibnamefont{and}
  \bibinfo{author}{\bibfnamefont{S.}~\bibnamefont{Torquato}},
  \bibinfo{journal}{PLoS Comp. Bio.} \textbf{\bibinfo{volume}{4}},
  \bibinfo{pages}{e1000152} (\bibinfo{year}{2008})\relax
\mciteBstWouldAddEndPuncttrue
\mciteSetBstMidEndSepPunct{\mcitedefaultmidpunct}
{\mcitedefaultendpunct}{\mcitedefaultseppunct}\relax
\EndOfBibitem
\bibitem[{\citenamefont{Knott et~al.}(2001)\citenamefont{Knott, Jackson, and
  Buckmaster}}]{KJB2001a}
\bibinfo{author}{\bibfnamefont{G.~M.} \bibnamefont{Knott}},
  \bibinfo{author}{\bibfnamefont{T.~L.} \bibnamefont{Jackson}},
  \bibnamefont{and}
  \bibinfo{author}{\bibfnamefont{J.}~\bibnamefont{Buckmaster}},
  \bibinfo{journal}{AIAA J.} \textbf{\bibinfo{volume}{39}},
  \bibinfo{pages}{678} (\bibinfo{year}{2001})\relax
\mciteBstWouldAddEndPuncttrue
\mciteSetBstMidEndSepPunct{\mcitedefaultmidpunct}
{\mcitedefaultendpunct}{\mcitedefaultseppunct}\relax
\EndOfBibitem
\bibitem[{\citenamefont{Kochevets et~al.}(2001)\citenamefont{Kochevets,
  Buckmaster, Jackson, and Hegab}}]{KBJH2001a}
\bibinfo{author}{\bibfnamefont{S.}~\bibnamefont{Kochevets}},
  \bibinfo{author}{\bibfnamefont{J.}~\bibnamefont{Buckmaster}},
  \bibinfo{author}{\bibfnamefont{T.~L.} \bibnamefont{Jackson}},
  \bibnamefont{and} \bibinfo{author}{\bibfnamefont{A.}~\bibnamefont{Hegab}},
  \bibinfo{journal}{J. Prop. Pow.} \textbf{\bibinfo{volume}{17}},
  \bibinfo{pages}{883} (\bibinfo{year}{2001})\relax
\mciteBstWouldAddEndPuncttrue
\mciteSetBstMidEndSepPunct{\mcitedefaultmidpunct}
{\mcitedefaultendpunct}{\mcitedefaultseppunct}\relax
\EndOfBibitem
\bibitem[{\citenamefont{Maggi et~al.}(2008)\citenamefont{Maggi, Stafford,
  Jackson, and Buckmaster}}]{MSJ2008a}
\bibinfo{author}{\bibfnamefont{F.}~\bibnamefont{Maggi}},
  \bibinfo{author}{\bibfnamefont{S.}~\bibnamefont{Stafford}},
  \bibinfo{author}{\bibfnamefont{T.~L.} \bibnamefont{Jackson}},
  \bibnamefont{and}
  \bibinfo{author}{\bibfnamefont{J.}~\bibnamefont{Buckmaster}},
  \bibinfo{journal}{Phys. Rev. E} \textbf{\bibinfo{volume}{77}},
  \bibinfo{pages}{046107} (\bibinfo{year}{2008})\relax
\mciteBstWouldAddEndPuncttrue
\mciteSetBstMidEndSepPunct{\mcitedefaultmidpunct}
{\mcitedefaultendpunct}{\mcitedefaultseppunct}\relax
\EndOfBibitem
\bibitem[{\citenamefont{Kolonko et~al.}(2010)\citenamefont{Kolonko, Raschdorf,
  and W\"{a}sch}}]{KRW2010a}
\bibinfo{author}{\bibfnamefont{M.}~\bibnamefont{Kolonko}},
  \bibinfo{author}{\bibfnamefont{S.}~\bibnamefont{Raschdorf}},
  \bibnamefont{and}
  \bibinfo{author}{\bibfnamefont{D.}~\bibnamefont{W\"{a}sch}},
  \bibinfo{journal}{Gran. Matt.} \textbf{\bibinfo{volume}{12}},
  \bibinfo{pages}{629} (\bibinfo{year}{2010})\relax
\mciteBstWouldAddEndPuncttrue
\mciteSetBstMidEndSepPunct{\mcitedefaultmidpunct}
{\mcitedefaultendpunct}{\mcitedefaultseppunct}\relax
\EndOfBibitem
\bibitem[{\citenamefont{Hume-Rothery et~al.}(1969)\citenamefont{Hume-Rothery,
  Smallman, and Haworth}}]{SMA1969}
\bibinfo{author}{\bibfnamefont{W.}~\bibnamefont{Hume-Rothery}},
  \bibinfo{author}{\bibfnamefont{R.~E.} \bibnamefont{Smallman}},
  \bibnamefont{and} \bibinfo{author}{\bibfnamefont{C.~W.}
  \bibnamefont{Haworth}}, \emph{\bibinfo{title}{The Structure of Metals and
  Alloys, 5th ed.}} (\bibinfo{publisher}{Metals and Metallurgy Trust},
  \bibinfo{address}{London}, \bibinfo{year}{1969})\relax
\mciteBstWouldAddEndPuncttrue
\mciteSetBstMidEndSepPunct{\mcitedefaultmidpunct}
{\mcitedefaultendpunct}{\mcitedefaultseppunct}\relax
\EndOfBibitem
\bibitem[{\citenamefont{Murray and Sanders}(1980)}]{MS1980a}
\bibinfo{author}{\bibfnamefont{M.~J.} \bibnamefont{Murray}} \bibnamefont{and}
  \bibinfo{author}{\bibfnamefont{J.~V.} \bibnamefont{Sanders}},
  \bibinfo{journal}{Phil. Mag. A} \textbf{\bibinfo{volume}{42}},
  \bibinfo{pages}{721} (\bibinfo{year}{1980})\relax
\mciteBstWouldAddEndPuncttrue
\mciteSetBstMidEndSepPunct{\mcitedefaultmidpunct}
{\mcitedefaultendpunct}{\mcitedefaultseppunct}\relax
\EndOfBibitem
\bibitem[{\citenamefont{Sikka et~al.}(1982)\citenamefont{Sikka, Vohra, and
  Chidambaram}}]{SVC1982a}
\bibinfo{author}{\bibfnamefont{S.~K.} \bibnamefont{Sikka}},
  \bibinfo{author}{\bibfnamefont{Y.~K.} \bibnamefont{Vohra}}, \bibnamefont{and}
  \bibinfo{author}{\bibfnamefont{R.}~\bibnamefont{Chidambaram}},
  \bibinfo{journal}{Prog. Mat. Sci.} \textbf{\bibinfo{volume}{27}},
  \bibinfo{pages}{245} (\bibinfo{year}{1982})\relax
\mciteBstWouldAddEndPuncttrue
\mciteSetBstMidEndSepPunct{\mcitedefaultmidpunct}
{\mcitedefaultendpunct}{\mcitedefaultseppunct}\relax
\EndOfBibitem
\bibitem[{\citenamefont{Denton and Ashcroft}(1990)}]{DA1990a}
\bibinfo{author}{\bibfnamefont{A.~R.} \bibnamefont{Denton}} \bibnamefont{and}
  \bibinfo{author}{\bibfnamefont{N.~W.} \bibnamefont{Ashcroft}},
  \bibinfo{journal}{Phys. Rev. A} \textbf{\bibinfo{volume}{42}},
  \bibinfo{pages}{7312} (\bibinfo{year}{1990})\relax
\mciteBstWouldAddEndPuncttrue
\mciteSetBstMidEndSepPunct{\mcitedefaultmidpunct}
{\mcitedefaultendpunct}{\mcitedefaultseppunct}\relax
\EndOfBibitem
\bibitem[{\citenamefont{Eldridge
  et~al.}(1993{\natexlab{a}})\citenamefont{Eldridge, Madden, and
  Frenkel}}]{EMF1993a}
\bibinfo{author}{\bibfnamefont{M.~D.} \bibnamefont{Eldridge}},
  \bibinfo{author}{\bibfnamefont{P.~A.} \bibnamefont{Madden}},
  \bibnamefont{and} \bibinfo{author}{\bibfnamefont{D.}~\bibnamefont{Frenkel}},
  \bibinfo{journal}{Nature} \textbf{\bibinfo{volume}{365}}, \bibinfo{pages}{35}
  (\bibinfo{year}{1993}{\natexlab{a}})\relax
\mciteBstWouldAddEndPuncttrue
\mciteSetBstMidEndSepPunct{\mcitedefaultmidpunct}
{\mcitedefaultendpunct}{\mcitedefaultseppunct}\relax
\EndOfBibitem
\bibitem[{\citenamefont{Eldridge
  et~al.}(1993{\natexlab{b}})\citenamefont{Eldridge, Madden, and
  Frenkel}}]{EMF1993b}
  \bibinfo{journal}{Mol. Phys.} \textbf{\bibinfo{volume}{79}},
  \bibinfo{pages}{105} (\bibinfo{year}{1993}{\natexlab{b}})\relax
\mciteBstWouldAddEndPuncttrue
\mciteSetBstMidEndSepPunct{\mcitedefaultmidpunct}
{\mcitedefaultendpunct}{\mcitedefaultseppunct}\relax
\EndOfBibitem
\bibitem[{\citenamefont{Eldridge
  et~al.}(1993{\natexlab{c}})\citenamefont{Eldridge, Madden, and
  Frenkel}}]{EMF1993c}
  \bibinfo{journal}{Mol. Phys.} \textbf{\bibinfo{volume}{80}},
  \bibinfo{pages}{987} (\bibinfo{year}{1993}{\natexlab{c}})\relax
\mciteBstWouldAddEndPuncttrue
\mciteSetBstMidEndSepPunct{\mcitedefaultmidpunct}
{\mcitedefaultendpunct}{\mcitedefaultseppunct}\relax
\EndOfBibitem
\bibitem[{\citenamefont{Cottin and Monson}(1993)}]{CM1993a}
\bibinfo{author}{\bibfnamefont{X.}~\bibnamefont{Cottin}} \bibnamefont{and}
  \bibinfo{author}{\bibfnamefont{P.~A.} \bibnamefont{Monson}},
  \bibinfo{journal}{J. Chem. Phys.} \textbf{\bibinfo{volume}{99}},
  \bibinfo{pages}{8914} (\bibinfo{year}{1993})\relax
\mciteBstWouldAddEndPuncttrue
\mciteSetBstMidEndSepPunct{\mcitedefaultmidpunct}
{\mcitedefaultendpunct}{\mcitedefaultseppunct}\relax
\EndOfBibitem
\bibitem[{\citenamefont{Cottin and Monson}(1995)}]{CM1995a}
  \bibinfo{journal}{J. Chem. Phys.} \textbf{\bibinfo{volume}{102}},
  \bibinfo{pages}{3354} (\bibinfo{year}{1995})\relax
\mciteBstWouldAddEndPuncttrue
\mciteSetBstMidEndSepPunct{\mcitedefaultmidpunct}
{\mcitedefaultendpunct}{\mcitedefaultseppunct}\relax
\EndOfBibitem
\bibitem[{\citenamefont{Widom and Mihalkovic}(2005)}]{WM2005a}
\bibinfo{author}{\bibfnamefont{M.}~\bibnamefont{Widom}} \bibnamefont{and}
  \bibinfo{author}{\bibfnamefont{M.}~\bibnamefont{Mihalkovic}},
  \bibinfo{journal}{J. Mater. Res.} \textbf{\bibinfo{volume}{20}},
  \bibinfo{pages}{237} (\bibinfo{year}{2005})\relax
\mciteBstWouldAddEndPuncttrue
\mciteSetBstMidEndSepPunct{\mcitedefaultmidpunct}
{\mcitedefaultendpunct}{\mcitedefaultseppunct}\relax
\EndOfBibitem
\bibitem[{\citenamefont{Hopkins
  et~al.}(2011{\natexlab{b}})\citenamefont{Hopkins, Jiao, Stillinger, and
  Torquato}}]{HJST2011a}
\bibinfo{author}{\bibfnamefont{A.~B.} \bibnamefont{Hopkins}},
  \bibinfo{author}{\bibfnamefont{Y.}~\bibnamefont{Jiao}},
  \bibinfo{author}{\bibfnamefont{F.~H.} \bibnamefont{Stillinger}},
  \bibnamefont{and} \bibinfo{author}{\bibfnamefont{S.}~\bibnamefont{Torquato}},
  \bibinfo{journal}{Phys. Rev. Lett.} \textbf{\bibinfo{volume}{107}},
  \bibinfo{pages}{125501} (\bibinfo{year}{2011}{\natexlab{b}})\relax
\mciteBstWouldAddEndPuncttrue
\mciteSetBstMidEndSepPunct{\mcitedefaultmidpunct}
{\mcitedefaultendpunct}{\mcitedefaultseppunct}\relax
\EndOfBibitem
\bibitem[{\citenamefont{Torquato and Jiao}(2010{\natexlab{b}})}]{TJ2010a}
\bibinfo{author}{\bibfnamefont{S.}~\bibnamefont{Torquato}} \bibnamefont{and}
  \bibinfo{author}{\bibfnamefont{Y.}~\bibnamefont{Jiao}},
  \bibinfo{journal}{Phys. Rev. E} \textbf{\bibinfo{volume}{82}},
  \bibinfo{pages}{061302} (\bibinfo{year}{2010}{\natexlab{b}})\relax
\mciteBstWouldAddEndPuncttrue
\mciteSetBstMidEndSepPunct{\mcitedefaultmidpunct}
{\mcitedefaultendpunct}{\mcitedefaultseppunct}\relax
\EndOfBibitem
\bibitem[{\citenamefont{Szeto and Villain}(1987)}]{SV1987a}
\bibinfo{author}{\bibfnamefont{K.~Y.} \bibnamefont{Szeto}} \bibnamefont{and}
  \bibinfo{author}{\bibfnamefont{J.}~\bibnamefont{Villain}},
  \bibinfo{journal}{Phys. Rev. B} \textbf{\bibinfo{volume}{36}},
  \bibinfo{pages}{4715} (\bibinfo{year}{1987})\relax
\mciteBstWouldAddEndPuncttrue
\mciteSetBstMidEndSepPunct{\mcitedefaultmidpunct}
{\mcitedefaultendpunct}{\mcitedefaultseppunct}\relax
\EndOfBibitem
\bibitem[{\citenamefont{O'Toole and Hudson}(2011)}]{OH2011a}
\bibinfo{author}{\bibfnamefont{P.~I.} \bibnamefont{O'Toole}} \bibnamefont{and}
  \bibinfo{author}{\bibfnamefont{T.~S.} \bibnamefont{Hudson}},
  \bibinfo{journal}{J. Phys. Chem. C} \textbf{\bibinfo{volume}{115}},
  \bibinfo{pages}{19037} (\bibinfo{year}{2011})\relax
\mciteBstWouldAddEndPuncttrue
\mciteSetBstMidEndSepPunct{\mcitedefaultmidpunct}
{\mcitedefaultendpunct}{\mcitedefaultseppunct}\relax
\EndOfBibitem
\bibitem[{\citenamefont{Filion and Dijkstra}(2009)}]{FD2009a}
\bibinfo{author}{\bibfnamefont{L.}~\bibnamefont{Filion}} \bibnamefont{and}
  \bibinfo{author}{\bibfnamefont{M.}~\bibnamefont{Dijkstra}},
  \bibinfo{journal}{Phys. Rev. E} \textbf{\bibinfo{volume}{79}},
  \bibinfo{pages}{046714} (\bibinfo{year}{2009})\relax
\mciteBstWouldAddEndPuncttrue
\mciteSetBstMidEndSepPunct{\mcitedefaultmidpunct}
{\mcitedefaultendpunct}{\mcitedefaultseppunct}\relax
\EndOfBibitem
\bibitem[{\citenamefont{Likos and Henley}(1993)}]{LH1993a}
\bibinfo{author}{\bibfnamefont{C.~N.} \bibnamefont{Likos}} \bibnamefont{and}
  \bibinfo{author}{\bibfnamefont{C.~L.} \bibnamefont{Henley}},
  \bibinfo{journal}{Phil. Mag. B} \textbf{\bibinfo{volume}{68}},
  \bibinfo{pages}{85} (\bibinfo{year}{1993})\relax
\mciteBstWouldAddEndPuncttrue
\mciteSetBstMidEndSepPunct{\mcitedefaultmidpunct}
{\mcitedefaultendpunct}{\mcitedefaultseppunct}\relax
\EndOfBibitem
\bibitem[{\citenamefont{Marshall and Hudson}(2010)}]{MH2010a}
\bibinfo{author}{\bibfnamefont{G.~W.} \bibnamefont{Marshall}} \bibnamefont{and}
  \bibinfo{author}{\bibfnamefont{T.~S.} \bibnamefont{Hudson}},
  \bibinfo{journal}{Contr. Alg. Geo.} \textbf{\bibinfo{volume}{51}},
  \bibinfo{pages}{337} (\bibinfo{year}{2010})\relax
\mciteBstWouldAddEndPuncttrue
\mciteSetBstMidEndSepPunct{\mcitedefaultmidpunct}
{\mcitedefaultendpunct}{\mcitedefaultseppunct}\relax
\EndOfBibitem
\bibitem[{\citenamefont{Demchyna et~al.}(2006)\citenamefont{Demchyna, Leoni,
  Rosner, and Schwarz}}]{DLRS2006a}
\bibinfo{author}{\bibfnamefont{R.}~\bibnamefont{Demchyna}},
  \bibinfo{author}{\bibfnamefont{S.}~\bibnamefont{Leoni}},
  \bibinfo{author}{\bibfnamefont{H.}~\bibnamefont{Rosner}}, \bibnamefont{and}
  \bibinfo{author}{\bibfnamefont{U.}~\bibnamefont{Schwarz}},
  \bibinfo{journal}{Z. Kristallogr.} \textbf{\bibinfo{volume}{221}},
  \bibinfo{pages}{420} (\bibinfo{year}{2006})\relax
\mciteBstWouldAddEndPuncttrue
\mciteSetBstMidEndSepPunct{\mcitedefaultmidpunct}
{\mcitedefaultendpunct}{\mcitedefaultseppunct}\relax
\EndOfBibitem
\bibitem[{\citenamefont{Cazorla et~al.}(2009)\citenamefont{Cazorla, Errandonea,
  and Sola}}]{CES2009a}
\bibinfo{author}{\bibfnamefont{C.}~\bibnamefont{Cazorla}},
  \bibinfo{author}{\bibfnamefont{D.}~\bibnamefont{Errandonea}},
  \bibnamefont{and} \bibinfo{author}{\bibfnamefont{E.}~\bibnamefont{Sola}},
  \bibinfo{journal}{Phys. Rev. B} \textbf{\bibinfo{volume}{80}},
  \bibinfo{pages}{064105} (\bibinfo{year}{2009})\relax
\mciteBstWouldAddEndPuncttrue
\mciteSetBstMidEndSepPunct{\mcitedefaultmidpunct}
{\mcitedefaultendpunct}{\mcitedefaultseppunct}\relax
\EndOfBibitem
\bibitem[{\citenamefont{Degtyareva}(2005)}]{Degtyareva2005a}
\bibinfo{author}{\bibfnamefont{V.}~\bibnamefont{Degtyareva}},
  \bibinfo{journal}{J. Synchrotron Rad.} \textbf{\bibinfo{volume}{12}},
  \bibinfo{pages}{584} (\bibinfo{year}{2005})\relax
\mciteBstWouldAddEndPuncttrue
\mciteSetBstMidEndSepPunct{\mcitedefaultmidpunct}
{\mcitedefaultendpunct}{\mcitedefaultseppunct}\relax
\EndOfBibitem
\bibitem[{\citenamefont{Hales}(2005)}]{Hales2005a}
\bibinfo{author}{\bibfnamefont{T.~C.} \bibnamefont{Hales}},
  \bibinfo{journal}{Ann. Math.} \textbf{\bibinfo{volume}{162}},
  \bibinfo{pages}{1065} (\bibinfo{year}{2005})\relax
\mciteBstWouldAddEndPuncttrue
\mciteSetBstMidEndSepPunct{\mcitedefaultmidpunct}
{\mcitedefaultendpunct}{\mcitedefaultseppunct}\relax
\EndOfBibitem
\bibitem[{\citenamefont{Barlow}(1883)}]{Barlow1883a}
\bibinfo{author}{\bibfnamefont{W.}~\bibnamefont{Barlow}},
  \bibinfo{journal}{Nature} \textbf{\bibinfo{volume}{29}}, \bibinfo{pages}{186}
  (\bibinfo{year}{1883})\relax
\mciteBstWouldAddEndPuncttrue
\mciteSetBstMidEndSepPunct{\mcitedefaultmidpunct}
{\mcitedefaultendpunct}{\mcitedefaultseppunct}\relax
\EndOfBibitem
\bibitem[{\citenamefont{Hudson and Harrowell}(2008)}]{HH2008a}
\bibinfo{author}{\bibfnamefont{T.~S.} \bibnamefont{Hudson}} \bibnamefont{and}
  \bibinfo{author}{\bibfnamefont{P.}~\bibnamefont{Harrowell}},
  \bibinfo{journal}{J. Phys. Chem. B} \textbf{\bibinfo{volume}{112}},
  \bibinfo{pages}{8139} (\bibinfo{year}{2008})\relax
\mciteBstWouldAddEndPuncttrue
\mciteSetBstMidEndSepPunct{\mcitedefaultmidpunct}
{\mcitedefaultendpunct}{\mcitedefaultseppunct}\relax
\EndOfBibitem
\bibitem[{\citenamefont{Kummerfeld et~al.}(2008)\citenamefont{Kummerfeld,
  Hudson, and Harrowell}}]{KHH2008a}
\bibinfo{author}{\bibfnamefont{J.~K.} \bibnamefont{Kummerfeld}},
  \bibinfo{author}{\bibfnamefont{T.~S.} \bibnamefont{Hudson}},
  \bibnamefont{and}
  \bibinfo{author}{\bibfnamefont{P.}~\bibnamefont{Harrowell}},
  \bibinfo{journal}{J. Phys. Chem. B} \textbf{\bibinfo{volume}{112}},
  \bibinfo{pages}{10773} (\bibinfo{year}{2008})\relax
\mciteBstWouldAddEndPuncttrue
\mciteSetBstMidEndSepPunct{\mcitedefaultmidpunct}
{\mcitedefaultendpunct}{\mcitedefaultseppunct}\relax
\EndOfBibitem
\bibitem[{\citenamefont{Filion et~al.}(2009)\citenamefont{Filion, Marechal, van
  Oorschot, Pelt, Smallenburg, and Dijkstra}}]{FMOPSD2009a}
\bibinfo{author}{\bibfnamefont{L.}~\bibnamefont{Filion}},
  \bibinfo{author}{\bibfnamefont{M.}~\bibnamefont{Marechal}},
  \bibinfo{author}{\bibfnamefont{B.}~\bibnamefont{van Oorschot}},
  \bibinfo{author}{\bibfnamefont{D.}~\bibnamefont{Pelt}},
  \bibinfo{author}{\bibfnamefont{F.}~\bibnamefont{Smallenburg}},
  \bibnamefont{and} \bibinfo{author}{\bibfnamefont{M.}~\bibnamefont{Dijkstra}},
  \bibinfo{journal}{Phys. Rev. Lett.} \textbf{\bibinfo{volume}{103}},
  \bibinfo{pages}{188302} (\bibinfo{year}{2009})\relax
\mciteBstWouldAddEndPuncttrue
\mciteSetBstMidEndSepPunct{\mcitedefaultmidpunct}
{\mcitedefaultendpunct}{\mcitedefaultseppunct}\relax
\EndOfBibitem
\bibitem[{\citenamefont{Bindi et~al.}(2009)\citenamefont{Bindi, Steinhardt,
  Yao, and Lu}}]{BSYL2009a}
\bibinfo{author}{\bibfnamefont{L.}~\bibnamefont{Bindi}},
  \bibinfo{author}{\bibfnamefont{P.~J.} \bibnamefont{Steinhardt}},
  \bibinfo{author}{\bibfnamefont{N.}~\bibnamefont{Yao}}, \bibnamefont{and}
  \bibinfo{author}{\bibfnamefont{P.~J.} \bibnamefont{Lu}},
  \bibinfo{journal}{Science} \textbf{\bibinfo{volume}{324}},
  \bibinfo{pages}{1306} (\bibinfo{year}{2009})\relax
\mciteBstWouldAddEndPuncttrue
\mciteSetBstMidEndSepPunct{\mcitedefaultmidpunct}
{\mcitedefaultendpunct}{\mcitedefaultseppunct}\relax
\EndOfBibitem
\bibitem[{\citenamefont{Levine and Steinhardt}(1986)}]{LS1986a}
\bibinfo{author}{\bibfnamefont{D.}~\bibnamefont{Levine}} \bibnamefont{and}
  \bibinfo{author}{\bibfnamefont{P.~J.} \bibnamefont{Steinhardt}},
  \bibinfo{journal}{Phys. Rev. B} \textbf{\bibinfo{volume}{34}},
  \bibinfo{pages}{596} (\bibinfo{year}{1986})\relax
\mciteBstWouldAddEndPuncttrue
\mciteSetBstMidEndSepPunct{\mcitedefaultmidpunct}
{\mcitedefaultendpunct}{\mcitedefaultseppunct}\relax
\EndOfBibitem
\bibitem[{\citenamefont{Toth}(1964)}]{TothRF1964}
\bibinfo{author}{\bibfnamefont{L.~F.} \bibnamefont{Toth}},
  \emph{\bibinfo{title}{Regular Figures}} (\bibinfo{publisher}{Macmillan},
  \bibinfo{address}{New York}, \bibinfo{year}{1964})\relax
\mciteBstWouldAddEndPuncttrue
\mciteSetBstMidEndSepPunct{\mcitedefaultmidpunct}
{\mcitedefaultendpunct}{\mcitedefaultseppunct}\relax
\EndOfBibitem
\bibitem[{\citenamefont{Leung et~al.}(1989)\citenamefont{Leung, Henley, and
  Chester}}]{LHC1989a}
\bibinfo{author}{\bibfnamefont{P.~W.} \bibnamefont{Leung}},
  \bibinfo{author}{\bibfnamefont{C.~L.} \bibnamefont{Henley}},
  \bibnamefont{and} \bibinfo{author}{\bibfnamefont{G.~V.}
  \bibnamefont{Chester}}, \bibinfo{journal}{Phys. Rev. B}
  \textbf{\bibinfo{volume}{39}}, \bibinfo{pages}{446}
  (\bibinfo{year}{1989})\relax
\mciteBstWouldAddEndPuncttrue
\mciteSetBstMidEndSepPunct{\mcitedefaultmidpunct}
{\mcitedefaultendpunct}{\mcitedefaultseppunct}\relax
\EndOfBibitem
\bibitem[{\citenamefont{Widom}(1993)}]{Widom1993a}
\bibinfo{author}{\bibfnamefont{M.}~\bibnamefont{Widom}},
  \bibinfo{journal}{Phys. Rev. Lett.} \textbf{\bibinfo{volume}{70}},
  \bibinfo{pages}{2094} (\bibinfo{year}{1993})\relax
\mciteBstWouldAddEndPuncttrue
\mciteSetBstMidEndSepPunct{\mcitedefaultmidpunct}
{\mcitedefaultendpunct}{\mcitedefaultseppunct}\relax
\EndOfBibitem
\bibitem[{\citenamefont{Torquato et~al.}(2000)\citenamefont{Torquato, Truskett,
  and Debenedetti}}]{TTD2000a}
\bibinfo{author}{\bibfnamefont{S.}~\bibnamefont{Torquato}},
  \bibinfo{author}{\bibfnamefont{T.~M.} \bibnamefont{Truskett}},
  \bibnamefont{and} \bibinfo{author}{\bibfnamefont{P.~G.}
  \bibnamefont{Debenedetti}}, \bibinfo{journal}{Phys. Rev. Lett.}
  \textbf{\bibinfo{volume}{84}}, \bibinfo{pages}{2064}
  (\bibinfo{year}{2000})\relax
\mciteBstWouldAddEndPuncttrue
\mciteSetBstMidEndSepPunct{\mcitedefaultmidpunct}
{\mcitedefaultendpunct}{\mcitedefaultseppunct}\relax
\EndOfBibitem
\bibitem[{\citenamefont{Torquato and Stillinger}(2001)}]{TS2001a}
\bibinfo{author}{\bibfnamefont{S.}~\bibnamefont{Torquato}} \bibnamefont{and}
  \bibinfo{author}{\bibfnamefont{F.~H.} \bibnamefont{Stillinger}},
  \bibinfo{journal}{J. Phys. Chem B} \textbf{\bibinfo{volume}{105}},
  \bibinfo{pages}{11849} (\bibinfo{year}{2001})\relax
\mciteBstWouldAddEndPuncttrue
\mciteSetBstMidEndSepPunct{\mcitedefaultmidpunct}
{\mcitedefaultendpunct}{\mcitedefaultseppunct}\relax
\EndOfBibitem
\bibitem[{\citenamefont{Filion et~al.}(2011)\citenamefont{Filion, Hermes, Ni,
  Vermolen, Kuijk, Christova, Stiefelhagen, Vissers, van Blaaderen, and
  Dijkstra}}]{FHNVKCSVBD2011a}
\bibinfo{author}{\bibfnamefont{L.}~\bibnamefont{Filion}},
  \bibinfo{author}{\bibfnamefont{M.}~\bibnamefont{Hermes}},
  \bibinfo{author}{\bibfnamefont{R.}~\bibnamefont{Ni}},
  \bibinfo{author}{\bibfnamefont{E.~C.~M.} \bibnamefont{Vermolen}},
  \bibinfo{author}{\bibfnamefont{A.}~\bibnamefont{Kuijk}},
  \bibinfo{author}{\bibfnamefont{C.~G.} \bibnamefont{Christova}},
  \bibinfo{author}{\bibfnamefont{J.~C.~P.} \bibnamefont{Stiefelhagen}},
  \bibinfo{author}{\bibfnamefont{T.}~\bibnamefont{Vissers}},
  \bibinfo{author}{\bibfnamefont{A.}~\bibnamefont{van Blaaderen}},
  \bibnamefont{and} \bibinfo{author}{\bibfnamefont{M.}~\bibnamefont{Dijkstra}},
  \bibinfo{journal}{Phys. Rev. Lett.} \textbf{\bibinfo{volume}{107}},
  \bibinfo{pages}{168302} (\bibinfo{year}{2011})\relax
\mciteBstWouldAddEndPuncttrue
\mciteSetBstMidEndSepPunct{\mcitedefaultmidpunct}
{\mcitedefaultendpunct}{\mcitedefaultseppunct}\relax
\EndOfBibitem
\bibitem[{\citenamefont{{\it et al.}}(2011)}]{Zeng2011a}
\bibinfo{author}{\bibfnamefont{Q.~Zeng} \bibnamefont{{\it et al.}}},
  \bibinfo{journal}{Science} \textbf{\bibinfo{volume}{332}},
  \bibinfo{pages}{1404} (\bibinfo{year}{2011})\relax
\mciteBstWouldAddEndPuncttrue
\mciteSetBstMidEndSepPunct{\mcitedefaultmidpunct}
{\mcitedefaultendpunct}{\mcitedefaultseppunct}\relax
\EndOfBibitem
\end{mcitethebibliography}

\end{document}